# Artificial Intelligence and Advanced Materials

## *Cefe López\**

*Instituto de Ciencia de Materiales de Madrid, (ICMM) calle Sor Juana Inés de la Cruz 3, 28049 Madrid, Spain. Consejo Superior de Investigaciones Científicas (CSIC) Email: c.lopez@csic.es*
*Also, Donostia International Physics Centre (DIPC), Paseo Manuel de Lardizábal 4, 20018 San Sebastián, España.*

Artificial intelligence is gaining strength and materials science can both contribute to and profit from it. In a simultaneous progress race, new materials, systems and processes can be devised and optimized thanks to machine learning techniques and such progress can be turned into innovative computing platforms. Future materials scientists will profit from understanding how machine learning can boost the conception of advanced materials. This review covers aspects of computation from the fundamentals to directions taken and repercussions produced by computation to account for the origins, procedures and applications of artificial intelligence. Machine learning and its methods are reviewed to provide basic knowledge on its implementation and its potential. The materials and systems used to implement artificial intelligence with electric charges are finding serious competition from other information carrying and processing agents. The impact these techniques are having on the inception of new advanced materials is so deep that a new paradigm is developing where implicit knowledge is being mined to conceive materials and systems for functions instead of finding applications to found materials. How far this trend can be carried is hard to fathom as exemplified by the power to discover unheard of materials or physical laws buried in data.

## 1. Introduction

In this section, we describe reasons why AI is a worthy topic and how it links with materials science. Then we describe the organization of the paper.

From as far as ancient history there are many accounts of devices that could be *prepared* to act as animated creatures or even resembling humans. Technical advances in the industrial age permitted practical applications to be realized.

The year 2021 commemorated the birth of *robot*, a Czech word of Slavic origin meaning *work* but with emphatic connotation of servitude (See **Fig. 1**). It bears a notion of artificial worker, conceived by the brothers Čapek for a play exploring the potentially negative influence of technological civilization on society. Čapek's allegory has been revisited in many fashions and with the foundational contribution by Isaac Asimov populated the culture and imagination of many generations.

Underlying this is the notion of *automaton* and on an advanced level that of artificial intelligence: that capacity to build complex ideas from simpler ones and determine life preserving behaviours by favourably handling the environment that humans possess and wish to instil in machines.

When questioned, many scientists choose *understanding how the brain works* as the ultimate goal for 21st century science. This is in close relation with the prevailing consensus that this century must see science address the immediate challenge of biocontrol.

### 1.1. Big data

Although the concept of *big data* —large sets of data that escape traditional data handling systems— is relatively recent, the generation of massive amounts of data can be traced to last century sixties and seventies when data centres and big relational databases were starting. A clear-cut definition is not available but most experts will agree to a threshold between some tens of terabytes and several petabytes. The complex nature of big data arises principally from its heterogeneity containing audio, image, text and raw data in an endless number of configurations.

In the last decades social networks contributed to a truly massive accumulation of data and to new frameworks to deal with its storage and indexing. The arrival of the internet of things and cloud computing have made this data generation rocket. The world per capita storing capacity has doubled every 40 months for last four decades.[1]

Data production is growing and its nature is becoming more heterogeneous, unstructured, so its management is getting out of hand; acquiring, storing, analysing, and sharing data are growing more difficult with ordinary means and the extraction of the relevant information from massive amounts of data more power hungry.

For all the foregoing and other reasons new computing methods and technologies are gathering renewed interest. However, apart from the limitations of technology— as the realization of conceptual contraptions— physics imposes limits to what can be achieved in terms of computing power, information storage, communication etc.

### 1.2. Computation limits

Computation in an abstract sense entails manipulating available data to extract hidden or implicit information about the system it refers to; usually in science, data from a system's current state is used to obtain the system's future state. Computation therefore depends on how much and under what shape information is available and

Prof. C. López, Instituto de Ciencia de Materiales de Madrid, (ICMM) calle Sor Juana Inés de la Cruz 3, 28049 Madrid, Spain. Consejo Superior de Investigaciones Científicas (CSIC)
Email: c.lopez@csic.es
Also, Donostia International Physics Centre (DIPC), Paseo Manuel de Lardizábal 4, 20018 San Sebastián, España.

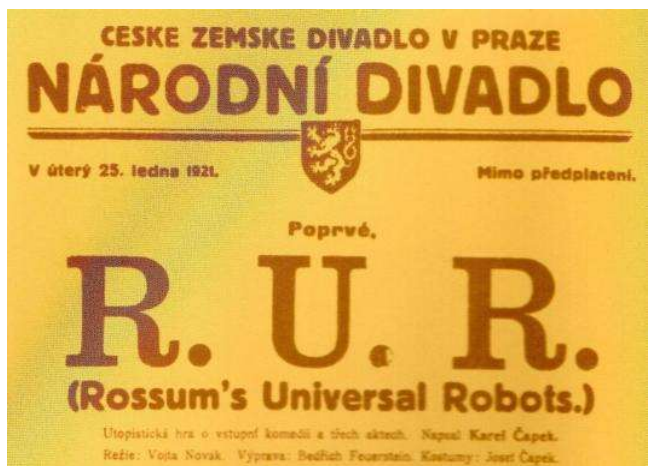

**Figure 1.** Robot, the most famous Czeck, celebrated his first Centenary. Adapted from https://www.mzv.cz/madrid/es/cultura_becas_cursos/cultura/robot_el_checo_mas_famoso_celebra_su.html

how efficiently it can be accessed and manipulated. All the processes involved are subject to physical bounds and limitations, some of which can be estimated regardless of the actual computing systems employed.

### 1.2.1. Information storage

The Bekenstein bound[2] limits the amount of information, $I$ —or entropy, $S$— that can be *stored* within a volume of size ~$R^3$ which has a finite amount of energy, $E$; or, conversely, the maximal amount of information required to perfectly describe a given physical system down to the quantum level. It can be inferred from the second law of thermodynamics and is expressed as $S/E = 2R/\hbar c$. Thermodynamics limits the data storage of a system based on its energy, number of particles and particle modes. In practice, it is a stronger bound than Bekenstein's.

### 1.2.2. Information transmission

Transmitting information costs energy because messages need a material embodiment. Answering in the negative the question whether the transmission cost can be arbitrarily lowered is already suggested by the Shannon's classic bound to how much information a communication channel of given bandwidth can transmit with a given signal to noise ratio.[3]

Indeed, by considering the maximum amount of entropy, hence information, and how it can be transmitted at a velocity smaller than $c$, one can arrive at $dI/dt < \pi E/h \ln 2$ as the maximum average rate of information transfer in bits per second for a given energy.[4]

### 1.2.3. Processing speed

Bremermann argued that there are three hurdles that limit computation speed: light, quantum[5] and thermodynamic barriers. The first one limits computation to the size of the computer as information cannot be transferred between parts of the computer at speeds higher that light's. The last one imposes limits to the effective use of energy since, owing to the second law of thermodynamics, heat must be generated at a rate of $k_B T \ln 2$ per bit.

The Bremermann's limit is the maximum processing speed of any material computing system in bits per second per gram, and is based on mass-energy equivalence ($E = m c^2$) and uncertainty principle ($\Delta E \Delta t > h$) constraints.[6] It limits speed to $c^2/h \sim 2.3 \times 10^{50}$ b/s kg. Incidentally, this implies that, if the whole mass of the universe were dedicated to computing, no more than $7 \times 10^{103}$ bits per year could be processed; a figure to be compared with $10^{120}$ possible move sequences in the game of chess.[7] Further discussion and references can be obtained in [8].

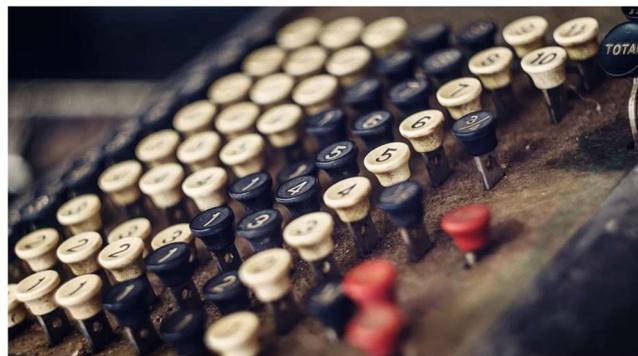

**Figure 2.** | A mechanical cash register computes at the expense of spring stored power drawn from the user hitting keys and pulling levers. It dissipates a fair amount of it in wear and friction.

### 1.2.4. Communication lag

The Margolus–Levitin theorem sets a bound to how fast a computing device can perform per unit of energy.

The result of a computation must involve a change from one state to a different one (or an orthogonal one for a quantum computing system). Thus, a measure of speed is obtained by establishing the maximum number of distinct states the system can pass through per unit time. This would be the maximum number of operations per second for a classical computer. In a system whose highest eigenstate has energy $E$ the maximum frequency at which states can change is expected to be bounded by $\nu \leq E/h$ or twice that if $E$ is the average energy.[9] This gives some $6 \times 10^{33}$ operations per second per joule. This bound, however, can be avoided if there is access to quantum memory. Computational algorithms can then be designed that require arbitrarily small amounts of energy/time per one elementary computation step.

### 1.2.5. Energy supply

It has been argued that mathematics requires a basis in physics.[10] What is not arguable is that information always has a physical embodiment, whether electrons in a trap, spins in atoms, bases in a DNA strand, holes punched in a card of pencil strokes on a paper it is impossible to get away from its physical reality. Therefore, computing, regardless of the implementation (an abacus, a mechanical cash register as **Fig. 2**, a supercomputer or a brain) is a physical process in which energy is exchanged. It is then fair to ask how much energy it takes to make a given computation and how efficiently different computing systems are in comparison to one another.

According to physical principles the limits to computing capacity are far away from current performances. Just as Carnot found the limits for thermal efficiency, and Shannon for information transmission, Landauer and Swanson established the limits imposed by physical laws for computing. Dissipation is a cause of loss of information as in "velocity and position of a mass attached to a spring contain all the information needed to describe (past and future) motion unless there is friction". Computing intrinsically entails dissipation as information is lost when an answer (= 4) is substituted for the question (2 + 2) because there are more questions (0 + 4, 1 + 3) that produce that answer. Obviously, erasing bits destroys information too but, in general, any logic gate that has more input than output lines —irreversible gate— inevitably does the same inasmuch as we cannot deduce the input from the output.

A reversible computation scheme can be devised using Fredkin gates, that have equal number of input and output lines and restoring the device to its initial state after each computing cycle. It is however the actual implementation that renders any computing ir-reversible since any physical element will involve dissipation or information discarding for error correction (due to thermal noise or chaos in classical systems or tunnelling in quantum).

Landauer's principle defines a lower theoretical limit for energy consumption: $E \geq k_B T \ln 2$ consumed per irreversible state change, where $k_B$ is the Boltzmann constant and $T$ is the operating temperature of the computer.[11] If we admit that, without spending more energy in cooling than is saved in computing, $T$ cannot be made lower than 3 K, the approximate temperature of the cosmic microwave background radiation, we might expect to be capable to perform $\approx 10^{23}$ operations per Joule. Different strategies have been discussed regarding the universe exponential cooling rate and the ultimate amount of computing per energy unit.[12]

Apart from fundamental limits, traditional Turing machine computing is about reaching its technological limits using von Neumann architectures as Moore's law[13] limit is approached with logic gate size nearing atomic scale and thermal dissipation capacity blocking practical use. At the time of writing 5 nm technology is state of the art.

Energy economy in the brain operation[14] has often been discussed and used as a benchmark. Its performance is remarkable in that it carries out many learning and computing tasks simultaneously. Certain artificial neural networks (ANNs) that perform tasks similar to natural language could run training activities for under 80 hours with the energy budget a brain would require for 6 years' activity.[15] In the computer, energy is spent in the central processing unit, the graphical processing units and random memory access. In the brain, energy consumption in adenosine triphosphate molecules can be tracked down to $K^+$ and $Na^+$ ions movements across cell membranes in spikes (60 pJ) and synaptic events (26 fJ).

### 1.2.6.    Materials needed

Not only because, arguably, information can be considered a material[16] but for the plausible reason that computation must be carried out in physical computers, all the aspects discussed so far entail a materials involvement.

Ordinary processors that run on CMOS technology materials will still evolve and adapt to the new tendency oriented towards neuromorphic computing. But, at the same time, a new wave of materials and technologies will be developed is pursue of better adapted materials and systems that work more like brains do. Current and future materials that serve these proposes and the (mathematical and physical) principles underlying their operation are the subject of this review.

### 1.3.    Structure of the review

The preceding considerations alone should explain the incessant search for a new computing paradigm where efficiency and massive parallelism[a] be the dominant features.

The prevailing answer to that quest —dispensing with the singular nature of quantum computing— is to follow the natural inspiration of the brain's operation to the inception of artificial intelligence through neuromorphic computing. In order to better understand the characteristics and features of artificial intelligence, conventional computing is first examined.

This rest of this review is divided in three major parts dealing with the concept of artificial intelligence and its mathematical aspects (part 3); the main hardware implementations (part 4); and two sections devoted to the main applications related to materials science(parts 5 and 6). A brief outlook section is included.

The general materials scientist is not expected to be concerned with each and every topic covered and, owing to the length of the review it advised to read those sections closer to their field of interest.

## 2.    Computing Concepts

Computing is about solving problems that involve mathematical manipulation of numerical data. In this task, there are several aspects to consider. The main considerations regard how "hard" a problem is, what resources are available and how efficiently they can be deployed.

### 2.1.    Computability

The concept of *computability* and its categorization is behind the development of computer science. It often involves profound mathematical elements such as calculability and the notion that the ability to pose mathematical problems not always runs abreast with the possibility to solve them (in the sense that even the truth of a proposition involving elementary number theory may not be assessed). Even deeper, lay the foundations laid by Gödel about *improvability* in axiomatic theories.[17] Very soon Church identified *effective[b] calculability* of a function of natural numbers with the existence of an algorithm for the calculation of its values.[18]

In an attempt to generate particular problem-solving strategies for classes of specific problems, Post formulated a solution involving a *symbol space* (in the form of a of row of boxes numbered with ordered integers, that can be marked or unmarked) which, with a fixed *set of directions* permitted a *worker* to carry out the instructions and move along the space one box at a time and marking/unmarking the box.[19] In much the same spirit, Turing defined "computable numbers as those whose decimals can be written down by a machine".

### 2.2.    Turing machine

A Turing machine is a conceptual computing machine consisting of a *data storage* device (let it be a tape holding zeroes and ones) a reading/writing head that can be in any of a number of *states*, can (over)write data on the tape and can move one step to the left or right on the tape to execute next instruction. A computing step comprises reading the datum on the tape and, following a (finite) table of *rules*, depending on the state of the head, set a new head state and move along the tape. Simple as it may seem, a Turing machine can be programmed to execute any computation a modern computer can. This is not to say that an answer to any problem can always be computed.[20] On the contrary, Turing proved the so-called *halting problem*, that a general algorithm cannot exist capable to determine for any pair (instruction set, input data) if the computation reaches the end. These seemingly different approaches at computability reached independently are reconciled in Church's thesis that "computable by a Turing machine is equivalent to effectively calculable" whose all-important implication is that any computing device can be simulated by a Turing machine. This greatly simplifies the formal study of computation reducing the analysis of an, in principle, infinite number of devices to a single type.

---

[a] Massively parallel is the common term used to describe the operation of a large number of computer processors (like a GPU with thousands of threads or a processor array with hundreds of thousands of central processing units) or separate computers (in a grid or a cluster) that simultaneously perform coordinated computations in parallel. These processors exchange tasks and require specialized software to exploit the vantages of parallel operation.

[b] *Effective* meaning systematic, mechanical, not its current everyday sense, implies that the procedure leading to a desired result can be carried out following a set of instructions in a finite number of steps, without need for insight or other means than, say, paper and pencil. In fact, in the years these concepts were being developed, the term *computer* implicitly referred to *human computers*, people employed by businesses and governments to perform, using *effective* methods, the kind of tasks electronic computers do nowadays.

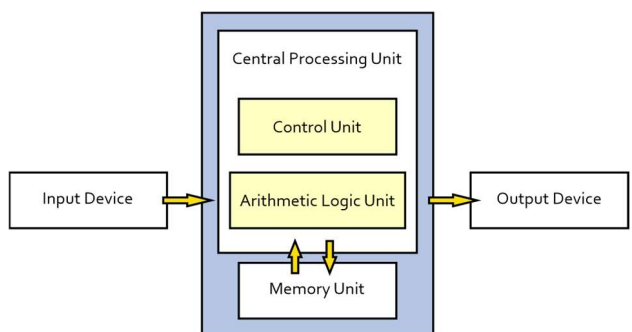

**Figure 3.** Von Neumann architecture.

Although the tape is, in principle, infinite the Turing machine evidenced a practical shortcoming modern computers solved resorting to random access memory (RAM).

## 2.3. Problem size

While the main concern of early mathematicians was about formal concepts such as absolute computability, the advent of practical devices made it necessary to describe the efficiency with which numbers or functions can be computed. In principle, the *size* of problems can be measured by the number of steps (or computing time), as a function of the input size, a good algorithm requires to arrive at the answer. For instance, determining if a number is prime is polynomial, finding all possible ways of ordering a set of $n$ elements takes $O(n!)$ steps while going through all subsets takes $O(2^n)$. The accepted reference is established at *polynomial* complexity where the computing time for an input of size $n$ is upper bounded by a polynomial of degree $n$. One of the reasons for this gauge is that if an algorithm makes a polynomial number of calls to routines that comprise polynomial numbers of steps, the complexity of the entire process is still polynomial. Because it has not been proved that true random numbers can be generated by an algorithm, for many applications a deterministic Turing machine is additionally equipped with a true random number generator.

The size of the task can be measured by the complexity of the problem to be solved. For this, different measures have proposed.

### 2.3.1. Algorithmic complexity

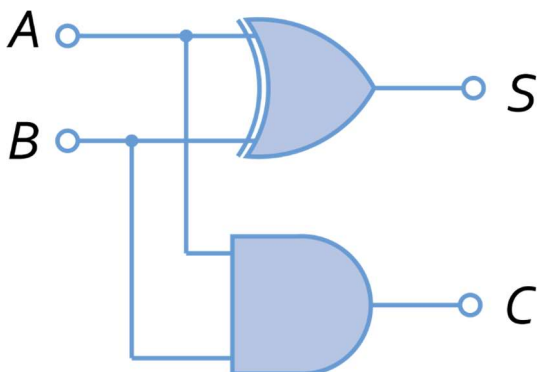

**Figure 4** . Addition hardware. A one-bit adder comprises one XOR and one AND gates. The former produces the sum and the second handles the carry.

Intrinsic measures of *complexity* have been issued in different contexts and, in particular, the accepted algorithmic complexity measure is Kolmogorov complexity which is the Shannon entropy[21] counterpart in information theory. Also known as Solomonoff-Kolmogorov-Chaitin,[22–24] this measure is given by the bit-length of the shortest computer program capable to generate a given sequence of bits and stop afterwards when running on a given machine. This length depends on the machine and the sequence but Kolmogorov proved that there exists a "universal" machine capable to carry out the same computation a program executes in a given machine but with a shorter, modified form of the program. The excess only length depends on the machine and as the length tends to infinity a *per unit* complexity is defined.[25]

### 2.3.2. Lyapunov exponents

The Lyapunov exponents were proposed in order to characterize sensitivity to initial conditions, $\delta x(0)$, in dynamical systems by describing the rate of separation of close trajectories, $x(t)$ in phase space as $|\delta x(t)| = e^{\lambda t}\delta x(0)$. In a chaotic (positive exponent) multidimensional system predictability can only be trusted up to a time related to the maximal exponent. In this way, chaotic dynamical systems may be seen as information sources that generate time series that cannot be arbitrarily compressed, that is, they are complex. These concepts surface in algorithmic complexity too when estimating the length of algorithms capable to produce sequences of symbols characterized by their Kolmogorov complexity.

## 2.4. The von Neumann architecture

The first lasting design of an electronic computer was that of stored-program computers (that keep data and instructions in memory) which constituted an advancement from the program-controlled computers of the 1940's (which were programmed by setting switches and connecting cables to route data between units).

### 2.4.1. Processor structure

Traditional computers as we know them nowadays are built on what's known as the conventional von Neumann architecture. The way in which electronic components should be assembled in a stored-program type of computer was outlined by John von Neumann in 1945 in an incomplete report, never properly published.[26] This report was the main result of the Electronic Discrete Variable Automatic Computer (EDVAC) project and contained the first description of the design of digital computers as envisaged then that became the underlying architecture till today. EDVAC was the *binary* sequel to earlier Electronic Numerical Integrator and Calculator (ENIAC) *decimal* project. For their speed of operation vacuum tubes (1 µs switching time) were used rather than relays (10 ms).

The von Neumann architecture, apart from input and output, comprises a single centralized control unit (responsible for deciding which instructions in a program should be executed) included in the central processing unit (CPU) and a separate storage section, the memory, where both data and instructions are kept. See **Fig. 3**. This memory is built such that any data register in it can be accessed without having to read all previous data as in a tape, hence its name: random access memory (RAM). Obviously, instructions are executed by the CPU therefore they must be brought there from the memory. The arithmetic and logic unit (ALU), in charge of operations (addition, multiplication, etc) and operands (comparison, Boolean, etc.), is also housed by the CPU so, again, data is brought there to be acted upon. The CPU has an additional section (register) where intermediate results and information about the state of an executing program can be stored and retrieved much faster than from memory.

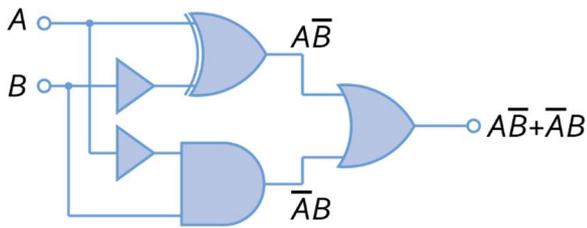

**Figure 5.** An XOR can be implemented with two AND, one OR and two NOT gates.

Instructions and data are thus physically separated from the computing units and transferred through a *bus* which consists of a collection of wires and some access control hardware. Instruction fetch and data operation must be sequential because they share a common bus. Thus, a von Neuman machine executes one instruction at a time. This separation is often called the von Neuman *bottleneck*. Nowadays CPUs can execute instructions more than one hundred times faster than they can fetch data from memory.

The Harvard architecture, at variance with von Neumann's, has two memory sections where instructions and data are kept separate.

### 2.4.2.   Bitwise logic; bitwise arithmetic

How hardware has come to be implemented can be understood as the result of how *logic* and *arithmetic* operations on bit words can be achieved with semiconductor devices called gates (the solid-state successors to vacuum valves).[a] These components are produced by the complementary metal-oxide semiconductor (CMOS) technology and the central element is the field effect *transistor* (MOSFET). In this prevailing technology, memory is based on charge stored in this kind of devices. In some cases, where leak current is null, no power is required to maintain the data as in conventional *flash memory*[27] whereas devices with larger leakage require a constant power supply as in random access memory (RAM).

The essence of how material devices operate can be grasped by examining a simple example of how to implement components in electronic chips that perform the addition of numbers expressed in base two. In order to sum binary numbers, one must follow the simple rule $0+0 = 0$, $0+1 = 1+0 = 1$ and $1+1=10$. All can be summarized by using the carry rule $A+B\rightarrow CS$ where $S$ is the result of the logic XOR binary operation $S = A$ XOR $B$ and $C$ the result of the logic AND binary operation $C = A$ AND $B$. See **Fig. 4**. In fact, this corresponds to the rule $0+0 = \underline{0}0$, $0+1 = 1+0 = \underline{0}1$ and $1+1=\underline{1}0$ where the carry bit (underscored) has been explicitly written.

Thus, in terms of gates, summation implies one AND gate (which requires both $A$ and $B$ being 1 to output 1; that is produces 1 only from two 1's) and one XOR gate (that requires either $A$ or $B$, but no both, to be 1; *e.i.* produces 1 only from two 0's) that produces [$A$ AND (NOT $B$)] OR  (NOT  $A$) AND  $B$]. See **Fig. 5**.

In order to implement an AND gate (See **Fig. 6a**) one can think of it as two switches in series (current would only flow if *both* are closed) and can be implemented with two diodes or transistors. Two MOSFET transistors in series act on the gate of a third one only when both are activated thus complying the AND condition. The only remaining component is a NOT gate. This can be implemented, as in **Fig. 6b**, with a single transistor so that when no voltage is applied at the gate ($A$ = 0) the output signal will be the supply voltage (1 = NOT $A$) whereas if a voltage is applied ($A$ = 0) the transistor with be shorted and current will flow producing a drop of voltage from the supply producing a zero in the output.

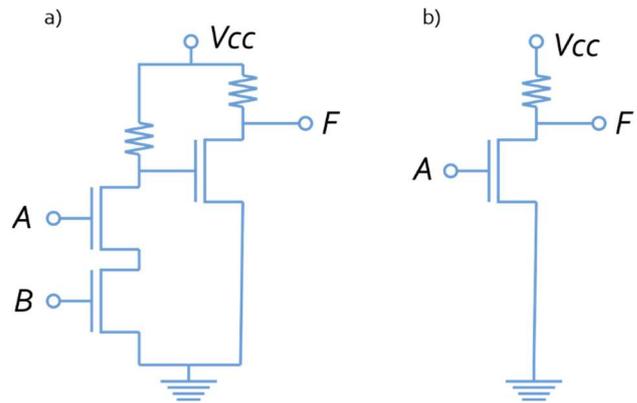

**Figure 6.** Metal oxide semiconductor implementations of the a) NMOS AND and the b) NMOS  NOT gates.

The structure of such a transistor relies on the switchable conductivity provided by a 2D electron (or hole) gas that can be accumulated in the depletion layer of a field effect transistor when biased by the electric field produced by a gate isolated from the conducting channel by a thin layer of oxide ($SiO_2$). See **Fig. 7**.

In this simple example a choice of components was made but it must be remarked that different combinations of elementary gates can lead to the same function and this choice depends on the technology used that may favour some over others. The essential message conveyed here is that a simple bit addition circuit can, as every other component of a silicon chip, be made of judicious combinations of transistors whose size has been decreasing, ever since they were first discovered, down to the few-nanometre scale reached nowadays. This decrease in size is accompanied by a corresponding increase in efficiency or computing power known as Moore's law.

### 2.5.   Quantum computing

Recent times have witnessed the surge in interest in a novel computing paradigm spurred by the suspicion that classical computing offers weak flanks in applications such as cryptography with the subsequent implications for sectors such as banking and defence.

### 2.5.1.   Computers

In its simplest form a Turing machine may be thought of using a set of symbols comprising zeros and ones, as modern real-world computers do. Manipulation of these symbols, performed according to classical mechanics, is enough to realize computing; and the state of any register in memory or the state of the machine can be represented with the use of only two states. The probabilistic character of wavefunctions and the inherent limits to accuracy imposed by uncertainty principles make it fair to wonder what Quantum computers can offer in comparison with classical Turing machines. Because classical mechanics, the physical realm of Turing machines, is a limiting case of more general quantum mechanics, it seems plausible to assume that the power of quantum computing should be superior. One of the first inroads into quantum mechanical computation potential already showed that unitary evolution at least guaranteed the power of a Turing machine if not more.[28,29] The quantum computer

---

could be embodied by a relay) any arbitrarily complex Boolean logic operation could be represented. Thus, his choice was *computationally complete* although just one among the various possible. It was adopted and had a great impact in digital logic in coming years.

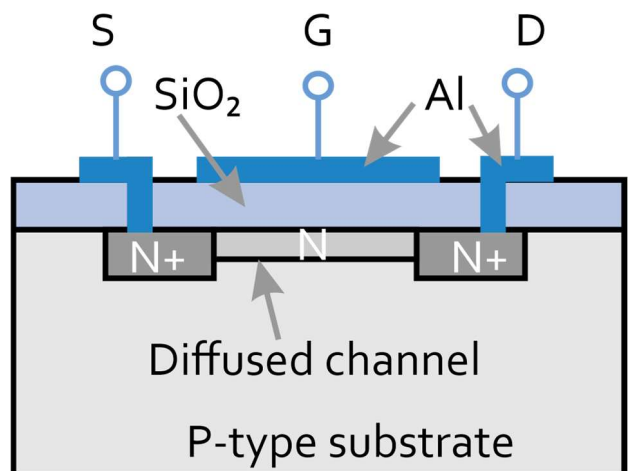

**Figure 7.** Metal oxide semiconductor or insulated gate field effect transistor. A positive (negative) bias at the transistor gate electrode creates and enhanced (decreased) density of electrons in the channel that augments (forbids) the current from source to drain in so called enhancement (depletion) mode.

is outstanding for the possibility of massive parallelism[a] in a single piece of hardware and differs from a stochastic classical machine (the output state is random within the probability distribution function depending on the input state) in that, while determined by the input, the output state is not an observable so that the user may not be able to read its label.[30] This relates to the fact that while a quantum system can be prepared in any desired permitted input state, all one can know about the output state is the result of measurement of some observable. From a general perspective and without entering into the details of the resolution of particular problems, the superiority of quantum has been shown to reside in that there are certain problems that quantum computers can solve quickly and exactly while classical computers can only do it with high probability and the aid of a random number generator.[31][32]

A major problem extensively studied and believed to be *hard* (not to be computable in polynomial time) relates to integer *factoring*. Finding the prime factors of large integers is so computation demanding that public key cryptosystems rely on it. This is the first practical problem for which it was proposed that an algorithm running on a quantum computer would find a solution in polynomial time with small probability of error. That is, it would take $T(n) \propto n^p$, rather than $T(n) \propto \exp(cn^{1/3})$, $n$ being the number of bits in the test number and $p$ and $c$ constants.[33]

### 2.5.2. Operation

In classical computers information is stored in transistors that can be in only two states, $s$, (say, $s = 0$ or $s = 1$) so that a system with $n$ such components can be described with $n$ bits. A similar quantum system of the same number of (two states, see **Fig.8**) components, qubits, requires, for a full description, $2^n - 1$ complex numbers since the state of the system can be in any linear combination $\psi = \Sigma \alpha_i S_i$ of the $2^n$ state vectors of the form $S_i = |s_{1i}\ s_{2i}\ \dots\ s_{ni}\rangle$ with $s_{ki} = 0,1$, $i = 1\dots2^n$; that span the Hilbert space associated to the compound system. Qubits being two-level physical systems can be embodied by spin ½ particles, two-level atoms, polarized photons, trapped ions or neutral atoms, superconducting junctions etc.[34] In any case they must be susceptible to external control by, for instance, electric or magnetic fields.

---

[a] Ordinary computers need to run twice to determine the output of a function $f$ when the output to two values, $x_1$ and $x_2$ are asked; whereas a quantum computer can be fed with an input state encoding both $x_1$ and $x_2$ in a quantum superposition and run only once. Furthermore, a superposition state encoding an exponential number of inputs can be processed in

Operating the machine requires being able to change its state which in turn entails changing qubits. This, in quantum mechanical terms, involves state transformations to be represented by unitary matrices (whose conjugate equals its inverse) for the probability of all possible final states to be 1. The evolution of a quantum state according to the Schrödinger equation, $i\hbar\ \partial\psi/\partial t = H\psi$, determines the evolution operator which needs to be applied to change the state of qubits, $U = \exp(iH/t)$, and, obviously, requires knowledge of the Hamiltonian, $H$, governing the qubit. Much like logical gates such as AND, OR, and NOT, are the elementary building blocks in ordinary electronic processors, quantum gates can be conceived as the elementary operations to manipulate qubits in quantum processors. For technical reasons, and thanks to the fact that quantum gates operating on just two bits at a time are sufficient to construct a general quantum circuit, transformations involving one or two qubits are most common just like gates operating on one or two bits are in electronics.[35]

Successive transformations, on just a few qubits, permit to think of the computer as a circuit much like electronic ones where the operation proceeds along it encountering gates, with one or two *wires* typically. The operation of the quantum processor must be observed only after the action of the last gate and that constitutes the output; observation in quantum mechanics collapses the wavefunction to one of the eigenstates and superposition disappears.

### 2.5.3. Decoherence

There is a consensus, summarized in the DiVincenzo criteria,[34] for reliable hardware operation that includes, for instance, the ability to initialize and manipulate qubits as well as to read the output. Paramount among them is the requirement that the system's qubits keep their coherence, that is, states remain unaltered at least as long as it takes for the computer to finish the computing. This is the crux of a single call. The catch is that measurement renders useful output only once owing to the disturbance imposed by Heisenberg's uncertainty principle.

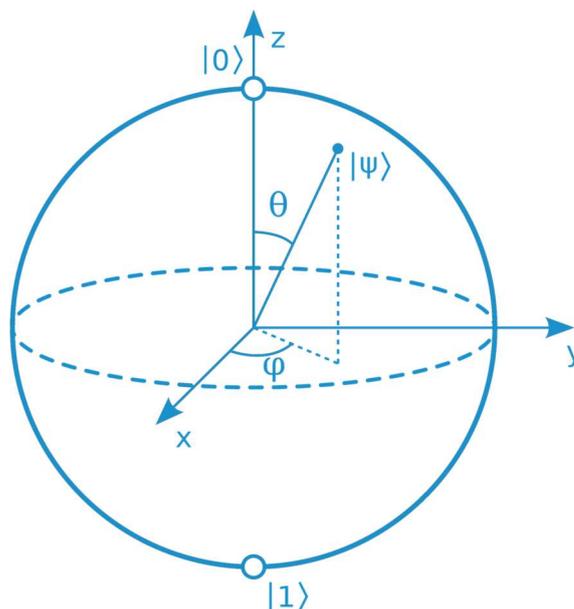

**Figure 8.** Bloch sphere. A qubit from a two-level system ($|0\rangle$, $|1\rangle$) is the superposition $|\psi\rangle = \alpha|0\rangle + \beta|1\rangle$ with $\alpha$ and $\beta$ complex, which might seem to have four degrees of freedom: However, normalization of the wavefunction requires $|\alpha|^2 + |\beta|^2 = 1$ which can be satisfied with $\alpha$ real positive using $\alpha = \cos(\vartheta/2)$ and $\beta = \sin(\vartheta/2)\ e^{i\phi}$, with $0<\vartheta<2\pi$, $0<\phi<\pi$, where $\vartheta$ accounts for the amplitude and $\phi$ for the relative phase.

the matter in building quantum computers. According to quantum mechanics, the evolution operator of the *world* is unitary and that concerns the quantum computer plus its environment. Processing involves manipulation of *parts* of the computer with unitary operators. Thinking in terms of matrices: since the world matrix must be unitary and so must the involved parts being manipulated, the only possible way is for the world's matrix to be block diagonal, each block corresponding to a part requiring manipulation. This directly implies that the parts only interact internally and are isolated from each other and from the environment. Unfortunately, in real world devices, many degrees of freedom not relevant to the computing operations (*e.g.* lattice vibrations, spurious electromagnetic fields, etc.) interact with those of the qubits preserving the world coherence but spoiling the qubit. One obvious implicit requirement is then that the qubits are embodied by atomic-scale objects.

*2.5.4.    Simulation*

One of the best-established application of quantum computing is in *simulation* of quantum systems where the nature of the systems, *e.g.* strong correlations, render classical computer methods largely inefficient for the exponential scaling of resources (time and memory) required. Following Feynman[29] lead, this was the second problem showing the superiority of quantum over classical computing.[36]

A quantum simulator is a piece of hardware subjected to tight external control which quantum mechanically reproduces the dynamical behaviour of a (quantum) target system. Furthermore, decoherence and thermal effects in the quantum computer can cease to be a hindrance to be exploited to faithfully mimic unavoidable decoherence and thermal effects in the simulated system.

The processing can be analogue where the continuous evolution of the quantum simulator maps the physical behaviour of the modelled system or digital where general purpose computing can be performed by changing states sequentially, at the strokes of a clock, by applying unitary operator. To this purpose, the typical procedure relies on convenient coding of the system along with a faithful mapping of the Hamiltonian in terms of unitary operations to be executed in sequence.

# 3.    Artificial intelligence: a panorama

Artificial intelligence (AI) is the ability of non-animal entities to show a behaviour similar to that of (mostly superior) animals such as mammals where memory, learning, communication, and other similar faculties are involved. Important components of artificial intelligence are learning, reasoning, problem solving, perception, and using language to manage abstract concepts. *See* a possible taxonomy in **Fig. 9**.

AI is often divided into *narrow* AI and artificial *general* intelligence where the former refers to the creation of systems capable to carry out certain intelligent behaviours in certain contexts. A change of context, for a narrow AI agent, will require human intervention in order to reconfigure or reprogram it so it can keep its performance. This can be seen as a clear limitation in comparison with the notion of human general intelligence. Thus, artificial general intelligence has grown into the antonym to narrow AI to refer to systems with a capability to generalize from the specific. This quality is rather different from the ability of solving specific tasks or problems; a generally intelligent system is expected to succeed in handling situations *qualitatively* different from these for which it was created.[37]

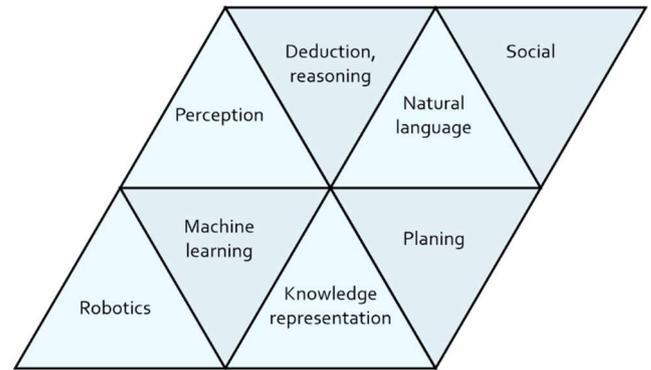

**Figure 9. |** Artificial Intelligence, a taxonomy.

Machine learning (ML) is one aspect of AI that concerns how automated processes carried out by machines can profit from experience to acquire the ability to interpret situations and thrive in future challenges.

Artificial neural network (ANN) is a branch of ML that has gained increasing popularity due to its capabilities of modelling complex correlations among large datasets. Its power arises from its network-based mathematical manipulation of data through high dimensional parametrized transformations that can account for large variations in data behaviour. The large number of parameters implicitly impose the need for large data sets for the methods to be reliable.

## 3.1.    Origins of AI

Also designated machine intelligence, artificial intelligence is the kind of intelligence possessed by machines at variance with natural intelligence demonstrated by humans and animals. The subjects exercising it are *intelligent agents* whose distinctive trait is their capability to perceive their environment and react by performing actions to thrive maximising the chances of achieving their goals. It is generally accepted that this essentially involves two capabilities: learning and solving problems. While artificial intelligence has many aspects, notably those concerning philosophical, ethical, or humanistic in general, the technical problems it faces include reasoning, learning, planning, perception, representation, etc.

Artificial intelligence was first mentioned in the 1955 Dartmouth Summer Research Project proposal by J. McCarthy and others in connection with the question whether machines can be made to simulate intelligence and generally at variance with analogue automata. The proposal was so visionary that aspects that are nowadays considered essential to the science and technology involved, were already mentioned in the proposal such as automatic computers (machine simulation), language usage (speech recognition), neuron nets (neural networks), calculation size (algorithmic complexity), self-improvement (machine learning), abstraction (symbolic calculation), randomness and creativity (intuition) and «other the participants might contribute»

Along its history, many controversies arose that, although not necessarily at odds, tended to congeal in two divergent approaches on how to achieve AI. On the one hand *computationism* (first proposed[a] by W. McCulloch and W. Pitts) maintains that the mind works as a computing machine by manipulating symbols according to rules dependent on states. Therefore, a mind's activity could be implemented exactly in some kind of computing machine. On the other hand, *connectionism* views the mind's operation as an *emergent* resulting from collective interaction of a multitude of neurons.

---

[a] Thomas Hobbes, a seventeenth century philosopher in "De Corpore", 1655, expressed that reasoning meant computation in the sense that to compute is to calculate the sum of many things added together, or to

reckon the remainder when one thing is taken from another; reasoning, therefore, is the same as adding or subtracting.

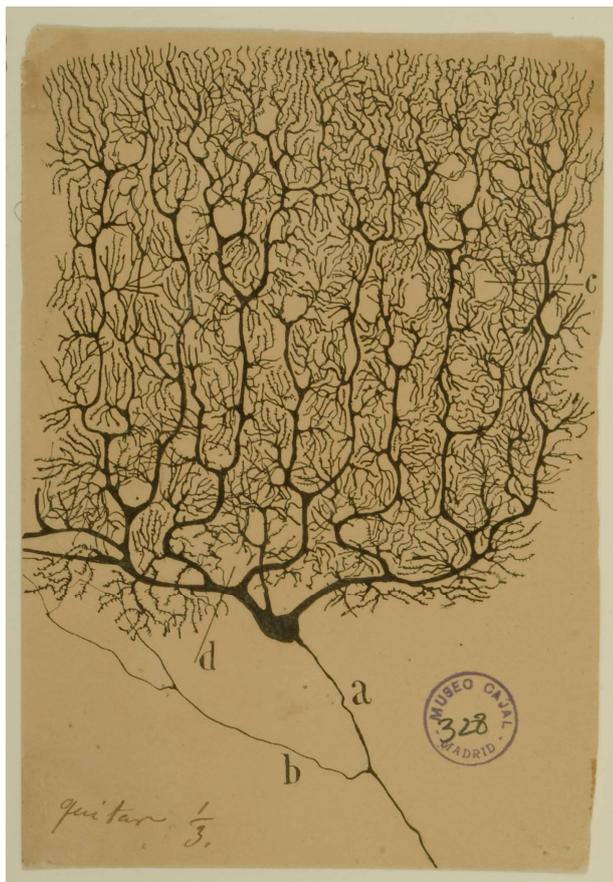

**Figure 10.** Purkinje cerebellum cells as drawn by Santiago Ramon y Cajal. (Museo Cajal, Madrid). With permission.

### 3.1.1.    Neurocomputing inspiration

Processing and memory density on the one hand and processing speed on the other imply that it is not possible to create ever more powerful individual processors; instead intelligence has to turn towards parallel systems where many elements work together (the human brain neatly demonstrates that even fairly slow and inefficient elements like neurons, as that in **Fig. 10**, can produce a very powerful computing system). The delay imposed by data traffic dictates smaller sizes as systems are made faster and faster. This hampers heat dissipation.

## 3.2.    Machine learning

Because one of the most notable characteristics of intelligence involves learning, *machine learning* also became one of the most prominent disciplines in AI. It is often defined as the ability of computer algorithms to improve through experience. It originates from the long-standing practice of pattern recognition and, relying on statistical learning theory, pursues building algorithms that improve on available data to predict about unknown data.

### 3.2.1.    Fundamentals

Learning is the phenomenon of acquisition of knowledge without explicit programming. It is performed by living organisms and can be emulated by machines.

#### 3.2.1.1    Learnability

Just as the advent of powerful models and computing machines made *computability* theory possible, the inception of artificial intelligence and associated methods originated the concept of *learnability*. The former deals with the ability of mechanical calculation to find solutions to mathematically expressed problems while the latter is concerned with how humans learn and how to replicate it in algorithms and machines. It should also offer insight regarding the limits of what can be learned as computability does on what can be calculated.

According to Valiant's formal approach to *concept learning*, a program for performing a task is acquired by *learning* if it was acquired by any means other than explicit programming.[38] Thus, learning is defined as the process of deducing a program for performing a task, from information that does not provide an explicit description of such a program.

Computational models of learning are usually restricted to procedures aimed at recognizing if a concept (predicate) is true or not for given data. Then the concept is considered as *learned* if the program that recognized it was deduced as opposed to explicitly written or imported from outside the program. A learning machine (*learner*) comprises a learning *protocol* (to determine the way in which information is obtained) and a *deduction* procedure whereby a recognition algorithm (*hypothesis*) for the concept learned is deduced. The learning protocol usually accesses data items that an *oracle* examines to tell the learner whether or not it constitutes a positive example of the concept.

In this framework a concept class is *strongly learnable* if, given access to examples of the unknown concept, the learner with high probability generates a hypothesis that is correct on all but an arbitrarily small fraction of the instances. For *weakly learnable* concepts the learner can produce hypotheses that are only slightly better than random guessing. Although the equivalence of these two notions was initially doubted it was eventually demonstrated.[39] This equivalence has strong implications in important technical aspects as algorithm efficiency, data compression, or setting bounds to the size of data needed to learn a concept.

#### 3.2.1.2    Classification

Pattern recognition or discrimination is the act of predicting the nature of an observation, an event registered by physical means. Typical examples (*e.g.* electrocardiogram) adopt the form of numerical measurements such as images, sounds, electrical signals or, in general, collections thereof in vector form or mixed formats, **x**. The nature of the observation, the class, $y$, is usually one value among a finite set of choices (*e.g.* mitral insufficiency). Mathematically the instrument representing this guessing act is a *classifier*—also *learner* or *hypothesis*—, a function mapping observations into their label or *class*: $h(\mathbf{x}) = y$. Classifiers can only be built for problems of simplicity and disorganized complexity.[40] However many phenomena in the intermediate region are often intractable, catalogued as truly complex and resisting this kind of classification.[41] At any rate, for many reasons (instance an observation may be subject to error or simply hidden variables are playing a function unknown to the classifier) it is best to accept that in general classification will be imperfect: so that $h(\mathbf{x}) \neq y$.

The process of constructing a classifier, $g$, with the least probability to cast wrong classes within a set of available data is what we understand by (supervised) learning.

#### 3.2.1.3    Explainability and Interpretability

AI has entered everyday life very deeply taking decisions for us every so often (shopping, personal relations, entertainment etc). In general, little care is placed on how algorithms arrive at the decision on those matters but life-changing choices (such as disease diagnosis, defence, finance, law, for instance) awakened an urge to know the reasons behind such critical decisions.

Thus, despite its extraordinary advancements, misgivings about the use of AI-based systems issue from the fact that they often lack *transparency*. A model's parameters discovery process is considered transparent when it is well described and motivated. Their black-box nature, especially ML's, allows powerful predictions, but it cannot be directly explained. This issue has triggered a debate and generated *explainable AI* (XAI).[42]

Such a debate gravitates around emerging concepts[43] like *interpretability* defined as the quality of a system to be explained in terms understandable to humans and *explainability*, associated with the notion of explanation as an interface between humans and a decision maker that, being comprehensible to humans, is a faithful intermediary.

### 3.2.1.4    Feature vector
The performance of ML methods strongly depends on how the information it feeds from is selected and presented and how it represents the system object of understanding. The way information is coded, called *representation*, is usually multidimensional and, as such, comprised in *feature* vectors. This does not mean that every dimension of the vector space is a continuous or digital variable, like space, time or bit depth, but it can itself contain global properties such as "contains a square" if dealing with image representations for instance. In general, features present observable properties of the objects the data points represent. The impact on ML efficiency is such that a very large effort is nowadays spent in the design of methods of data manipulation that results in representations of the data that helps improve the effectiveness of ML process.[44] Feature engineering requires insight and is the fruit of experience and human ingenuity opened the area of representation learning where ML is also gaining traction.[44]

### 3.2.1.5    Feature selection
Features added to the objects under study, depending on their origin, may turn out to be irrelevant or redundant resulting in biased or even incorrect models or in overly complex (excess of dimensions). This makes feature selection an important data preparation step that caused an expansion of the area.[45] Feature selection can be performed on data sets, depending on the information they contain, by supervised (that assess the importance of features through the correlation between features and class labels), unsupervised[46] and intermediate methods.

### 3.2.2.    Machine learning forms
Machine learning, the dominant current in AI, in its pursuit of automating problem solving, can adopt essentially two different, and to some extent opposing, methods. The *symbolic* ("top-down" also known as *good old-fashioned AI*) approach attempts to simulate reasoning independent of any biological instance (brains); the *connectionist* (or "bottom-up") drawing inspiration from the connections forming the network structure of the brain gives rise to the artificial neural networks.

### 3.2.2.1    Symbolic
*Symbolic* ML, based on first-order logic and truth tables, uses symbols that can occur as components of another type of entities called expressions in an algebra (that set the operating rules) to extract information from data or reach conclusions logically derived therefrom. A crude description would be: input a question (data) and an algorithm (that contains the knowledge) to obtain an answer (*e.g.* testing an input image against all possible descriptions of a written character to interpret hand writing).

The algorithms used in this kind of computing follow a top-down method structured following the same steps of problem-solving humans often follow for building from the data to reach the answer if direct decision tree strategies are available or using trial and error otherwise.

The earliest successful AI program was written in 1951 by Christopher Strachey. His draughts program ran on a primitive computer whose main (random access) memory used cathode ray (Williams) tubes (typically 8) to store 20-bit words as rows of dots of charge (typically 64) in the phosphor and a magnetic drum as secondary memory; other memory was registered in several thousand additional tubes. The operation cycle was 1.2 ms, that is, the computer run at 850Hz and could play draughts at acceptable pace.

These programs were endowed with the ability to learn from experience and, by confronting successive generations of the program, improved it in evolutionary computing. These methods derived in what is now known as *genetic algorithms*. In 1997, Deep Blue, a chess computer built by the International Business Machines Corporation (IBM), beat then-reigning world champion.[47]

### 3.2.2.2    Connectionist
*Connectionism* derived from the intuition that imitating the assumed operation of neurons in the brain could be implemented in artificial computing environments. Their most distinctive feature being the large number of connections between them gives rise to the term. In this model many different implementations are included enveloped under the artificial *neural network* umbrella. Typically, in this approach, a large set of questions (data) and their answers (labels) are used to obtain the rule of association (algorithm) in a training stage and, then, use the algorithm to process unknown data (inference stage).

The first treatise dates back to 1943 and assumes that each neuron works as a processor and the brain as a Turing machine.[48] A decade later, following this approach B. Farley and W. Clark run the first computer simulation of an artificial neural network with 128 neurons. Neurons were assumed to be nodes receiving input from and sending output to other neurons through the network synapses. Each neuron fired an output only if combined input received was over its particular threshold. Each connection bore a weight that could be trained by repeatedly feeding "known" patterns through the input to the network while watching the output for the desired result. Subsequently, when an unknown pattern is fed to the trained network a likely response is obtained. When many layers of neurons are involved an optimised procedure to establish the weights was proposed by Rosenblatt[49] that came to be known as backpropagation error correction.

Supervised learning relies on expert data which imposes a limit on the performance. Reinforced learning computing improves by challenging themselves and in principle can exceed the abilities of any external trainer. Based on these techniques AlphaGo, an AI program engineered by DeepMind, defeated the Go world champion in 2016[50] and AlphaZero is capable to learn, from scratch, in a few days to beat the best expert programs developed so far.

### 3.2.2.3    Probabilistic or Bayesian
By conceiving learning as the capacity to infer plausible models that account for the observed data, a *probabilistic* framework is built where ML can be founded. Because finite data —let alone limited data— can be consistent with different models, doubt about the model adequacy naturally arises and predictions on future data are subjected to uncertainty. This framework dictates how to characterise and handle uncertainty about models and predictions.[51]

The distinctive feature of Bayesian models is that the uncertain character of actual data is considered as a principal component in the learning process. If we assume that data can be consistent with several hypotheses, then which model is appropriate, given the data, becomes uncertain. This model is based on Bayesian statistics[52] conditional probabilities and networks. It avoids the rigidity of symbolic models by assigning conditional probability distributions to relationships but lacks in reasoning capability.[53]

The mathematical basis for Bayesian ML is Bayes' theorem from probability and statistics theory connecting the probability of an event with *prior knowledge* of conditions that might be related to the event.[54] Since the joint probability $P(x,y) = P(y,x)$ can be expressed as a function of conditional probabilities, $P(y|x) P(x) = P(x|y) P(y)$, we have $P(y|x) = (P(x|y)P(y))/(P(x))$. By identifying $y$ with the unknown parameters of a model ($\theta$) and $x$ with the available data ($D$) we can express the *posterior* probability of parameters $\theta$ based on given data $D$ as:

$$P(\theta|D) = \frac{P(D|\theta)P(\theta)}{P(D)}$$

where $P(\theta)$ is the *prior* probability of $\theta$ (the estimated likelihood of parameters in the model before any data is available, assigned by belief or intuition); $P(D|\theta)$ the probability of data in the model under test (expressing the compatibility of available data with the model); and $P(D)$ is the *marginal* likelihood (or model evidence that expresses the preference for simpler models) but as it does not depend on the parameters being optimized, it can be disregarded. The process starts by enumerating all reasonable models of the data and assigning your prior belief $P(\theta)$ to each model. Then, upon observing the data $D$, the probability of the data under each of these models is computed: $P(D|\theta)$. The product of this likelihood and the prior yields the posterior probability over models, $P(\theta|D)$.

### 3.2.2.4 Analogistic
*Analogistic* ML uses distance analysis in the descriptor vectors (feature space) to detect analogies and differences between data. The most relevant examples are support vector machines and *k*-nearest neighbour methods.

### 3.2.2.5 Evolutionary
*Evolutionary* ML uses computational methods inspired by evolution whereby improvement is achieved through competition between models generated with mutation and combination of previous generations. Notable examples are *genetic* algorithms (genetic improvement of program parameters) and *neuroevolution* (genetic improvement of neural networks' structures).[55]

### 3.2.2.6 Possibilistic
*Possibilistic* ML is based on classical symbolic ML but extended to treat data as ambiguous in a way that considers facts only partially true. A notable example is fuzzy logic.[56]

### 3.2.3. Machine learning theory
Current interest in AI in general and ML in particular has directed attention heavily towards its numerical methods and algorithmical aspects because of the stunning success achieved in many applications. This should not distract attention from the fact that many efforts have been devoted to, and many advances gained in, the theoretical foundations and mathematical bases that describe the requirements regarding hypotheses and limits of applicability along with boundaries of what can be learned. Theory has essentially formed around knowledge stemming from the general *computational learning theory* by Valiant[38] on the one hand and the *statistical* approach contributed by Vapnik and Chervonenkis on the other.[57] The former investigates the performance of optimal learning algorithms for precisely defined situations. The latter deals with uncharacterized classes to base the statistical theory for learning and generalization from the perspective of learning as a function estimation problem for empirical data.

### 3.2.3.1 PAC: Probably approximately correct
Computational learning theory is a rigorous mathematical formulation of supervised learning that can adopt several models. The most notable is known as *probably approximately correct* (PAC). Other models include agnostic learning model,[58] or membership query.[59]

Briefly, let there be a function $f$, the *concept*, that univocally assigns correct labels to a set of data; PAC operates by finding a hypothesis, $h$, that, for a sample of a given size, $N$, within the overall data set, issues tags for the test data *approximately* (the average distance between the tag $h(\mathbf{x})$ and the correct label, $f(\mathbf{x})$, the error, is not null). In this model the *probability* that the error is less than a given ceiling, $\epsilon > 0$, can be made as close to unity as desired simply by expanding the sample size. That is for any $\epsilon, \delta > 0$, there exists $N$ that makes such a probability $P > 1 - \delta$. The dependence of $N$ on the data set, the error and the probability margin, is known as the *complexity*. This concept describing the difficulty to issue a good hypothesis for a learning problem can be put in a more general perspective with the definition of optimization *hardness*[60] and the demonstration of so-called *No Free Lunch* theorems[61] stating that a high performance of an algorithm over a class of problems does not guarantee similar performance over another. Essentially, no single learner can learn everything. This calls for some kind of user intervention in selecting algorithms well adapted to the problem in hand based on knowledge of the problem.

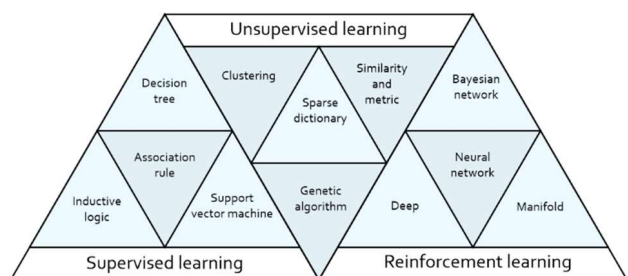

**Figure 11.** Machine learning, a taxonomy.

### 3.2.3.2 Vapnik and Chervonenkis
The *statistical learning formalism* of Vapnik and Chervonenkis is based on the concept of risk (equivalent to the error defined in the PAC formalism) defined as the total error of a given hypothesis over all data. However only the empirical error is available (that evaluated over the training data). Thus, the objective is to evaluate the performance of hypotheses over data beyond the training set. In other words, known the empirical risk for a hypothesis, that is, the error on the finite training set, the theory shall provide an estimate of the unknown risk on future data. In pursuing this challenge, the theory produces the parameter *complexity* that, in a way, acts as the conjugated magnitude in uncertainty principles. Intuitively this means that the more complex the model is the greater the chances are to perfectly label all training data and, consequently, the more probably future data will be misclassified. This will cause a low *generalization*. In fitting terms, imagine a cloud of statistical data obeying a low degree polynomial law. Fitting with a polynomial of higher degree can produce a perfect fit but will fail on additional data.

In this context the VC dimension, that somehow measures the complexity of the learning problem, is a parameter that determines the error for which the probability that the total risk differs from the empirical one approaches unity and is, in a way, the number of degrees of freedom.

### 3.2.4. Learning Models
Two principal categories are usually considered: supervised and unsupervised, although biological inspiration invites to consider reinforced learning too.[62] See a possible taxonomy in **Fig. 11**. Although these are the canonical models, actual implementations often profit from mixing characteristics from several of them. In order to unify language, all ML protocols can be set in a paradigm where AI is exhibited by an *agent* with the capacity to act, placed in an *environment* from which stimulus is obtained. This view is continuously forming the modern definition of AI around the agent-environment paradigm where ML is just one methodology. Obviously, the agent can be as simple as a computer program or as complex as a humanoid robot.

### 3.2.4.1 Supervised
*Supervised* learning is the process of learning from data represented as many instances of pairs $(\mathbf{x}_i, y_i)$ where the vectors $\mathbf{x}$ represent data —often multidimensional— and $y_i$ is a collection of labels (typically binary or real) and the goal is to find the rule that attributes the label to the datum so that an unseen $\mathbf{x}$ is labelled $y$ with some certainty or more specifically finding the conditional probability $P(Y = y | X = \mathbf{x})$.

Here learning leads to finding correlations between data and label as in probability to suffer a given condition based on DNA data after learning from a large set of studied patients. The trivial case is that where the data obey a deterministic law where all variables and their relations are known. One of the most popular approach is that of neural networks (NN) which are used for unsupervised learning too.

The idea behind is to generate a (not unique) hyperplane in the X space separating $\mathbf{x}_i$ associated to positive and negative labels. Such simple task can already be performed by perceptrons on certain data sets (linearly separable) but the aim of SVMs is to generalize this strategy to all kinds of data by nonlinear mappings to higher dimensions where they become separable.

### 3.2.4.2    Support vector machines

*Support vector machines* are widely used supervised learning models for classification when data $(\mathbf{x}, y)_i$ has binary labels, that is, when $y_i \in \{-1, 1\}$.

Support vector machines constitute one of the most widely used classification method in supervised learning in which the vector character of the data is used to represent data in a hyperspace such that a hyperplane exists that separates data according to their (typically binary) label much like perceptrons do. The real power of the support vector machine approach resides in its extension to sets that are not separable in this way by means of nonlinear mapping to higher dimension.[63] Finding the hyperplane that maximizes the separation of nearest data optimizes the classification and defines the support vectors (nearest data on both sides).[64]

### 3.2.4.3    Unsupervised

*Unsupervised* learning, as performed by biological organisms, arises from the advantageous use, with a limited processing capacity, of the massive amount of unfiltered stimuli constantly being perceived, through its essential feature: *redundancy*, the excess of information received over that beneficially processed. Because information is carried by unpredictable events pure noise is the only form of nonredundant stimuli. Thus, redundancy separates sensory experience from noise and so the perception of regularities —patterns— is conjectured to drive unsupervised learning.[65] To benefit from redundancy is thus essential to extract the non-random features of stimuli contained in their statistical distribution and their correlations.

In unsupervised learning no labels are provided and the algorithm is expected to find the structure of the data by, for example, discovering classes where data are grouped by some definable distance which is minimized on average within each class while distance between groups is maximized. Class membership may be viewed as a label in this case, but it is assigned *a posteriori* as a result of the learning process. The principal advantage of unsupervised learning is the lower effort required to prepare training sets because no labels are provided which often entails human intervention or use of previous supervised learning tools. Additionally, it can provide previously undetected patterns not properly identified by the experts in charge of preparing training data or even detect erroneous assigned characters.

#### 3.2.4.3.1    Clustering

*Clustering* is a process, not specific to ML disciplines,[66] whereby objects in a set are grouped such that objects in the same group are *more similar* to other objects in the group that to any objects outside the group. This goal can be achieved through numerous means where *small distances* (with a proper metric defined in the objects feature space), *dense areas* (regions of said feature space where many objects accumulate), or particular *statistical distributions* are often used. Depending on the nature of the training data, different approaches with different parameters may adapt better to the task; but it usually involves prior knowledge and requires fine tuning through trial and error.[67]

#### 3.2.4.3.2    Dimensionality reduction

The vast amounts of data nowadays available to ML methods tends to be almost equidistant in feature space and consequently hard to deal with for clustering methods. In certain practical cases data sets may present several thousand or millions of dimensions which reduce the generalization performance of the classifier algorithms.

Mathematically *dimensionality reduction* transfers the original data from its natural feature-rich space to a lower dimension space preserving the essential features, those truly descriptive of the object. Let data consist of $q$ available vectors $\mathbf{x}$ of dimension $D$ that can be cast as a matrix $\mathbf{X}$, with elements $X_{ij}, i = 1, 2, \ldots D; j = 1, 2, \ldots q$ with $D > d$ the *intrinsic* dimensionality, that is, the minimum number of parameters needed to account for the properties of the data. The goal is to transform $\mathbf{X}$ into $\mathbf{X'}$ with dimensionality $d$. This assumes that data are lying on or near a manifold of dimension $d$ embedded in the $D$-dimensional data space. As an unsupervised technique, in general neither $d$ nor the geometry of the manifold are known which makes the problem an ill posed one and requires further assumptions on the nature of the data (notably $d$). Methods devised to carry out this reduction are numerous and can be classified according to different aspects. Feature selection (finding the relevant features) versus feature extraction (generating new combined features better representing data),[68] linear versus nonlinear; convex versus nonconvex[69] are just two taxonomy examples.

Principal component analysis is one of the most widely used such algorithms. Briefly, the method essentially takes the $\mathbf{X}$ matrix as a distribution of points is a $D$-dimensional space to find $d$ mutually orthogonal unit vectors such that the $i$-th vector points in the direction that best fits the data to a line. Best fitting is taken in terms of a minimum square distances, for example. Start by finding $\mathbf{u}^1$ such that the distribution of $\{\mathbf{u}^1 \cdot \mathbf{x}\}$ have maximum variance; next find $\mathbf{u}^2$ such that $\mathbf{u}^2 \cdot \mathbf{u}^1 = 0$ and $\{\mathbf{u}^2 \cdot \mathbf{x}\}$ have maximum variance and so on so that $\mathbf{u}^k \cdot \mathbf{u}^i = 0$ is orthogonal to all $i < k$ preceding vectors and $\{\mathbf{u}^k \cdot \mathbf{x}\}$ distributes with maximum variance. The vectors found in this way are the *principal components* and there can be as many as $D$ but, in general, it is hoped that $d \ll D$ will suffice to account for the variations in $\mathbf{x}$. This procedure is aimed at finding the directions along which data *aligns*, so that its projection on the corresponding vector is more spread. For all $\mathbf{u}^k$ directions with $d < k \leq D$ data lies on a thin volume that signifies the irrelevance of that dimension. The actual procedure to find the principal components relies on the fact that, if $\mathbf{\Sigma}$ is the matrix of covariance between components of the $q$ vectors representing the data or a representative sample taken among them, $\Sigma_{ij} = \text{cov}\{x_i, x_j\} = \langle (x_i - \langle x_i \rangle)(x_j - \langle x_j \rangle) \rangle$, it so happens that $\mathbf{u}^k$ is the eigenvector of $\mathbf{\Sigma}$ corresponding to the $k$-th largest eigenvalue which is given by $\lambda^k = \text{var}\{\mathbf{u}^k \cdot \mathbf{x}\}$.[70]

#### 3.2.4.3.3    Autoencoders

An *autoencoder* is an unsupervised ANN that learns how to efficiently compress and encode data (reduce dimensionality, thus disregarding noise) and then learns how to reconstruct the data back from the reduced encoded representation to a representation that is as faithful to the original input as possible.[71] The ANN contains a part, *encoder*, that reduces the dimensionality; a layer that contains the least expression of the input data, *bottleneck*; and a part where the compressed representation is expanded, *decoder*. The reconstruction loss (distance measure between original and reconstructed data) is minimised often through backpropagation.[72]

If all features in the input data were relevant (each independent of one another), compression would be impossible. However, if data exhibit some structure (*i.e.* correlations between input features), this structure can be learned and consequently leveraged when channelling the input through the network's bottleneck. The uses of such networks that expand what they have just compressed can vary from detecting outliers in classification tasks, to denoising data such as images, audio, etc.

#### 3.2.4.4    Reinforced learning

This is a reward-driven learning category where an agent interacts with an environment through *percepts* (received from that environment) and actions (exerted on the environment) which eventually lead to rewards that represent the goal of the interaction. *Reinforcement* can assume the combinations of positive (supplying reward) or negative (removing penalty) with reinforcement (when correct) or punishment (when incorrect). This model, the essential underlying strategy in the AlphaGo project,[50] is closest to animal learning and can be described as the attempt to learn how to behave in an unknown environment. This is closely related to Markov decision processes,[73] where an actor executes an action over a system (environment) in a state to place the system in a new state and receive a reward. The goal is to find a policy that results in maximum rewards. This basic description can be altered and tuned in the size (number of states) and structure (connectedness) of the environment, the way rewards accumulate, etc. to simulate different kinds of learning processes.

#### 3.2.4.5    Transfer learning

Usually machine learning techniques work under the assumption that training and test data belong to the same feature space and obey the same statistics. If the distribution changes, statistical models have to be rebuilt from new data. The profusion of data sources like sensors in all kinds of devices (cell phones, cars, public transportation, etc.) and locations (buildings, public places, etc.) generates immense amounts of virtually free but unlabelled data that highlights the cost of making it useful by attaching complementary annotation. This need spurred the exploration of solutions that facilitated learning methods based on data from the same domain but lacking labels or labelled data but from a different domain. In this way, data gathered in similar conditions as the targeted data or data from different conditions or different provider can be made useful. A typical example could be speech recognition for an unknown language based on data collected from a previous one or image or video recognition adapted to novel environment, novel purposes different from those where available data was obtained. A case where data changes with time or simply gets outdated is an important example. These methods are collectively called *knowledge transfer* and *transfer learning*.[74]

Depending on the abundance, scarcity or absence of labelled and unlabelled source and target data, different authors have adopted different, sometimes inconsistent, terminology for supervised, unsupervised and semi-supervised, informed and uninformed, inductive and transductive transfer learning.[75] An important process in transfer learning is *domain adaptation* whereby information from a source domain is altered so that its statistical distribution comes closer to that of the target domain thus improving the performance of the target learner.[76]

### 3.3.    Neural networks

Neural networks, a kind of structure where many ANs placed in a network[77] architecture that provides connections between them, are arguably the most widely used and best understood supervised ML model.

The efficiency of *neuromorphic computing* as in the brain is deemed to be some seven orders of magnitude better that the best conceivable digital processor.[78] More recent estimates still confirm.[79] Inspired by the brain, ANNs were developed expecting them to enjoy many characteristics not available in von Neumann's architectures such as massive parallelism, distributed computation, learning and generalization ability, fault tolerance or low energy consumption, among others. Perhaps the most important feature, a distinctive additional advantage, is the fact that the brain is not hardwired but enjoys *plasticity*, the capacity to adapt to solving new tasks using the same structures by changing neurons' conductance. In terms of ML this is equivalent to changing the weights in ANNs.

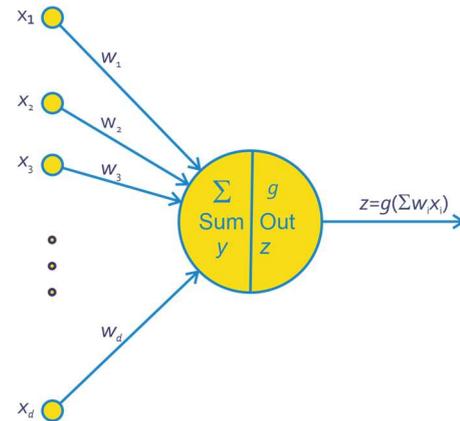

**Figure 12.** Schematics of the multiply and accumulate operation of one neuron.

For an introductory review of concepts, taxonomy and historical perspective see [80].

#### 3.3.1.    Components

ANN are composed by ANs that are multiply connected, sometimes forming layers. Depending on the arrangement and the directions of the communications between them, different architectures can be described.

##### 3.3.1.1    Neurons

Artificial *neurons* are the basic components of ANNs in analogy to the biological nervous systems where nervous cells take signals from and send signals to each other through the synapses. See **Fig. 12**. As biological cells, neurons in ANNs are modelled to receive input through many dendrites and upon certain conditions (threshold) send signals through many others. In mathematical terms this is expressed by means of a typically sigmoidal trigger function representing the response of the cell which in its simplest form can be expressed as:

$$g(\mathbf{w} \cdot \mathbf{x}) = g\left[\sum_j x_j w_j\right]$$

where $\mathbf{x}$ is a vector containing input stimuli ($x_i$), $\mathbf{w}$ is a vector containing the weights ($w_i$) with which these inputs affect the neuron in question and $g$ is the *activation function* that measures the response of the neuron to the sum of weighted stimuli. Sometimes an explicit bias is included that, for the sake of algebraic simplicity, can be thought of as an extra weight applied to an auxiliary component of the data vector set equal to 1.

In a training stage, such a neuron is fed with numerous data vectors and their tags $\{\mathbf{x}, y\}_i$ and the weights $\mathbf{w}$ and activation function $g$ are optimized to fit the labels by maximizing certain figure of merit. When the optimization is properly carried out an unknown vector fed to the AN will very likely be properly tagged. A trivial training routine will consist of: initializing the weights vector $\mathbf{w}^1$ to small random numbers; presenting the first data vector $\mathbf{x}^1 = (x_1, x_2 \dots x_n)^1$ and evaluate $g(\mathbf{w}^1 \mathbf{x}^1)$ and compare with target $y^1$; update weights according to $\mathbf{w}^2 = \mathbf{w}^1 + \eta(y^1 - g(\mathbf{w}^1 \mathbf{x}^1))\mathbf{x}^2$ with $0 < \eta < 1$; repeat to the end of available data.

##### 3.3.1.2    Perceptrons

*Perceptrons* are a special class of neuron where the activation function $g(x) = \theta(x)$ the Heaviside step function.[48] Even such a simple neuron can work as a binary classifier which can decide whether or not an input, represented by a vector $\mathbf{x}$, belongs to some specific class.

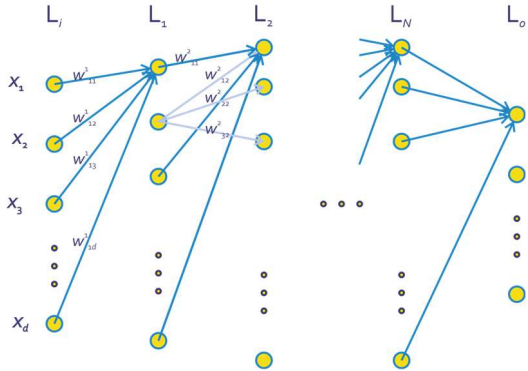

**Figure 13**. Architecture of a deep feedforward ANN.

### 3.3.1.3   Layers

In certain cases, ANs are organized in groups called layers all of whose members communicate only with ANs belonging another layer. See **Fig. 13** where an input layer, (L$_I$), several (*N*) hidden layers (L$_k$) and an output layer (L$_o$) are shown. If the signals from one layer only go to the next one the ANN is called *feedforward*, whereas ANNs in which ANs in one layer communicate with the next as well as with the previous, or no layers can be defined, are called *recurrent*.

### 3.3.2.   Deep ANN

ANNs can adopt many different organizations as can be seen in **Fig. 14**. Typical ANN have their neurons arranged in layers, the first of which is the input layer, the last one serving as output layer. All intermediate layers usually interact only inside the ANN and are called *hidden* layers. When the ANN has more than one hidden layer it is called *deep* neural network (DNN).

While single ANs only have the capacity to linearly classify, once the ANN contains three layers it already can approximate any continuous function if the activation function is sigmoidal[81] or even simply monotonic and bounded. This is known as the *universal approximation theorem* and means that a three layer ANN with sufficient neurons can be trained to learn a data set, that is, there exist weights and activation function that represent the data.[82] It is found that the same number of ANs are more efficient if arranged in several (deep) layers than in a large shallow one.

DNN's capacity to generate complex functions —*expressivity*— derives from the abundance of parameters and, while it constitutes an advantage in that it allows to learn any data set, it also leads to a decrease in *interpretability* — the ease to produce generalization understandable to humans. This is a serious issue that creates misgivings in the use of ML in critical systems (*e.g.* self-driving cars).

### 3.3.2.1   Information Flow

As discussed, it is useful to broadly classify the network architectures of ANNs as *feedforward* or *recurrent* and either is more convenient depending on the type of computation intended. Feedforward ANNs are better adapted to static data processing where, even if data is fed to the network in succession, each input is dealt with independently. Conversely recurrent ANNs are best adapted for *dynamic* (in the temporal sense) data processing. This is because the retroactive flow (by circling back from deeper layers) can establish interplay between time-delayed data.

### 3.3.2.2   Activation function

Following the inspiration of natural neurons, the signal produced by an AN depends on the weighted sum of signals received from ANs in previous layer filtered with an *activation function*. Typically, activation functions take the form of $f(s) = \text{atan}(s)$, sigmoidal: $g(s) = 1/(1+e^{-s})$, etc. and can include a bias term. Since the operation of the computation elements comprises two steps: weighted sum of vectorial input into a scalar (input function), and output of the result to the next layer (activation function), different names are sometimes given to each part of a compound *transfer* function combination of the aforementioned ones.

The choice of activation function, *f*, can have an impact on the performance of the ANN and as a consequence activation functions themselves can be improved through learning. Such activation functions are usually referred to as *trainable*, *learnable* or *adaptable* and their value derives from their economy as many of the proposed approaches are equivalent to adding AN layers with non-trainable activation functions.[83]

Broadly one can classify activation functions into *fixed*-shape (essentially meaning they use no parameters) and *trainable* depending on whether the function itself gets optimized or not during the ANN training.

### 3.3.2.2.1   Fixed shape activation functions

Among the former we can include the *classic* ones such as sigmoid, tanh, step functions etc. and the *rectifier-based* that are inspired by the response of such electronic devices. Most examples of classic activation functions are bounded, asymptotically tending to a finite limit for both positive and negative bias, whereas rectifier-based ones tend to zero on the left and grow unbound for positive input. Bounded activation functions have shown excellent results in shallow ANN but may bring on *vanishing gradient* problems in training where minimisation of the merit figure stalls.

Among the rectifier-based functions, the rectifier linear unit (ReLU) $g(s) = \max(0,s)$ has gained pre-eminence for its performance and is a *de facto* standard. The fact that it is nondifferentiable at zero is a problem that leads to *dying* ANs when a large negative bias is learned triggering no response from the AN. This issue motivated variants such as leaky or truncated ReLU, sofplus, and many others that often marginally or only in specific cases improve over simple and reliable ReLU. Two examples are shown in **Fig. 15**: a sigmoidal activation in a) and a ReLU in b).

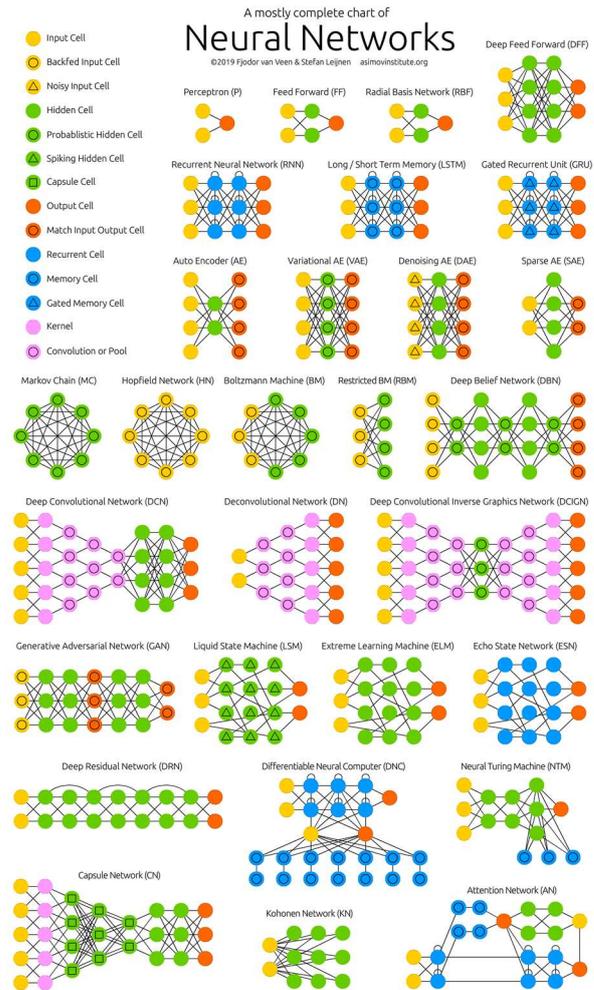

**Figure 14**. A mostly comprehensive representation of ANNs. (Adapted from www.asimovinstitute.org).

### 3.3.2.2.2    Trainable activation functions

Trainable activation functions are similar in shape to fixed ones but are flexible and contain parameters that can be learned during the training process. Two main methods are used to train activation functions: parametrization and ensemble methods. Parametrization is based on fixed-shape functions but including parameters that can be tuned to finely adapt the shape like for instance an adjustable sigmoid: $g(s) = \alpha/(1+e^{-\beta s})$ or the Swish $g(s) = s/(1+e^{-\beta s})$ with $\alpha$ and $\beta$ to be trained. The same can be done with other fixed shape functions *e.g.* tanh, ReLU etc. There seems to be little advantage in using trainable activation functions since it is possible to model them with a very shallow neural subnet which evidences that they offer no qualitative advantage over fixed-shape ones.

Alternatively, ensemble techniques have been developed that search for activation functions by using a set of basis functions[84] or by searching through combinations that often consist of linear combinations of one-to-one functions $g$ as in $g(s) = \sum \alpha_i g_i(s)$ where coefficients $\alpha_i$ are trained. It is worth noting that mere polynomials yielding $g_i(s) = s^i$ should be avoided since ANN with polynomial functions are not universal approximators.[85]

A seemingly different approach is taken with the so-called *Maxout* approach.[86] Usually, in classic models, the input to an AN is a scalar, $a$, resulting from the linear combination, $a = \mathbf{w}\mathbf{x}+b$, of signals, $\mathbf{x}$, from ANs in the previous layer with weights given by $\mathbf{w}$ (and, here explicit, bias $b$). In the Maxout approach instead, it is given by the largest component of a $k$-dimensional vector $\mathbf{a}$ whose components $a_i = \mathbf{w}^i\mathbf{x}+b^i$ , $i = 1... k$, are produced by a set of $k$ weights, $\mathbf{w}^i$ and $k$ biases, $b^i$. That is, Maxout($\mathbf{a}$) = max($a_i$). The number of trainable parameters is thus largely increased: where each AN involved a single set of weights $\mathbf{w}$ (one component for each AN from which it receives signal) now there are $k$ sets of weights. It can be shown that Maxout units are a generalization of classic rectifier-based units.[83] Several variations of this initial approach have appeared but one should not lose sight of the fact that "choosing the largest from a list" can be effected by an extra $k$-neuron hidden layer with one output neuron, with a fixed-shape activation function and, as such, could be implemented in the architecture of the ANN.

### 3.3.2.3    Feedforward

In most typical ANN signal only proceeds from ANs in layers closer to the input to those farther away from it. Thus, deeper ANs' states have no impact on shallower ones.

#### 3.3.2.3.1    Input layer

The *input* layer is a group of ANs that serve as interface with the exterior in order to receive the input data. Each AN is connected with the next layer of ANs and passes modulated signal according to a matrix, $\mathbf{W}_{in}$, whose dimension is $N_i \times N_1$ if there are $N_i$ input ANs and the first layer contains $N_1$ ANs.

#### 3.3.2.3.2    Hidden layers

In a simplified example, a number of *hidden* layers are connected, in a layer to layer fashion so that AN $j$ in layer $k$ is connected and passes signal to AN $l$ in layer $k+1$ according to element $(j,l)$ of weight matrix $\mathbf{W}^k$.

#### 3.3.2.3.3    Output layer

The operation of the ANN is probed through a set of AN that act as *output*. In principle, the dimension of the output vector can be any number and different methods can be devised to interpret the answer provided but in general there will be a number of AN whose signal will form a vector, $\mathbf{y}$, which, to simplify, will be the weighted $\mathbf{y} = g_{out}(\mathbf{W}_{out}\mathbf{x}_N)$ where $\mathbf{W}_{out}$ is the output matrix and $\mathbf{x}_N$ holds the signals emerging from the ANs in the last layer.

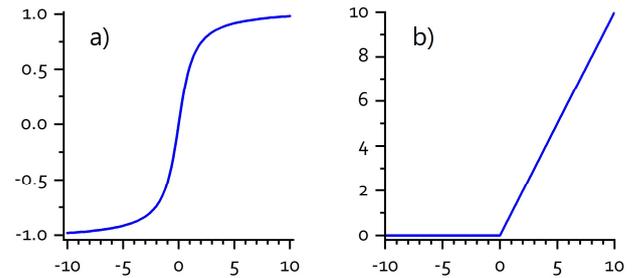

**Figure 15.** Classic activation functions: a) Sigmoid and b) ReLU.

### 3.3.2.4    Recurrent ANN

Not all ANN are feed-forward; when deeper layers' output can reach shallower ones, that is, signals *back-propagate*, the ANN is considered *recurrent*. See **Fig. 16**. Back-propagation acquires importance in sequential processes where data is fed to the ANN after the network has already been acted upon with some (earlier) data. RNN process an input sequence one element at a time, keeping record of their state hidden in a *state vector* that implicitly contains information about the history of all the past elements of the sequence.[87]

In these circumstances, input received by an AN from a deeper one (back-propagated) is short-term memory of its state (because it was acquired in previous processing cycle and is altered in the present one) that enables the network to process temporal context information.[88] The non-linear transformation of the input sequence, is useful when processing data with time serial character such as natural language.

At variance with the more widely used feedforward ANNs, in this connection topology, loops are formed. As a consequence, RNNs may develop self-sustained activation dynamics for signals traveling along its pathways, even in the absence of input which renders RNNs dynamical systems, whereas feedforward networks are, merely, functions.

The formalism distinguishes the non-temporal or static and temporal or dynamic (for which the RC is best suited) cases.[88] In the former, data $u(n)$, and target labels, $\mathbf{y}_t(n)$, take the form of list — number with index $n$, but with no particular order— of independent vectors (of different dimension) and the purpose of the method is to learn the functional relationship $\mathbf{y}(n) = y(\mathbf{u}(n))$ so that the error, for instance, the normalized "distance"

$$|\mathbf{y}, \mathbf{y}_t| = \sqrt{\frac{\langle |\mathbf{y}(n) - \mathbf{y}_t(n)|^2 \rangle}{\langle |\mathbf{y}_t(n) - \langle \mathbf{y}_t(n)\rangle|^2 \rangle}}$$

where the average is taken over all available pairs, is minimized.

In the temporal case, data come in a succession of iterated vectors —akin to a discrete time sampled signal— and the goal is to learn the rule of succession $\mathbf{y}(n) = y(... \mathbf{u}(n-1), \mathbf{u}(n))$ with the same minimization of error. At variance with the former, in this case the function *y* can be viewed as having memory and, after being trained on *past* data, being able to predict coming data (future of the signal). The assumption that a function exists that relates the data to the labels is equivalent to assuming that the labels are only unpredictable but for a noise term, $\xi(n)$, that limits the learning precision $\mathbf{y}_t(n) = y_t(\ldots \mathbf{u}(n-1), \mathbf{u}(n)) + \xi(n)$.

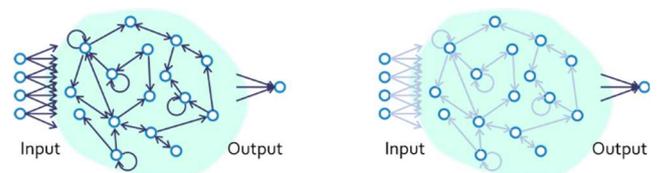

**Figure 16** Recurrent neural networks can operate in two ways depending on how the learning process is carried out. Traditional gradient descent tweaks the weights in *every* connection within the network and the input in order to minimize the loss function (left diagram, dark arrows). Reservoir computing selects only a few connections between the network and the output (readout weights) to minimize the loss function. Connections within the reservoir (light arrows) are left untouched.

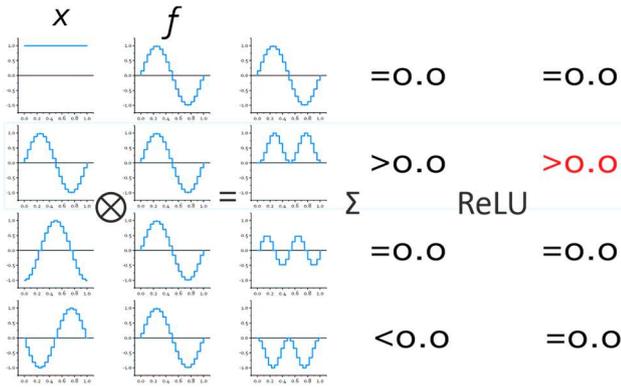

**Figure 17**. Feature detection in CNN. The feature to detect, *f*, is convoluted with the data, *x*, and passed to a ReLU filter that only yields a non-zero when the data contains the feature.

### 3.3.2.4.1    Expansion

Most often, tasks cannot be solved using linear relations between $\mathbf{u}$ and $\mathbf{y}_t$ in the form $\mathbf{y}(n) = \mathbf{W}\,\mathbf{u}(n)$ with $\mathbf{W}$ a $\dim(\mathbf{y}) \times \dim(\mathbf{x})$ matrix and it is customary to resort to non-linear expansions of the input data $\mathbf{u}(n)$ to higher dimension feature vectors $\mathbf{x}(n)$ so that the new problem is $\mathbf{y}(n) = \mathbf{W}_{out}\,\mathbf{x}(n) = \mathbf{W}_{out}\,x(\mathbf{u}(n))$ with $\mathbf{W}_{out}$ a $\dim(\mathbf{y}) \times \dim(\mathbf{x})$ matrix where $\dim(\mathbf{x}) \gg \dim(\mathbf{u})$. The functions $x(\mathbf{u})$ are called *kernels* or *expansions* depending on context.[88] This process of expansion is carried out thanks to the multiple connections between the neurons in the network which, taking the data fed through some input neurons, establish a larger dimension state which is, for convenience, probed through fewer output neurons. This expansion may include the raw data and additionally consider some bias.

In the temporal case the function to be learned depends on the data history and, therefore, so must the expansions: $\mathbf{x}(n) = x(\ldots \mathbf{u}(n-1), \mathbf{u}(n)))$ which may depend on a large (in principle infinite) amount of previous data. In order to avoid functions with an intractable number of parameters, practical implementations choose the last two, $\mathbf{x}(n) = x(\mathbf{u}(n-1), \mathbf{u}(n))$, and treat the new problem as a non-temporal one.

Neglecting bias, once initialized, a typical scheme of RNN, will take the current ($n$) system neurons' input, $\mathbf{u}(n)$, weigh it with input matrix, $\mathbf{W}_{in}$, and feed it to the network; add the immediately previous state, $\mathbf{x}(n-1)$, weighted with internal reservoir matrix ($\mathbf{W}$) and pass it through the activation function, $g$, to produce the current ($n$) state data vector: $\mathbf{x}(n) = g(\mathbf{W}_{in}\mathbf{u}(n) + \mathbf{W}\,\mathbf{u}(n-1))$. The output is read from current state with the expression $\mathbf{y}(n) = g_{out}(\mathbf{W}_{out}\,x(\mathbf{u}(n)))$ using a sigmoid or similar function as $g_{out}$.

More complicated recurrent feedback schemes can contain output feedback to include influence of the previous output in current state through ad hoc weight matrix $\mathbf{W}_{ofb}$.

### 3.3.2.4.2    Hopfield networks

*Hopfield* networks are comprised of layers in which all neurons send their output to every other. Transmission can be made to alter the state of neurons one by one or in rounds where all are altered synchronously. For the adequate definition of an energy functional, this ANN can be made to emulate an Ising model. If the ANs are perceptrons and the network is initialized with binary values ($s = \pm 1$) the system can converge to configurations ($s_i$) that minimize the energy.[89]

### 3.3.2.4.3    Boltzmann machines

Technically, the correspondence between Hopfield networks and the Ising model holds only at $T = 0$, something that can be solved by including a *Boltzmann* type of probability that the state is $s_i = 1$ (or $-1$) depending on the energy.

### 3.3.2.4.4    Long short-term memory networks

Long short-term memory networks are designed so that the ANN can be trained to keep memory for arbitrarily long intervals.[90] This is at variance with simple RNN where memory lasts for one cycle and learning takes too long. This technology is especially apt for speech recognition used in modern mobile phones and machine translation in web applications.

### 3.3.2.5    Convolutional neural networks

*Convolutional* neural networks (CNN or Conv-Nets) are ANNs that use convolution operations in place of general matrix multiplication (as in standard ANNs) in at least one of their layers.[91] CNNs are designed to process data that come in the form of (multiple) arrays as, for example, colour images captured by modern digital megapixel cameras, composed of three 2D arrays containing pixel intensities in the three colour channels. Data dimensions can rise to tens of millions.

Supervised learning can be crudely described as a high-dimensional interpolation problem. The usual $h(\mathbf{x})$ guessing deals with data where $\dim(\mathbf{x})$ is very large and volumes in the sample space grow exponentially with the dimension while available data only grows linearly. Therefore, a problem is often faced where data is scarce and finding the relationship between data and labels $y = h(\mathbf{x})$, (real for regression, indices for classification), is only possible if $h$ has some regularity that acts to reduce the dimension of the problem so that if some feature-detecting tool is useful somewhere in the data, it is likely to be useful elsewhere too. To solve it, data may be subject to *separation* by finding an intermediate function, $\boldsymbol{\eta}(\mathbf{x})$ such that if two data belong to different classes, *i.e.* carry different labels $h(\mathbf{x}_1) \neq h(\mathbf{x}_2)$, then they also carry different $\eta$-labels: $\boldsymbol{\eta}(\mathbf{x}_1) \neq \boldsymbol{\eta}(\mathbf{x}_2)$. The $\boldsymbol{\eta}$ function is a *projector* that reduces the dimensionality of the data, $\dim(\boldsymbol{\eta}) < \dim(\mathbf{x})$ and finding it is the crux of the problem since it, in fact, dispenses with the irrelevant components of the data vectors, those that make $h(\mathbf{x}+\mathbf{z}) = h(\mathbf{x})$ for every $\mathbf{z}$ belonging to the space orthogonal to that onto which $\boldsymbol{\eta}$ projects. Effectively this is finding the direction along which $h$ is constant.

Alternatively, a *linearization* can be tackled with variable change into a potentially much larger dimensional space than $\mathbf{x}$'s to subsequently find a projection along directions of constant $f$. For a mathematical framework and description of the foundations see [92].

In practice the implementation of learning in CNNs requires a special disposition of the ANN where layers fulfilling different tasks are connected. This responds well to the reduction of dimensions required and operates by creating feature detectors (kernels) that scan the data in a convolutional way shifting along the data and storing detection in a *feature map*. Different kernels detect different features so that each feature map contains the locations of each feature. Let data be a collection of time-varying signals storing some events like acceleration in a smart wrist-band fitness tracker. Take data sample represented by the array $\mathbf{x}$ with components $x_i (i = 1 \ldots d$, data dimension, recording length divided by sampling frequency). In order to detect features characterized by certain patterns of acceleration one defines filter vectors $\boldsymbol{\phi}^\beta$ with components $\phi_j^\beta$ ($\beta = 1 \ldots n$, number of features; $j = 1, \ldots, s$, filters size matching feature size and desired detail). When the filter is dot-multiplied by the signal it produces a map $\mathbf{m}^\beta$ with components $m_i^\beta = \sum_{j=1}^{j=n} x_{i+j-1}^\alpha \phi_j^\beta$ that is positive where the signal contains the feature describe by the filter. When passed through a *rectifier*, it outputs a value to the feature map. See **Fig. 17**. In this way, whenever a fluctuation similar to that stored in the filter is located in the data a positive value is stored in the corresponding feature map location. If the signal exactly contains an instance of the feature the convolution will produce the largest output in exactly the place the feature occurs as any displacement (during the sweeping) will shift the coincidence and reduce the convolution much like in the interpretation of cross correlation. A full processing of the signal will require as many filters as features are wanted or expected. Let's take a sixteen samples data vector as example: make **x** = [0 0 0 0 1 2 3 -2 -1 0 0 0 0 0 0 0] to be searched for a burst. The dimension five oscillation filter **ϕ** = [0 1 0 -1 0] when scanned step by step through **x** produces **x** ⊗ **ϕ** = [0 −1 -2 -2 4 4 -2 -1 0 0 0 0] which has only twelve components that passed through a ReLU produces the output [0 0 0 0 4 4 0 0 0 0 0 0] marking the occurrence of an oscillation around the fifth position in the signal. This is schematized in **Fig. 18**.

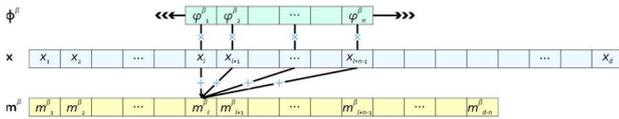

**Figure 18.** Convolution of data. For each data vector, **x**, every filter, **ϕ**, is scanned over the data (multiplying and accumulating) to produce a corresponding map vector components containing the information relative to the feature represented by the filter.

The sweeping step needs not be, and seldom is, one, which works towards dimension reduction. And the size of the kernel should be adapted to the kind of elemental featured sought/expected. The feature map is down-sampled in a *pooling* layer by averaging over neighbouring components. The convolution and pooling pair just described can be repeated so that hierarchically more complex features are detected until a full identification is achieved.

In Fig. 17 an example of a filter detecting oscillations is schematized. A filter to detect bursts could be similarly designed and, in a deeper convolutional layer, another could be used to detect burst-next-to-oscillation to identify exercise characterized by that kind of acceleration patterns. Obviously, care must be exercised in padding data so that kernel overlaps with data boundaries are properly dealt with. In actual applications three or four convolutional layers can recognize handwriting and twenty layers can distinguish human faces. At this level CNNs contain millions of parameters.

The economy in CNN arises from reusing the same parameters as kernels slide unaltered across data which results in a not fully connected network. Only the last layer is often fully connected.

### 3.3.2.5.1 Adversarial NN
A particularly interesting configuration using CNNs is the generative *adversarial* network (GAN).[93] It consists of two CNNs that are trained simultaneously in a competitive fashion.

One of the networks, the *generator*, is an inverse convolutional network that aims to produce random noise vectors —fake images— that the discriminator classifies as legitimate or counterfeit. The generator's training is inverse in the sense that, instead of improving the weights based on inputs, it improves the (synthesized) data to achieve a wrong classification. The second one, *discriminator*, is a standard CNN normally trained to take a true image as input and classify it, with a probability distribution, and then compete with the generator to identify input as fake or real.

The competition between the two contributes to improve the ability of the discriminator to tell fake from real data and the ability of the generator to create better fakes.

### 3.3.3. Deep learning training
Deep learning is the kind of learning executed by DNN. The term *deep* derives from the high number of layers often characterizing such ANN which, apart from the input and output interfaces, comprise a number of hidden layers. A very thorough overview in ANNs with a historical overview with close to one thousand references can be found in [94].

The training stage is essentially the same for all ANN and consists of providing data to the network and adjusting its parameters according to the success of the network prediction. Every ANN topology can have its own peculiarities but in order to describe the training a deep ANN with feedforward can provide the essential concepts. Let's assume one input layer, one output layer and several hidden layers; and, to facilitate formulation, let's assume that every AN receives input only from one layer. Assume further that data items come in the form of many pairs of vectors, $\{\mathbf{x}, \mathbf{y_t}\}^q$, with dim(**x**) the number of input neurons, dim(**y_t**) the number of output neurons; $q = 1, \dots N$, numbering the elements in the training data set and assuming the function associating data inputs to target labels as $y(\mathbf{x}) = \mathbf{y_t}$. Data is then fed to the *input* layer that transforms it according to $\mathbf{z}^1 = \mathbf{w}^1\mathbf{x}$ (with dim(**z**$^1$) the number of neurons in hidden layer 1) that is passed to the activation function $g^1$ which produces $\mathbf{y}^1 = g^1(\mathbf{z}^1)$ that constitutes the input to the next layer. Bias can be thought of as an extra weight applied to an auxiliary component of the data vector set equal to 1. The dimension of matrix $\mathbf{w}^1$ is the number of input neurons by the number of neurons in the first hidden layer. This process is iterated to next layer where $\mathbf{y}^1$, through $\mathbf{w}^2$ produces $\mathbf{z}^2 = \mathbf{w}^2\mathbf{y}^1$ and then activation $g^2$ giving $\mathbf{y}^2 = g^2(\mathbf{z}^2) = g^2(\mathbf{w}^2\mathbf{y}^1)$. Now $\mathbf{w}^2$ is a matrix whose dimension is the number of neurons in hidden layer 1 by the number of neurons in hidden layer 2. In general, a layer $l$ receives input from the preceding layer so that the vector of inputs to its neurons is $\mathbf{y}^{l-1}$ has as many components as neurons in the layer that summed through the $l$-th weights matrix $\mathbf{w}^l$ by $\mathbf{z}^l = \mathbf{w}^l\mathbf{y}^{l-1}$ or explicitly $z_i^l = \sum_j w_{ij}^l \, y_j^{l-1}$ and passed through the activation function $y_i^l = g^l(z_i^l) = g^l(\sum_j w_{ij}^l \, y_j^{l-1})$. This propagates forward to the output layer where the output vector **y** is compared to the target $\mathbf{y_t}$ for each data pair. For the sake of simplicity all the $g$ activation functions in a hidden layer were assumed equal which needs not be the case. See the flow diagram in **Fig. 19**.

#### 3.3.3.1 Gradient descent
Initially the network is created with random values in all of the weights and biases and the training aims at tuning them based on the difference between the target labels and the output of the ANN. How this difference, the error or cost function, is measured depends on the nature of the output but it can essentially be reduced to a Euclidian distance $E = |\mathbf{y}-\mathbf{y_t}|^2$. Thus, the procedure to update the weights is iterative and typically sets a learning rate, $0 < \eta < 1$, and for round $t+1$ of data input makes $\mathbf{w}^l(t+1) = \mathbf{w}^l(t) - \eta \, \nabla_\mathbf{w} E(t)$.

Minimization of the error can be achieved by backpropagating[95] the error through the layers' weights so that each new round of data (epoch) walks the weights space descending the distance function along the steepest slope by computing the gradient.

#### 3.3.3.2 Backpropagation
In order to bring **y** nearer $\mathbf{y_t}$ it is necessary to nudge the weights so that for the next input **x** the ANN is closer to the corresponding target and the error is smaller. To do this it is necessary to know how the cost function varies as every weight varies, that is $\nabla_\mathbf{w}E$ (and possibly, the same with the activation functions) starting with the last layer. If no biases are considered (or if they are included in the data) this reduces to computing $\partial E/\partial \mathbf{w}^L = \partial E/\partial \mathbf{y}^L \cdot \partial \mathbf{y}^L/\partial \mathbf{z}^L \cdot \partial \mathbf{z}^L/\partial \mathbf{w}^L$. These functions depend on the actual structure of the ANN (number of layers, number of ANs in each layer and their connections) and the activations functions and biases.

Weights and activation functions for the last layer depend solely on the cost function but for previous layers further use of the chain rule is necessary to find the gradient components as in $\partial E/\partial \mathbf{w}^{L-1} = \partial E/\partial \mathbf{y}^L \cdot \partial \mathbf{y}^L/\partial \mathbf{z}^L \cdot \partial \mathbf{z}^L/\partial \mathbf{y}^{L-1} \cdot \partial \mathbf{y}^{L-1}/\partial \mathbf{z}^{L-1} \cdot \partial \mathbf{z}^{L-1}/\partial \mathbf{w}^{L-1}$ although the first two factors can be taken from the previous calculation. So, every new layer we reach updating backward, we reuse two factors up to the first layer

$$\mathbf{x} \otimes \mathbf{w}^{(1)} \to \mathbf{z}^1 > g^1 \to \mathbf{y}^1 \otimes \mathbf{w}^{(2)} \to \mathbf{z}^2 > g^2 \to \mathbf{y}^2 \otimes \mathbf{w}^3 \to \mathbf{z}^3 > g^3 \to \mathbf{y}^3 \dots$$

$$\dots \mathbf{y}^{l-1} \otimes \mathbf{w}^l \to \mathbf{z}^l > g^l \to \mathbf{y}^l \dots$$

$$\dots \mathbf{y}^{L-2} \otimes \mathbf{w}^{L-1} \to \mathbf{z}^{L-1} \to \mathbf{y}^{L-1} \otimes \mathbf{w}^L \to \mathbf{z}^L > g^L \to \mathbf{y}^L \, \overset{?}{=} ? \, \mathbf{y}_t$$

**Fig. 19.** Flow and processing of the data towards the output and testing. In each layer input from preceding layer is weighed, summed, filtered and passed to the next layer. In the output layer, the result is tested against data target and a *cost* estimated.

$$\frac{\partial E}{\partial \mathbf{w}^l} = \frac{\partial E}{\partial \mathbf{y}^L} \frac{\partial \mathbf{y}^L}{\partial \mathbf{z}^L} \dots \frac{\partial \mathbf{z}^{l+2}}{\partial \mathbf{y}^{l+1}} \frac{\partial \mathbf{y}^{l+1}}{\partial \mathbf{z}^{l+1}} \frac{\partial \mathbf{z}^{l+1}}{\partial \mathbf{y}^l} \frac{\partial \mathbf{y}^l}{\partial \mathbf{z}^l} \dots \frac{\partial \mathbf{z}^2}{\partial \mathbf{y}^1} \frac{\partial \mathbf{y}^1}{\partial \mathbf{z}^1} \frac{\partial \mathbf{z}^1}{\partial \mathbf{w}^1}$$

In this way it is possible to work out the derivatives all the way to the first layer hence the *backpropagation* term.

### 3.3.4. Reservoir computing

With common activation functions, RNNs have shown to be Turing equivalent so they can, in principle, complete complex temporal ML challenges. However, training RNNs by gradient descent-based methods, may suffer from slow convergence rates that limit their practical applicability. It is also known that RNNs suffer from unstable dynamics due to expected evolution of the network eventually leading to bifurcation;[96] they present heavy computation demand brought about by the fact that the impact of single parameter update is amplified through backpropagation; moreover, strong gradient information dilution in time occurs owing to the inherent back and forward spreading of dependencies. Reservoir computing (RC), arrived at from different independent approaches as a solution for this problem —Liquid State Machine[97] and Echo State Network[98] — is based on separating the RNN into a dynamic *reservoir* that operates as a nonlinear temporal expansion and a, usually linear, *readout* often not subject to recurrence and has been proposed as a unifying denomination for three learning methods.[99]

RC profits from a particular learning protocol where optimization is carried out only on a reduced number of weights directly related to the output. In this procedure an RNN called the reservoir, is randomly created and kept unchanged throughout the training. Input is fed through a reduced number of neurons that create a *state* in the reservoir as a nonlinear transformation of the input. Output is generated as a linear combination of a reduced number of neurons' signal. The (few) linear combination coefficients are optimized by simple regression. Briefly, a reservoir computing system consists of a block of ANs for mapping inputs into a high-dimensional space and a readout for pattern analysis from the high-dimensional states in the reservoir. The governing equations look like:[100]

$$\mathbf{x}(t+1) = f(\mathbf{W}_r^r \mathbf{x}(t) + \mathbf{W}_{in}^r \mathbf{u}(t) + \mathbf{W}_{out}^r \mathbf{y}(t) + \mathbf{w}_b^r)$$
$$\hat{\mathbf{y}}(t+1) = f(\mathbf{W}_r^{out} \mathbf{x}(t) + \mathbf{W}_{in}^{out} \mathbf{u}(t) + \mathbf{W}_{out}^{out} \mathbf{y}(t) + \mathbf{w}_b^{out})$$

where $\mathbf{x}(t)$ represents the state of the reservoir, $\mathbf{u}(t)$ the vector of inputs and $\mathbf{y}(t)$ the output at iteration time $t$. See **Fig. 20**. The weight matrices with superindex 'r' refer to the reservoir acting on the state, input or output vectors that modify the reservoir state; and those with superindex 'out' act on state, input and output to modify the output vectors. The reservoir is fixed ($\mathbf{W}$) and only the readout ($\mathbf{W}^{out}$) is trained with a simple method such as linear regression and classification. Thus, the major advantage of reservoir computing compared to general RNN is fast learning, resulting in low training cost. Another advantage is that the reservoir without adaptive updating is amenable to hardware implementation using a variety of physical systems, substrates, and devices.[101] Notable hardware implementations such as waves excited in a bucket[102] or cat's brain[103] highlight the power of such simple approach.

The vast success of RC may be grounded in several features like its computational universality for continuous time varying signals relying on bounded resolution both in time and signal. Its power compared to previous RNN results was shown to be many orders of magnitude superior in predicting chaotic dynamic behaviour for instance.[104] Both in liquid state machines and echo state networks the readout, performed from a few ANs only in the back of the ANN, makes the operation of the reservoir similar to that of the kernel in other RNNs by projecting the input data into a high dimensional space that facilitates separability. A comprehensive review on RC methods and taxonomy can be found in [88].

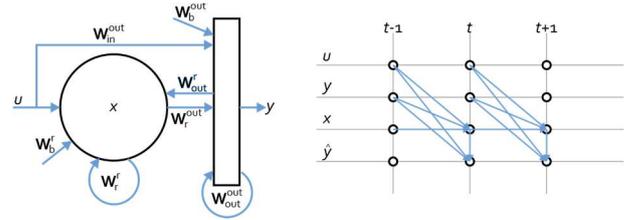

**Figure. 20.** Reservoir computing elements. Adapted from ESANN 2007 Proceedings - 15th European Symposium on Artificial Neural Networks

#### 3.3.4.1 Liquid state Machine
*Liquid state* machines were proposed[105] with a computational neuroscience inspiration in that they were aimed at understanding properties of neural circuits. Understanding computation in an RNN fed with time-varying in general and integrate-and-fire in particular, signal constituted a phenomenal challenge.[89]

Owing to their origin, more oriented to emulate biological processes, LSMs used sophisticated biological models of dynamic synapses and realistic *spiking* integrate-and-fire neuron connections. In this model the reservoir is called the *liquid* in reference to the medium sustaining ripples on its surface that can be excited by short wave pulses. The readout often extracts information through a feedforward multilayer ANNs just like echo state networks.

Although not requiring sequential transitions between well-defined internal states like those in the Turing machine, an LSM is based on equally rigorous mathematical framework that guarantees, under idealized conditions, universal computational power but adapted to real-time processing of time varying signal as in biologic settings.[97] The astonishing finding was that a readout neuron that receives signals from immensely many neurons in a *liquid* can learn to extract the relevant information from the high-dimensional ephemeral states of the reservoir and produce stable readouts, that is, it can build a notion of equivalence between reservoir dynamical states and therefrom interpret unexperienced novel inputs—invariant readout. Furthermore, multiple readout units can be trained to extract different kinds of information from the same reservoir state dynamics thus achieving parallel real-time computing. **See Fig. 21**.

#### 3.3.4.2 Echo state networks
Echo state networks (ESN), the other pioneering model, were proposed[98] based on the observation that for certain RNN satisfactory results were achieved by training just a linear readout, thus avoiding the trouble of training the internal weights. In this case the reservoir states were termed echoes. An ESN consists of a random RNN driven by a one- or multi-dimensional time varying signal and acting as a complex nonlinear dynamic filter. Obviously, like LSMs, they can operate on a single input to simultaneously solve several tasks by using as many readouts with different specialized training. For reasons that can be intuitively explained ESNs, like LSM on the edge of stability, work optimally on the edge of chaos where internal states diverge for infinitesimally differing inputs thus producing maximally different readouts.[106]

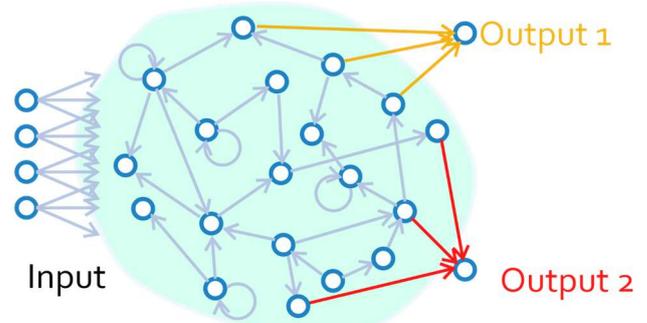

**Fig. 21.** Liquid state machine processes one input and extracts two kinds of information at the same time by training different sets of output weights.

### 3.3.4.3    Backpropagation decorrelation
A further proposal suggested an alternative learning rule called back-propagation decorrelation.[107] Its origin was the lack of a simple learning rule for RNNs similar to backpropagation in feedforward ANNs. The main difference with previous models is the feedback from the readout neuron into the reservoir which renders it effectively nonlinear.

### 3.3.4.4    Extreme Learning machines
The learning speed of feedforward ANNs is, in general, much slower than desirable. This constituting a major bottleneck entailing an efficiency and sustainability problem. The principal reasons behind this fact are the slow gradient-based learning algorithms used in the training process and (connected to it) the need to tune all parameters (weights and biases). Extreme learning machines, similar to RC in randomly initialising hidden ANs' parameters and not optimising them, differ in that the reservoirs of RC architectures may and often does contain recurrent connections, giving it a short-term memory, whereas ELMs, often single-layer, are feedforward architecture with no short-term memory.

The controverted extreme learning machine (ELM) for single-hidden layer feedforward ANNs was proposed based on the proven fact[108] that the input weights and hidden layer biases of such networks can be randomly assigned if the activation functions in the hidden layer are infinitely differentiable. The reason is that for any randomly chosen set of weights and biases the output weights can be analytically determined. The method can learn thousands of times faster than conventional learning algorithms for feedforward neural networks because it reduces the problem to a linear one of matrix inversion.[109]

### 3.3.5.    Other models
While ANN and SVM are the most widely used learning approaches other models can also be considered.

### 3.3.5.1    Function approximation
Most often classification assigns discrete or even binary labels as classes. When the labelling is continuous, say real numbers in an interval, the classification is designated *function approximation* and tackled using regression techniques like minimization of the sum of squares of residuals (difference between predicted and actual label) also called *method of least squares*. This technique is close to the familiar fitting of parametrized functions in overdetermined systems (many more data than parameters).

### 3.3.5.2    k-nearest neighbours
The *k-nearest neighbour* algorithm takes a decision on label assignment to a new datum by averaging (for regression) or majority voting (for classification) among its $k$ nearest neighbours in the training data. The reason is that the most intuitively appealing classification procedures are those for which sample observations nearer an unclassified datum should have the greatest weights.[110]

### 3.3.5.3    Decision trees
In *decision trees*, used for classification, the data features (or their vector components) are examined in sequence, each time taking a decision and advancing to the next component which usually leads from general to particular class clusters until the final assignment like following a plant classification manual. Best known applications of decision trees are in data analysis of particle accelerators.[111]

### 3.3.5.4    Boosting algorithms
Most often the search for classifiers is restricted to choosing the *best* element in a family or function space. However constructing a good classifier can benefit from the finding[112] that under certain conditions, there exists a linear combination of *poor* hypotheses —that are better than random in certain parts of the data— which is better performing overall and can be a strong learner.

Briefly, this can be carried out by feeding a weak learner selected data samples from the training set so that different weak hypotheses are obtained that are successful in different regions of the data space. Then, weights of the weak learners in the composite space are optimized on the complete data space. That this is possible was eventually shown not to be surprising after all when it was demonstrated that a weak learner is as strong as one whose error can be made arbitrarily small.[39]

## 3.4.    Spiking ANNs

It is generally accepted that, in their evolution, ANNs have gone through three generations. A first generation based on McCulloch-Pitts neurons,[48] with step-function activation functions (threshold gates) that produce *digital* output and can compute any *boolean* function. A second one, usual ANN (both feedforward and recurrent) are characterized by static activation functions and use static or continuous varying signals as information carrier with a continuous range of output values in response to weighted sums of inputs. The latter, being able to compute functions with analogue input and output— they are *universal* in fact— can also compute boolean functions, and do it more economically than threshold gates circuits. A third generation of networks, *spiking* ANNs (SNNs),[113] are comprised of neurons that operate on an *integrate-and-fire* process using information encoded as pulses.[114] Inspired by biological nervous systems, SNN neurons use discrete pulses —action potentials— to compute and transmit information carried by spike times (*pulse* encoding) or spike rates (*rate* encoding) rather than by intensity.[115] Pulse encoding is more powerful for the wide range of information that may be encoded by the same number of neurons.

Unlike previous generations' ANNs —where all signals input to the neuron contributed to its response, implicitly meaning they arrive simultaneously, only intensity being relevant— SNNs operate in a way that neurons response depends on signal *time of arrival*. Thus, SNNs are a closer match to actual brains and the systems to study if one wants to understand brain operation. Their value resides in their economy: SNNs are, for the same number of ANs, computationally more powerful than the previous generations of ANNs and owing to the sparse character of the signals (pulsed) they are also less power hungry. That, in addition to the fact that any function computable by a small (sigmoidal type) ANN can be computed by a SNN too, is at the basis of their worth. They still present challenges though, like training. Usual transfer functions are not differentiable which precludes the use of backpropagation; additionally, spike trains are formally sums of Dirac deltas lacking derivatives. Thus, in many SNN implementations, learning is restricted to one layer, leaving multi-layer network training an open area.[116]

SNNs can adopt both feedforward and recurrent architectures and also hybrid ones that include synfire chains[117] and reservoir computing nets. As truly dynamic networks useful for computation purposes and to understand brain operation, SNNs require and inspired modelling that necessarily must range from single neuron through mesoscopic to the macroscopic scale.[118]

### 3.4.1.    Bioinspiration
SNNs' mathematical models date back to early XX century (see [113] and references therein) notably with the Hodgkin–Huxley model, or conductance-based model.[119] The latter accounts for the *action potential* (local fluctuations of voltage across the neural cell membrane like in **Fig. 22**) considering linear (from *voltage gated ion channels*) and nonlinear conductances (from *leak channels*), voltages (from *electrochemical gradients*) driving ions flow and currents (from *ion pumps*) in an electrical model.

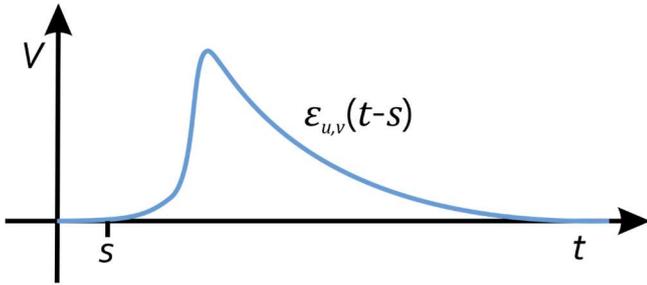

**Fig. 22** Pre-synaptic potential. Incoming spikes from presynaptic neurons are weighed and then integrated. When their integral overcomes the threshold, a postsynaptic spike is generated.

### 3.4.2. Operation

Formally such networks consist of a finite set of neurons among which synapses (connections mediated by neurotransmitters in biological cells) are established, characterized by (trainable) *weights*, $w$, and *response* functions, $\epsilon$. A spiking neuron $\nu$ will fire whenever its potential $P_\nu$ (representing the membrane electric potential at the trigger zone) reaches certain threshold $\theta_\nu$. Such a potential is the sum of excitatory (that raise) and inhibitory (that lower) postsynaptic potentials resulting from the firing of upstream neurons. The distinguishing feature of SNNs arises from the fact that response of AN $\nu$ at time $t$ depends on the firing of neuron $u$ at previous time $s$ through the term $w_{u,\nu}\,\epsilon_{u,\nu}(t-s)$ in a sum over all presynaptic neurons with $w \geq 0$ (often a constant) and $\epsilon$ positive or negative depending on their *excitatory* or *inhibitory* character. Response functions of biological neurons present a fast rise, slow decay pulse shape that can be modeled by, for instance $\epsilon = \frac{t}{\tau}\,e^{1-t/\tau}$. In a typical biological neuron, the resting membrane potential is around –70 mV; it changes temporarily by at most a few mV in response to a postsynaptic action potential and has a firing threshold of around –50 mV. Thus, the time dependent potential for neuron $\nu$ at time $t$ will be given by $P_\nu(t) = \sum_u \sum_{s<t} w_{u,\nu}\epsilon_{u,\nu}(t-s)$ where the $u$-summation runs over all presynaptic neurons and the $s$-summation runs over all $u$'s firing times prior to $t$. Delays can also be considered so that the response becomes $\epsilon(t - s - d)$ and in some cases several connections with different delays are considered in a single synapse.[120]

The time dependence is intrinsic to SNNs since spiking neuron response presents periods in which response to presynaptic action potential behaves differently. Once a neuron $\nu$ has fired, it enters a *refractory period* in which it will not fire again for a few ms, no matter how large its current potential $P_\nu(t)$ gets. Then, for a few further ms it is still reluctant to fire, requiring a larger value of $P_\nu(t)$ than usual (*relative refractory period*). See **Fig. 23**. This fact can be modelled by using a time-dependent threshold function $\theta_\nu(t-t')$ where $t'$ is $\nu$'s time of most recent firing. A useful $\theta_\nu(t-t')$ is such that $\theta_\nu(\infty) = \theta_\nu(0) \geq 0$ and diverging positive for small $t-t'$ (which reflects the refractory period).

In a practical computing SNNs, for a certain set of ANNs, selected as input, firing times are not determined by the preceding protocol but are set according to the coding of input data. From the rest of ANs a set is selected as output from whose spike trains computation results are extracted. In simulations, stochastic or noisy networks, $P_\nu(t) - \theta_\nu(t-t')$ gives the probability for neuron $\nu$ to fire at time $t$ having fired at $t'$ last time. It is also possible to consider delays

$\Delta_{u,\nu}$ to model the time it takes for an action potential from ANN $u$ to have an effect on $P_\nu(t)$. For biological neurons modelling, these delays may be different depending on axon length and other factors and their specifics can be tuned by the learning algorithm.

A comprehensive overview of structures and working principles of neurons and synapses of the biological nervous system, and hardware implementations of spiked computing inspired by them, can be found in [121].

### 3.4.3. Coding

In order for the SNN to work, the input (analogue) signal must be encoded into spikes. This can be achieved by several methods including *rate* based, *temporal* coding or *population* coding. A sequence of firing times (spike train) can then be described with a sum of delta functions $S(t) = \sum_f \delta(t-t_f)$ if spikes occurring at times $t_f$ are labeled by $f = 1, 2, \ldots$ and $\delta$ is the Dirac delta function. An electronic equivalent using binary spikes is depicted in **Fig. 24**.

Initial observations in frog cutaneous cells that responded with more spikes to higher mechanical pressure led to the assumption that information was encoded in the firing rate[122] which has been the dominant paradigm for many years. However, more recent results suggest that, at least in some neural systems, efficient processing is more likely the result of precise timing of action potentials. In biological systems behavioral response is too quick to rely on the estimation of firing rates (they require many firing events to determine the rate; the more the higher the desired precision). Consider that typical spike duration is in the ms range while some systems have shown reproducible sub-millisecond response and reliable discrimination of time intervals of $10^{-8}$ s have also been observed. In order to compute with rate codes in short time scales it is believed that biological neural networks resort to *pooling*, where a large number of spiking neurons are probed to achieve fast, reliable, averaged spiking rate estimations. The *resonant burst model*, that uses membrane potential oscillations through a resonance phenomenon, was suggested to excite selected downstream neurons whose frequency is tuned to that of short bursts of spikes.

Temporal coding can be implemented in several ways usually referring to how the coded response relates to a provided stimulus. It is strongly associated to neurosciences and how different biological neurons actually operate. Not all implementations offer practical inspiration for computing. In *time of first spike*, information is carried by the lapse of time between the start of the stimulus and the first spike. *Rank-order* coding places the information in the order a population of neurons produce their first (and usually only) spike. *Latency* code, is a scheme, akin to rate coding, that conveys information in the distances between successive pulses. Coding by *synchrony* assumes different neurons encoding partial information on the same stimulus fire synchronously. *Phase* coding uses a periodic reference baseline signal to encode the information in the relative phase with respect such oscillation.

### 3.4.4. Training and learning

Given that information is contained in firing times—whatever the coding scheme—the goal the SNN will pursue is to learn a set of target firing times $\{t_j^f\}^q$ on output neurons $j = 1,2\ldots N_o$ when a set of input firing times $\{t_i\}^q$ is fed at the input neurons $i = 1,2,\ldots N_i$ when $q$ such patterns are used for training. One way to implement a back-propagation strategy[120] resorts to an error or cost function that can be defined by computing the added differences between that *actual* and the *target* firing sequences at the output, $E = \sum_j (t_j^a - t_j^f)^2$, from which the partial derivatives with respect to weights can be obtained.

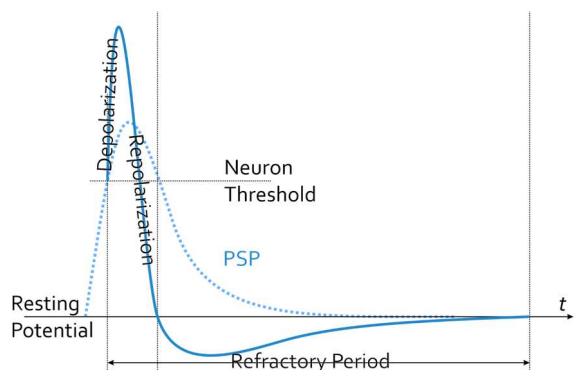

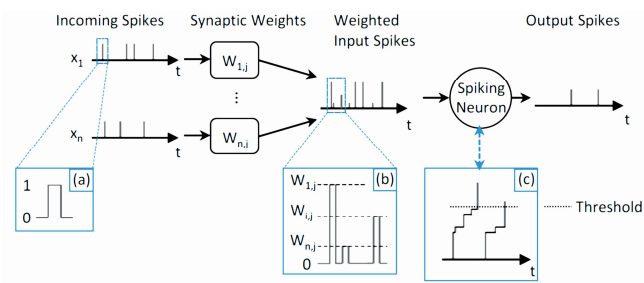

**Fig. 24.** Incoming spikes from presynaptic ($x_1, ... x_n$) neurons (inset a) are weighed ($w_{1,j}, ... w_{1,j}$) and then integrated (inset b). When their integral overcomes the threshold (inset c), a postsynaptic spike is generated. Adapted from M. Bouvier, A. Valentian, T. Mesquida, F. Rummens, M. Reyboz, E. Vianello, E. Beigne, *ACM J. Emerg. Technol. Comput. Syst.* **2019**, *15*, 1.

**Fig. 23.** Action potential (continuous line) in spiking neurons as response to presynaptic potential (dotted line). Absolute refractory period is shown. Relative refractory period only covers the time when potential goes below the resting potential.

Neuroscience has identified multiple variations of a learning rule of the type spike-timing-dependent plasticity (STDP). Its essential feature is that the weight connecting two neurons across a synapse (*efficacy*) is adjusted according to their relative spike times. In unsupervised learning, when a presynaptic neuron fires shortly before the postsynaptic one, their connecting weight is strengthened (long-term potentiation, LTP). If the order is the opposite the relationship cannot be causal and therefore the weight is weakened (long-term depression, LTD). This is considered one of the principal mechanisms of neural *plasticity* (the capability of synaptic connections to change their strength) and believed to be the basis of learning and memory.

SNNs have proved their potential with every learning model—unsupervised, supervised and, in a lesser way reinforcement—spanning a broad gamut of applications[123] ranging from image or odour recognition to spatial navigation and path planning through decision making or rehabilitation. But they have also become useful in helping achieve a deeper understanding of actual biological neural circuits and the study of many brain regions.

Most common supervised learning methods tune weights via a gradient descent of a cost function resulting from the *distance* between the target label and the computed result. But in SNNs the derivatives leading to updated weights, involving activations functions and cost function gradients present problems that have generally been addressed by using substitute or approximate derivatives. Additionally, issues emerge with the use of weight in the backward propagation that are experiencing progress using new feedback strategies including random weights. A major restriction in training SNNs is the nature of available data: the ideal situation would be input in the form of spike trains with information encoded in the timing whereas actual data is in the form of static-image datasets. Such data needs to be transformed prior to use.

Training adopts two major strategies: *conversion-based* and *spike-based* approaches.[124] In the former the goal is to obtain a SNN that offers the same *mapping* of input data onto output result that a given DNN yields.[125] This requires weight rescaling and normalization methods leading to a good matching of the leak time, refractory period, membrane threshold etc. In this way, training the SNN in the temporal domain is avoided at the expense of having to test the DNN on transformed static data. SNNs trained in this manner suffer from some issues such as the fact that timings are positive whereas DNN handle signals that can be both positive and negative (impact on accuracy); the difficulty to optimize firing rates with minimal loss (impact on performance) etc. While conversion-based SNNs have reached state of the art accuracy, they still operate slowly and with low energy efficiency.

In the spike-based approach two major roads are used adopting supervised or unsupervised learning. Early work with single layer *supervised* SNNs led to multilayer networks by adapting loss minimization techniques to spike-based data through global backpropagation with approximate differentiable functions or stochastic gradient descent. So far, although more computationally efficient than conversion-based techniques, supervised SNNs have failed to match them. *Unsupervised* SNNs are more easily adaptable to near brain-like computing both in algorithmic and hardware aspects and the former have shown very successful results with written character recognition even with a single processing layer.[126] The so-called vanishing forward-spike propagation is a serious hurdle in the development of deeper SNNs which constitutes current research focus.

Although both theoretical and experimental results indicate that *deep* ANN (many layers of few ANs) perform better than *wide* ones (many ANs in few layers) the performance of deep SNNs is poorer in general. The reader can find a comprehensive summary of recent DL models developed for SNNs in [116].

## 4.     Hardware implementations of AI

The von Neumann architecture is well suited for scientific computing where high precision data representation is essential and Boolean calculations must be accurate. This sequential kind of processing is adequate to deploy computing procedures closely adapted to mathematical sequential logic and reasoning but not very efficient in AI applications where *parallel* computing is essential. Beside its serial character, the *memory wall* (bottleneck produced by the data being shuttled from memory to processor and back) and difficulty in heat dissipation prompted a search for alternatives resorting to different technologies.

Although most ANN applications in commercial use are still performed by software, some entail such processing demands in real time (*e.g.* video streaming) that they impose truly parallel processing hardware. Current literature is abundant in hardware implementation reviews and it is a common feature to mention the difficulty in presenting a clear-cut classification because the many factors to consider (network topology, structure, information carrier, material, algorithms, number of layers, activation function, number of inputs, number of neurons, synaptic connections, desired application, digital/analogue operation, precision, etc.). [127]

We review the hardware implementations by breaking architectures down in two broad categories. The first one comprises those close to the classical computer —whose materials aspects are highly evolved and optimised in the CMOS technology. The second category encompasses architectures where memory is used also to compute embodying a neuromorphic inspiration —which are more in their infancy as a technology— and still needing much progress in research and technology. These architectures use different building blocks that, for information storage and processing, rely less on charge and more on other physical magnitudes and very often interweave with classic electronics.

## 4.1.    Computer-science oriented hardware

Apart from the current massive availability of heterogeneous data provided by modern internet connectivity and widespread sensing devices capabilities, the recent surge of ML techniques can be ascribed primarily to a substantial growth in computing power and the ability to train networks on graphics processing units (GPUs). Increasing complexity and improved accuracies demand larger and deeper NNs which translates in more computing power.

Normal computers' CPUs often resorted to complementary dedicated processors for special tasks dubbed *accelerators*. These include video, sound, graphic cards, digital signal processors, etc. Some of these were adopted when AI computing demands rose in recent years leading to heterogeneous computing where systems or even chips comprised specialized processors aimed at specific tasks.

Computer science proposes special-purpose hardware optimized to perform neural-network inference and training.[128] Here the approach mainly involves explicit algorithms executed on CMOS technology chips. Of these algorithms, the prevailing non-spiking artificial neural networks —inspired in part by the cortex in terms of spatial complexity— have made substantial progresses in dealing with specific tasks, such as image classification, speech recognition, language processing and game playing.[129]

A systematic literature review that focuses on exploring the available hardware accelerators for the AI and ML tools is out of the scope here and can be found elsewhere.[130]

### 4.1.1.    CPU

The CPU is the heart of the traditional computing scheme. Its architecture is optimized to a general-purpose computing strategy, an ability that endows the systems with the capability to tackle any solvable problem and depends only on programming. At the same time, its efficiency in certain tasks is penalized. For this reason, ordinary computers devoted to high power processing often make use of special or additional processing accessories.

### 4.1.2.    Accelerators

Traditionally, computing algorithms are *ex professo* designed for specific problems, implemented in high-level languages and only so often are they threaded so that some parts of the computation occur simultaneously in different parts of the processor. Massive parallelization however was difficult to achieve because of Amdahl's law.[131] For the massive amounts of data processing required in big data handling, the standard instruction set architectures are not best suited and AI accelerators, high-performance parallel computation devices specifically designed for the efficient processing in neural networks, are preferred.

### 4.1.3.    GPU

Graphics processing units (GPU) enhance parallelism by using many cores (sometimes more than 100), each of which may enjoy a dedicated high-throughput connection with the memory.[132]

The mathematical backgrounds for neural networks and graphical image manipulation are so similar: highly parallel and based on matrix algebra, and multiply-accumulate (MAC) operations, that graphics processing units gained high popularity in AI computing, both in DNN and SNN applications. Some manufacturers now incorporate neural network specific hardware in optimized hardware with towards 20 pJ/MAC operation in terms of energy consumption.

### 4.1.4.    FPGA

Field programmable gate arrays are reconfigurable devices that can be configured by the user to suit a specific task thus allowing the adaptation of systems hardware and software together.

At a comparably lower cost, they offer the flexibility of software but suffer from a lower circuit density as compared to high performance processors, which typically limits to thousands the numbers of neurons that can be implemented. Since they are programmable —the hardware is prepared to be *written* by the user before the software is deployed— they can operate in any algorithmic approach.

### 4.1.5.    ASIC

ASICs are application-specific integrated circuits and have been specifically designed to implement ANN computations. Their architecture is optimised for high-speed matrix multiplication and fast bidirectional memory access. For their nature they can be designed directly to carry out AI algorithms so they can be categorised as neuroscience oriented.[127] They are most often developed by internet and computer companies such as Intel,[133] Google,[134] IBM,[135] etc. and tend towards 1pJ/MAC operation in terms of energy consumption.

## 4.2.    Neuroscience oriented hardware

Neuromorphic ANN circuits try to make a close parallel to how biological neurons are networked and operate mostly in analogue mode. Often the visual systems of superior animals are taken as inspiration. As these systems generally handle enormous amounts of information in parallel, they are liable to *flooding* a phenomenon that stimulated the *selective attention* mechanism whereby only relevant fractions of the information are taken into account and serially processed. Less general categories attending to how signal is coded, where data is kept, different operations are executed, etc. can be considered which also vary in how appropriate they make the implementation depending on the learning strategy, neural processing, etc. and the application problem tackled.[127]

Microelectronic chips with a neuroscience inspiration are characterized by in-memory computing and find their predecessors in the early attempts with mechanically acted potentiometers for weights[136] or with variable electrical resistors.[137] Naturally these kinds of devices only became practical with the advent of microelectronics technology and chip fabrication with very large scale integration but remained largely governed by Kirchhoff's laws and semiconductor physics.

The von Neumann bottleneck, created by the incessant back and forth transfer of data between memory and processor, has no parallel in the way the brain operates. Instead, neuroinspired computing circuits should implement physical neurons interconnected by synapses that perform processing and memory storage *in situ* in a nonvolatile way. This is complicated to implement using CMOS technology alone because many transistors are needed to emulate each neuron and additional memories are needed to implement synapses. Transistors are a few nm yet CMOS-based artificial neurons and synapses are several micrometres wide which severely hampers large scale integration. Consider, for instance, that typical image recognition algorithms comprise millions of neurons and billions of synapses. This makes current neuromorphic systems bulky because many chips need to be assembled together reaching, in the most advanced cases, cubic metres.[138] An additional difficulty CMOS technology faces in trying to emulate the brain operation is matching the number of synapses per neuron, which is of the order of ten thousand. This in a 2D chip is all but impossible.

To solve these obstacles, new materials and new physical phenomena are called for and various approaches are considered such as realizing physical systems that *map* efficient AI algorithms (previously run in classical CMOS chips) or adding new properties and *dynamics* to enrich computing.[139] The paradigmatic example of the former is DNNs that can be implemented with hybrid CMOS/memristive or photonic systems for instance.

It is widely accepted nowadays that, no matter how complex or deep, mere nonlinear functions mapping between input and output as in DNN is not all there is in the operation of the brain. Dynamics and, in particular, spikes are considered the key to energy efficiency and extended functionality. This realization nourishes the notion that other properties of neurons should be copied into artificial implementations. They comprise a varied gamut of properties biological neurons present related to their leakage, memory effects, randomness, synchronizability, function compartmentalization, selective response, or even the existence of cells specialized in certain tasks.[139]

### 4.2.1.    Electronic neurons
The neuroscience-oriented approach to AI tries to carefully imitate the brain operation in essentially three points: a close interaction between memory and computing, a rich spatiotemporal dynamics, and spike coding and processing. This is in clear contrast with computer-oriented operation where explicit rules are encoded in a more or less rigid protocol.[140] An advanced implementation of such neuroscience-oriented should be capable to integrate large ANN embodying complex temporal and spatial associations and provide support for flexible network topologies along with the possibility to establish cooperative procedures between them. It should be capable to run varied models, algorithms, and coding strategies. Early attempts at combining both philosophies are already show prowess with successful interoperation of ANNs and SNNs with their peculiar modes and needs.[129]

### 4.2.2.    Analogue neuron
In analogue neurons, weights can be implemented with a variety of devices with a memory capacity such as resistors, capacitors, charge-coupled devices or other semiconductor devices that sometimes suffer from quantization for the short channel lengths characterizing the transistors. Currents and voltages are typically used to encode the signals. For the activation function, the inherent nonlinearity of the transistors is used. Because they rely on a certain physical effects, they are liable to malfunction due to environmental conditions (*e.g.* temperature, voltage fluctuations).

In this architecture, apart from fabrication variability and materials uniformity, the main challenges are making the synapse work in a wide range of conditions and achieving an efficient storage of weights.

Implementing the nonlinear activation function may not be difficult as many devices present some property or other that can be used. However, achieving a complete set of properties is not always easy. Other operations such as multiplication and accumulation can be obtained with simple physical effects like current and voltage sums in Kirchhoff circuits. Analogue devices tend to be simpler than digital ones; however, they suffer from response variability and dependences on temperature, voltage etc. and may require sophisticated design.

### 4.2.3.    Digital neuron
In digital neurons, weights are stored in registers, latches or memories. Different kinds of memory include up to six transistors. Adders, multipliers and other logic operation circuits are readily available and the activation functions can be implemented with look-up tables

(taking up space in the chip) or using the former to carry out the calculation (taking up computing time and energy).

A full digital implementation has the advantages of cost, that every operation is easily attainable from the gallery of CMOS components, and that it is noise resilient and modular although the nonlinearity required for the activation functions is a major issue in digital large scale integration for the large cost in time and silicon area.[127]. In addition, it has some disadvantages like a slower operation or the unavoidable need for digital to analogue conversion at some stage (input or output). They can be realized as FPGA chips and can integrate different learning models offering software flexibility and lower cost at the expense of lower circuit density.

### 4.2.4.    Hybrid chips
*Hybrid neurochips* combine the speed of analogue processing and the economy of digital for storing weights to achieve the best of both worlds. Some designs separate the circuit architecture in two parts dedicated to analogue (properly neural processing) and digital (clock, error correction) tasks.

### 4.2.5.    Spiking
Electronic spiking devices are directly inspired by biological neurons operation where short electrical membrane potential pulses are transmitted through the synapses. *Integrate and fire* is the usual process responsible for its dynamics. This kind of operation permits to deal with time varying signals.

Spiking neurons can be embedded in complex integrated chips where the various functions[141] of the ANN are implemented, such as the weight × input and summation that are on the chip whereas stored weights, activation functions, neuron state etc. can be off the chip or performed by external means. Early designs, capable to generate varied functionalities, tended to necessitate silicon real estate in the order of several $mm^2$ per neuron and presented power requirements of the order of $\mu W$,[142] and led to complex modern structures comprising billions of transistors[143] offering high performances.[144]

Although their hardware implementations can be shared by other ANNs, spiking ANNs have some characteristics that set them apart.[124] For instance, asynchronous address event representation is at odds with conventional chip design, where all parts intervening in the processing operate in synchrony under the control of a central clock. On the contrary, operation in a SNN is sparse, taking place only when a spike is generated and therefore asynchronous computing is preferred. Another differential feature is the need for network-on-chip (NOC) modules, network-based communications subsystems inside integrated circuits in charge of routing digital information packets. This is largely brought about by the restricted access between parts in an essentially 2D architecture.

Non-volatile in-memory computing schemes, and other neuromorphic inspired implementations, although mostly focused on DLNs, have found application in SNNs too. And it appears that combinations of low precision memristors and high precision transistors hold promise of improved versatility like mixed-precision in-memory computing.[145]

## 4.3.    Non-volatile memory components

In the search for alternatives to the von Neumann concept, apart from improvements in processors, changes to the traditional memory configurations have emerged as the *non-volatile memory* concept that pursues to bring storage closer to processing.[146] Such memory devices are loosely grouped under the *resistive switching*[147] concept that, unlike ordinary RAM, do not rely on *charge* to store information. They are two- or three-terminal[148] devices susceptible to voltage driving so that application of voltage can change a material's property and thus its memory content. An additional distinguishing characteristic is their ability to perform analogue computing. More on materials, mechanisms and functionalities can be found in ref. [149].

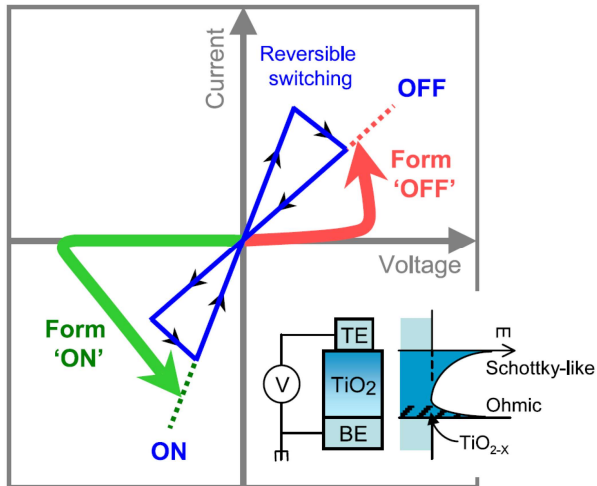

**Figure 25** Electroformation and typical response of a memristive device. Adapted from J. J. Yang, F. Miao, M. D. Pickett, D. A. A. Ohlberg, D. R. Stewart, C. N. Lau, R. S. Williams, *Nanotechnology* **2009**, *20*, 215201.

Regarding performance, in comparison with CMOS memories, they offer similar values in many aspects such as packaging, DNN inference accuracy etc. However, in state-of-the-art experimental demonstrations, the reported energy efficiencies (≥10 TOPS $W^{-1}$) in reduced-precision matrix-vector multiplication is still well behind CMOS technology. In addition, special training strategies are indispensable in this technology to compensate for the negative effect of abundant materials' defects and actual devices fabrication.[150]

### 4.3.1. Memristive devices
Memristive devices are mostly bipolar, voltage-actuated devices characterized by the unique property that their *resistance* can be changed (between a high and a low value) by applying a suitable voltage; the resistance value will not change afterwards even if the voltage is turned off. Since Chua predicted their existence in 1971,[151] more than thirty years elapsed until the HP Labs identified one such device[152] that was based on an effect already observed in the 1960's.[153] They offer the great benefit of direct access by interconnect lines but their greatest advantage is the capability to electrically reconfigure the device.

A typical memristive gate can be built from a piece of material contacted with electrodes on top and underneath. The material presents a doped region close to one end whose length can be changed by the applied voltage so that the device acts as a variable resistance. One of the biggest advantages of memristors is miniaturization, the possibility to reduce their size to that of state-of-the-art MOS transistors.[154] In addition, they can be three dimensionally arranged by stacking[155] or vertical formation[156] thus reducing the footprint of the computing chips.

#### 4.3.1.1 Operation
The idealized electrical operation of initial memristors (with Schottky top and Ohmic bottom electrodes) as depicted in **Fig. 25** is based on its hysteretic electrical response once the device has been subjected to an *electroforming* process. Such a process can be achieved by the application of a large positive (or negative) voltage that converts the pristine, almost insulating material into a switchable one where two states characterized by low (OFF) or high (ON) conductivity are established. In a typical forming process, green path in Fig. 25, a large, initial, negative voltage sweep is done to assure that formation takes place. At a certain voltage, a sudden, several orders of magnitude *increase* in conductivity takes place that *sets* the device in its low-resistance ON state. Alternatively, a positive forming, red path in Fig. 25, can be achieved with a positive voltage sweep for which, at a certain voltage, a similar sudden *increase* in conductivity places the device in an OFF state characterized by a conductivity similar to but *lower* than that of the ON state. The polarity of switching is controlled by the asymmetry of the device determined by the conductive nature of the contacts. The electroforming process in these devices is an electroreduction of the $TiO_2$ that produces oxygen vacancies that create high conductance filamentary channels.[157] See **Fig. 26**. The electroforming process was revealed *unnecessary* when the devices were shrunk to the nanoscale (a few nm thick) which also mitigates certain deformation induced by the neutralization of oxygen ions that drift towards the anode while their vacancies move towards the cathode. Sweeping the driving voltage between the *setting* and *resetting* values keeps the memristor in its ON or OFF state and only exceeding these values *changes* the state.

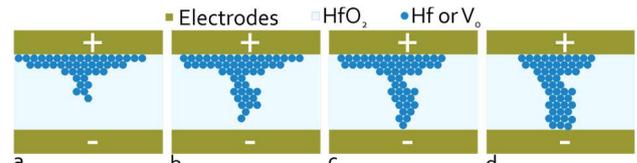

**Figure 26.** Set takes place through (a) and (b) [1] D. Ielmini, *IEEE Trans. Electron Devices* **2011**, *58*, 4309.the top electrode, where a positive voltage is applied. Ions might consist of oxygen vacancies or metallic cations. Reset is achieved by the opposite sequence, namely, (d) and (c) conductive filament narrowing and (b) and (a) final disconnection under a negative voltage applied to the top electrode. The reservoir of ions at the top electrode side might result from the previous reset operation. Adapted from D. Ielmini, *IEEE Trans. Electron Devices* **2011**, *58*, 4309.

The insulating material in memristors is typically an oxide such as $HfO_2$,[158] $TaO_2$,[159] or $SiO_2$,[160] $WO_x$.[161] Other materials comprising ternary oxides such as $LiSiO_x$[162] and quaternary oxides such as $Pr_{0.7}Ca_{0.3}MnO_3$[163] have also been proposed.

### 4.3.2. Processing with resistive switching
Whether digital or analogue, memristive systems integrate this kind of components into CMOS circuitry where information flows as electrical currents. This implementation uses electrical resistors as valves that regulate the flow of information mimicking neurons and synapses. Memristors are resistive switching devices that can be miniaturized to the nanoscale and whose analogue conductance, once tuned by an applied voltage, will remain until further acted upon. The setting, resetting, and reading processes and corresponding voltages for an idealized device are schematically presented in **Fig. 27**. In a crossbar configuration,[164] they serve as non-volatile memory where the weights between successive layers in the network are stored.

#### 4.3.2.1 Digital processing
Shannon's choice of logic operations[165] was a major determinant in the development of digital logic and its microelectronics implementation once the transistor was discovered. However, this choice is not unique as shown many years earlier by logicians Whitehead and Russell who considered the *material implication*, $A$ IMP $B$ that only returns 0 when $A = 1$ and $B = 0$ (much like negating the consequence requires the hypothesis being false).[166] They had found that IMP, was a powerful operation that, along with the FALSE operation (that always returns *false*, 0), form a computationally complete logic basis. The material implication can be naturally realized in a simple electronic circuit with one resistor and two memristors. This alone signifies that the whole logic can be implemented with simple combinations of gates based on memristors.[167] Because the result of the logic operation is stored in the gate itself the IMP logic is said to be

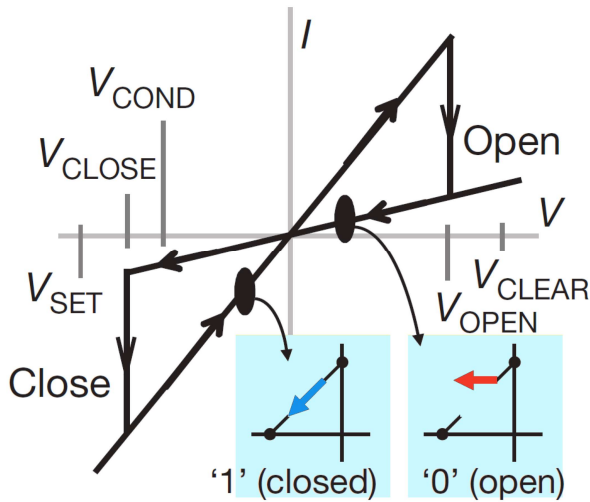

**Figure 27.** A given device may be 'set', that is, assigned a logic value of 1 (logic operation TRUE) by applying a large negative voltage, $V_{SET}$ which is slightly larger than $V_{CLOSE}$ to compensate for the other circuit resistance. Similarly, the device may be 'cleared', that is, assigned a logic value of 0 (logic operation FALSE) by applying a positive voltage, $V_{CLEAR}$, again, slightly larger than $V_{OPEN}$. These states are verified by applying a small reading voltage and measuring the resulting current to confirm that the desired values were stored in the switches. Adapted from J. Borghetti, G. S. Snider, P. J. Kuekes, J. J. Yang, D. R. Stewart, R. S. Williams, *Nature* **2010**, *464*, 873.[164]

*stateful.*[a] The respective voltages applied constitute the input while the output is automatically stored as the resistance created. Reading it involves simply applying a small voltage and measuring the current. This clearly shows that the computing and the storage is performed by the same element. Since NAND is known to be universal too, synthesizing it with IMP and FALSE circuits suffices to demonstrate that *any* logic operation can be performed with the memristor kit.[164]

### 4.3.2.2 The IMP gate

A practical implementation of the IMP gate comprising two memristors and one resistor is shown in **Fig. 28**. By applying the conditional toggling property of the crossbar latch[168] to the circuit, it can be confirmed that the truth table produced is that of $q' \leftarrow p$ IMP $q$ where $q'$ is the logic state of memristor Q after the operation. If P is open ($p = 0$) applying $V_{COND}$ will have no effect on its state (so it will remain open, $p = 0$) and therefore it won't have any influence on the state of Q which will be determined solely by the voltage it is subjected to. That is, application of $V_{SET}$ will *set* it to closed ($q' = 1$) regardless of its initial state ($q = 0, 1$). In contrast, when P is closed ($p = 1$) application of $V_{COND}$ won't change its state but will *short* the circuit so that $V_{SET}$ on Q won't be able to change its state ($q' = q$).[167]

### 4.3.2.3 Configurations

This kind of operation, that can be called digital computing by binary resistive switching, can be implemented in different configurations depending on the magnitudes used as input and output. In *V-R* gate implementation of IMP logic the result is stored in the resistance; *V-R* is not fully in-memory because some cascaded operations require additional circuits outside the memory to complete. In the *V-V* logic both input and result are held by voltages; can be viewed as a one-layer ANN and is also not fully in-memory. Only *R-R* logic where both input and output are held in resistance values is truly in-memory.

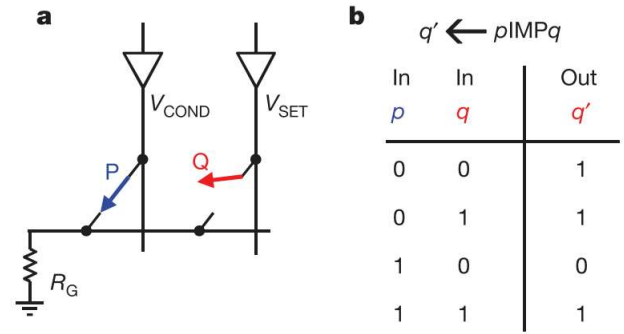

**Figure 28.** a) Implementation of material implication gate with two memristors and one load resistor. The state of Q holds the output. B) The truth table of the corresponding gate. Adapted from J. Borghetti, G. S. Snider, P. J. Kuekes, J. J. Yang, D. R. Stewart, R. S. Williams, *Nature* **2010**, *464*, 873.

A general key limitation of in-memory digital computing is the time (typically ≈ 10 ns) and energy burden (typical current and voltages of ≈ 10 µA and ≈ 1 V) incurred due to the physical process inside the device (ions, vacancies drift involving Joule heating) which impose an energy of ≈ 1 pJ per operation.[b] In addition to these disadvantages, degradation and eventual breakdown due to the relatively high electric fields and temperatures limits the device lifetime to some ≈ $10^{16}$ switching cycles.[169]

### 4.3.2.4 Other computing schemes

The binary character of memristive devices can be extended if additional levels of resistance can be considered, for instance, by increasing (or decreasing) memristors conductance with successive electrical pulses. This allows gradual changes enabling computing by *cumulative resistive changes* which in turn can be made to operate like analogue computing. If a known number of pulses is required to transition from the high to the low conductance state, simple arithmetic can be performed by counting pulses like in a nano-abacus.[170] This summation can also be extended to the integration of analogue pulses or spikes, something essential in integrate-and-fire neurons of SNNs.

### 4.3.2.5 Stochastic operation

Resistive memory devices, typically depending on temperature activated ion migration mechanisms (apart from electric field, naturally) can suffer from stochastic variations even in the set (and reset) transition. This causes random fluctuations, even from cycle to cycle, in the high (and low) conductance values as well as in the concerned voltages with repercussion in the reliability of the computing process. However, this inconvenience can be turned into a random number generator. Although such a stochastic *random bit generation* might not be state-of-the-art in terms of speed it can constitute an enabling technology in probabilistic SNN where noise is required to emulate the stochastic operation of synaptic junctions.[171]

### 4.3.2.6 Analogue operation

The *V-V* logic is a perfect configuration to implement *analogue computing with crosspoint arrays* that constitute a one-layer ANN, as in **Fig. 29**. Here, the intersections between orthogonal row and column electrodes are filled with a memristor maximizing the ease of writing and reading and reducing the AN size and packing. The current $I_j$ flowing through a given electrode of column $j$ of the output layer is the memristor conductance-weighted sum of the input voltages, $V_i$, in every row of the input layer, $I_j = \sum G_{ij} V_i$, in a direct implementation of MAC operations or matrix-vector multiplication. A load resistance is used to recover voltages and a rectifying stage (*e.g.* a comparator) to obtain a digital output. If signed weight synapses are to be considered, two memristors per weight are required. Feeding the output to a next layer opens the possibility to increase the depth of the ANN arbitrarily.

---

[a] Memristive stateful logic refers to the form of computational logic in which memristors both store logic values and perform logical operations on these values.

[b] For the purpose of comparison consider that that is the energy budget for a 32-bit integer addition processor[377] in no so state-of-the-art 45 nm CMOS technology (which involves many Boolean logic operations, see section on bitwise arithmetic).

### 4.3.3.    Other resistive switching mechanisms
Although still too slow, with limited data band-width, or depending on technologies that are too expensive to significantly contribute to overcome the memory bottleneck, memristors are but one among several approaches where resistance switching is the enabling effect that can be derived from the material property. Widely varying effects make use of different materials' properties under different kinds of stimuli (mostly electrical but also magnetic or optical) which makes a simple classification difficult. More so because some materials can work in response to various stimuli. Notably, for instance, 2D materials such as van der Waals compounds can operate in two (diode) or three (transistor) terminal modes but also through various mechanisms.[172] Reference [173] presents an extensive classification.

### 4.3.3.1    *Electrical*
Electrically driven devices respond to electrical pulses by changing their conductivity like memristors but can do so by many effects.

#### 4.3.3.1.1    Ion migrating
*Ion migrating* devices are the common memristors described above. Electrical pulses drive the movement of metal ions or oxygen vacancies leading to the formation and breaking of conductive filaments. Most of the materials involved are oxides but organic materials[174] are gaining pull owing to their low cost and processing advantages and versatility.[175]

#### 4.3.3.1.2    Ferroelectric
Electric charges separate under the effect of an applied electric field and in *ferroelectrics* remain displaced after the electric field is removed creating a hysteresis loop which constitutes the memory basis[176] and can provide a wide range of resistance.[177]

#### 4.3.3.1.3    Phase change
The change in resistance experienced by a material upon *phase change* can be leveraged as the varying magnitude to operate a memristor. This phase change is triggered by the heat generated by a sufficiently large electrical pulse and can be morphological (amorphous/crystalline),[178] of metal/insulator character[179] etc. The electrical, magnetic, optical, morphological, dynamical, and other response properties can be used in order to apply these materials for in-memory devices.[180]

#### 4.3.3.1.4    Magnetoresistive RAM
*Magnetoresistance* memories are composed of two magnetic materials (one of them permanently set to one polarity) separated by a thin insulating layer forming a magnetic tunnel junction. The tunnel current through the junction (hence magnetoresistance) depends on the relative orientation of the two magnetizations. Orienting the non-permanent one with the magnetic field created by a current driven by an auxiliary transistor can switch the device. Using spin-aligned electrons directly can torque the domains into alignment[181] with a much lower current which outperforms classical magnetoresistance cells both in power demand and miniaturization.[182]

#### 4.3.3.1.5    Carrier capture/release
Unlike ion drift, the common mechanism in memristors, in which the carriers accumulate to form conductive filaments, the *capture and release* mechanism operates by moving protons, electrons or holes and pinning them in trap centres.[183] The energy budget for this mechanism is expected to be more contained than for ion migration and phase change. Embedded nanoparticles can be used as traps demonstrating different synaptic functions like facilitated learning.[184]

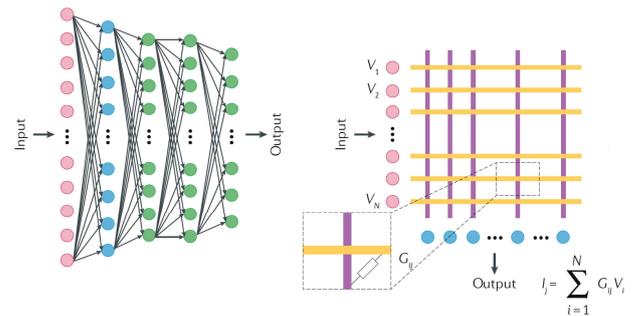

**Figure 29.** Implementation of one set of synapses between two AN layers in a DNN. The crossbars provide connections between every AN in a layer and every AN in the next layer and every junction is endowed with a resistive element that acts as weight. Stacking so these layers can create DNNs any number of layers deep. Adapted from D. Marković, A. Mizrahi, D. Querlioz, J. Grollier, *Nat. Rev. Phys.* **2020**, *2*, 499.

### 4.3.3.2    Optical
Devices based on *optical* memristors rely mainly on the ionization and dissociation of oxygen vacancies,[185] enabling synaptic functions through the inherent persistent photoconductivity characteristic of amorphous oxide semiconductors; capture and release of carriers by traps, and light-induced phase transition[186] like their electrically driven counterparts.

Electrically assisted optical stimulation and optically assisted electrical stimulation combine in different ways the effects produced by electrical and optical pulses. In the former, the synergistic effect can be used to overcome a shortcoming: that erasure of trapped carriers caused by the optical stimulus; in the latter, light is employed as a supplementary way to regulate the device.

### 4.3.4.    Memristor materials
Although candidate materials for this technology will be judged for their ability to be integrated with CMOS technology, current investigation has tested a wide variety of fabrication techniques comprising epitaxial approaches (atomic layer deposition, sputtering, pulsed layer deposition, chemical vapour deposition, molecular beam epitaxy), chemical techniques (sol-gel techniques, chemical solution deposition) etc. A recent, comprehensive catalogue of materials and techniques can be found in ref. [187].

Of the three layers (two electrodes and a resistive switching layer) the device is generally comprised, the electrodes present little variation and the smallest technological challenge. They are often made of inert metals such as Pt, and TiN is being used to solve etching issues in mass production.

### 4.3.4.1    Oxides
*Oxides* are the most common materials employed as resistive switching layer for their simple fabrication and safe compatibility with CMOS technology. Binary oxides in particular embody these characteristics to perfection. depending on the projected use, different oxides are chosen for their simple fabrication, well known properties, special features, multiple oxidation states, forming-free operation, thermal stability, switching speed, multilevel capabilities, etc.[188] Additionally, the oxide/metal combination also has importance since it may result in a different response: combinations with abrupt response are adequate for binary memory use while gradual response combinations are apt for multilevel storage. A current direction of research is the improvement of performance by doping (*e.g.* nanoparticles, metal or semiconductor dopants that improve carrier trapping) and inclusion of a capping layer to the resistive switching layer (to improve tuning linearity, electric field and temperature response).

### 4.3.4.2    2D materials

*2D materials*, for their intrinsic nature, are well adapted to both electrode and resistive switching layers. More generally, their wide variety of properties places great expectations in these materials. In this category[189] we find the graphene type of materials (including boron hBN, graphene oxide, etc.), 2D chalcogenides (metal and semiconductor dichalcogenides, layered semiconductors, etc.); and 2D oxides (silicates, complex oxides, layered oxides, perovskites, etc.).

For its conductive properties, graphene has been used not only as electrode but also as a stop layer to prevent conductive filaments penetrating the bottom electrode; transition metal dichalcogenides offer additionally photonic properties; The ample variety of properties of 2D materials enable them to be used under different conditions and for different functions.[190]

In some cases, large area preparation, monolayer stability, stoichiometry, doping and substrate compatibility still constitute a challenge before 2D materials are fully integrated in in-memory computing technology.

### 4.3.4.3    Ferroelectrics

Thin layers of *ferroelectric* materials between metal electrodes forming a tunnel junction have been realized mostly with titanates. These materials have a long research history, so the focus of research is place more on electrode optimization (*e.g* use of heavily doped semiconductors); substrate improvement (flexible, transparent) and doping.

### 4.3.4.4    Solid electrolytes

Memristors based on *solid electrolytes* use one active electrode (that can be oxidized and reduced in a redox reaction with the resistive switching layer material) and one inert electrode. The ions resulting from the reaction can drift and diffuse through the memristive layer. Silver dendrites were the first observation of such an effect.[191]

### 4.3.4.5    Organic

The wide variability of *organic materials* properties may offer novel alternative switching mechanisms (potentially less stochastic and more energy economic) that, added to a generally inexpensive production, favours their study and use.[192]

The kinds of organic materials comprise polymers, small molecules, complexes, etc. and operate under the same principles as the inorganic counterparts: filamentary conduction, ionic charge transfer, electromigration, etc. They can be operated in binary, multivalued or analogue modes and their most valued characteristics are their low-energy switching, excellent tunability, low cost and biocompatibility.

## 4.4.    Photonics hardware

Although swiftly developed in the last decade[193] initial attempts at *optical neural computing* date back to last century's eighties,[194] were desk-top type and essentially analogue.[195] Photonic integration in Si[196] and InP[197] constituted a boosting of the technology. The former benefited from a high refractive index and the availability of a mature CMOS technology capable to mass produce reliable and cost-effective devices; mostly passive owing to an indirect bandgap. The latter brought the huge advantage of an active material (emitting) enabling monolithically integration of light sources and electrooptic components.[198]

Optical technology enjoys the virtues of massively parallel, free-space beam processing, energy inexpensive, virtually instantaneous, free from the carriers' interactions technologies and, at the same time, susceptible to multiplexing in time, space, polarization and wavelength. A major advantage offered by optical implementations is that any linear transformation (discrete finite-dimensional unitary operator) can be performed by passive optical components without power consumption and virtually no time delay,[199] and can be then detected nowadays at rates of tens of Gb/s. Let's just remember that the Fourier transformation of an image that is mathematically involved and computationally costly, can be performed in no time by a simple lens placed a one-focal length from the image. However, some components such as optical switches and memories, well solved in electronic systems, pose a serious challenge in pure photonic implementations. Unlike electronic devices, photonic ones are ultimately limited in size by the optical wavelength.

Many implementations tackle in different ways the pervasive matrix-vector multiplication tasks, so often encountered in ML. These include diffraction techniques in free-space, wavelength division multiplexing, and Mach-Zehnder interferometry in integrated circuits.[200]

All-optical implementation of logic avoids inefficient optoelectronic conversion and has been implemented through Mach-Zehnder interferometer structures relying on the ultrafast non-linear properties of semiconductor optical amplifiers.[201] The mechanism in this application is the use of the phase shift induced through the polarization sensitivity of semiconductor optical amplifiers. This birefringence rotates the polarization of light as it travels the device allowing an optical control and all-optical switching. The variation of the birefringence with input power can be used to change the polarization of a strong beam with two weak ones carrying the input bits so that the output polarization state provides the result of the logic operation.[202] With this technology XOR, NOR, OR and NAND gates have been realized sufficient to build a complete set of logic gates.[203]

Photonic ANN can be classified according to multiple criteria like a division into *stateless* (comprising multilayer perceptrons and convolutional NN) and *stateful* (spiking NN and reservoir computing)[204] but it is perhaps better to break them down into categories that respond not to the operation principle but to the technology used in the computing process.

As suggested by the preceding paragraphs, photonic ANNs can be separated in three major blocks according to the media in which the information carrier is transported and manipulated: free space, optical fibres or integrated circuits.

### 4.4.1.    Free-space optics

The first kind of optical matrix-vector multiplication was implemented by the diffraction of light in free space.[205] A feature vector contained in an array of beams or sources along the *x* axis was expanded (and replicated) along the *y* axis with a cylindrical lens before hitting a diffracting plane (containing the matrix elements) and later combined and summed with another cylindrical lens (perpendicularly oriented) to generate the product vector (along the *y* direction).

In a recent implementation using bidimensional *diffractive* structures constructed by 3D printed gratings, manipulation of 0.4 THz sub-millimetre waves is used for computing in all-optical DNN.[206] These gratings are disposed in stacks. Each element in a layer, which presents a complex response to incident light, can be thought of as a neuron. Each AN is connected to the neurons of the following layer through a diffracted wave modulated in amplitude and phase by both the input interference pattern created by the preceding layers and the local transmission or reflection coefficient at that point. Input data (*e.g.* written characters) are projected on the first layer which, through successive diffractions, create an intensity pattern in the output layer (*e.g.* pockets of high intensity associated to each digit). In practice the training is carried out by numerically computing the combined response to training data and once done it is fixed and can only perform the computing it is programmed for.

A similar arrangement with two SLMs in a 4f reconfigurable implementation is capable to diffractively couple multiple Gaussian laser beams encoding the feature vector and carry out any linear transformation.[207]

An SLM and a digital camera connected by a computer can act as a reservoir computing system[208] including *nonlinearity*. It operates in the following manner. A constant laser beam is sent onto the SLM whose pixels (binned for the purpose of noise reduction, signal uniformity and stability) independently shift the phase of each segment of the incident beam by an amount $\phi_i(t_k)$. Two polarizers (before and after the SLM) with rotated axes perform a $\sin^2(\cdot)$ nonlinear function on the intensity emerging from the SLM so that the intensity captured from the $i$-th pixel imaged by the digital camera in the $k$-th round is given by $I_i(t_k) = I_0 \sin^2(\phi_i(t_k)/2)$. The set $I_i$ forms the state vector $\mathbf{x}(t_k)$. In each round of processing a computer executes a linear combination of the recorded intensities by multiplying by a reservoir weights matrix and adds input data $\mathbf{u}(t_k)$ to generate a new reservoir state $\mathbf{x}(t_{k+1}) = I_0 \sin^2[\mathbf{W}^r \cdot \mathbf{x}(t_k) + \mathbf{W}^i \cdot \mathbf{u}(t_k)]$ (a new mask to send to the SLM). A readout weight matrix extracts the output from the reservoir $\mathbf{y}(t_k) = \mathbf{W}^o \cdot \mathbf{x}(t_k)$ which should be minimized against the corresponding target label associated in the data set $\{\mathbf{u}(t_k) \rightarrow \mathbf{y}_t(t_k)\}$.

A similar set up using binary phases $(0,\pi)$ on the wave front of an amplitude-modulated laser beam with a detection and feedback method was shown capable to optically calculate the low-energy ferromagnetic spin configuration of a spin glass.[209] The set up comprises a pixelated mask —that sets field amplitudes— and a spatial light modulator —that sets the phases— of a laser wave front. The phase pattern in the SLM encodes the spins; the amplitude modulation encodes their interactions and propagation (through a lens placed one focal length from the SLM to a CCD camera) maps the Ising Hamiltonian. Minimizing the loss function with respect to a target intensity pattern on the CCD camera, $|I(x,y) - I_t(x,y)|$, is equivalent to minimizing the Hamiltonian.

An SLM placed at the back focal plane of a lens can perform the linear vector matrix multiplication occurring at each NN layer because each point $z_j$ in the front focal plane (Fourier transform of the SLM configuration) receives light from *every* SLM input pixel ($x_i$). In order to impart multiple weights to every input pixel, multiple phase gratings split the said pixel signal in multiple directions with the desired weights ($w_{ij}$) so that $z_j = \sum w_{ij} x_i$. The non-linear activation function is carried out by nonlinear electromagnetically induced transparency employing cold Rb atoms.[210]

Free-space optics often relies on 3D printed gratings which are not reconfigurable or SLMs that are slow; ICs on the other hand suffer from a large footprint and difficult scalability. In this context the *coherent matrix multiplier*[211] proposes a scheme based on coherent detection where both inputs and weights, encoded in optical signals, are fanned out, and combined by a beam splitter. Synapses, that translate into matrix-vector products, are realized by the quantum photoelectric multiplication process in the homo-dyne detectors. Input data, $\mathbf{x}$, is encoded as a train of pulses (each pulse corresponding to a component of the feature vector, $x_i$) as are the weights, $w_{ij}$, that enter the unit in parallel. Both are fanned-out into as many copies as neurons and superimposed, by means of a beam splitter onto so many homodyne detectors. If the signals originate from the same coherent source and the path-lengths differ by less than the coherence length, the detectors produce the real part of the product of the electric fields. These products are subjected to a nonlinear filter, converted to optical signals and fed to the next layer.

#### 4.4.1.1 Diffractive coupling

An array of vertical cavity surface emitting lasers can be connected to the array of reflecting elements of a spatial light modulator by means of a *diffractive-optical element*. In this way, individual lasers in the 8×8 array become mutually coupled to their neighbours through different diffractive orders of the diffractive-optical element. Since the spatial light modulator is reconfigurable so are the coupling weights.[212]

Following this scheme, and using a single laser source, a network of up to 2025 modes, each pixel of an SLM, forming a large-scale recurrent neural network was realized.[213] See **Fig. 30**.

#### 4.4.1.2 Multiple scattering

Multiple light *scattering* in static diffusive media is a complex linear problem that involves many optical modes whose coupling to incoming and outgoing fields can be mathematically represented with high-dimensional random matrix multiplication (transmission matrix). For this very reason a reservoir computing system can be based on disordered scattering materials where input modes are encoded with a micro-mirror device or a phase-only spatial light modulator.[214] The cross section of an expanded laser beam is divided in segments corresponding to (binned) pixels of a spatial light modulator that imparts specific relative phase differences. The beam thus modulated is scattered by the diffusive material and the speckle collected by a CCD camera is processed and fed back to the spatial light modulator. The diffusive material carries out the vector-matrix multiplications at the speed of light and virtually no energy cost. In a practical implementation one third of the spatial light modulator is reserved to contain the input data, $\mathbf{u}(t)$, coded as phases through a function $p$; one third for the reservoir state, $\mathbf{x}(t)$ and one for the bias, $\mathbf{b}$. In each computing step the reservoir state is $\mathbf{x}(t_{k+1}) = (1-a)\,\mathbf{x}(t_k) + a\,g[\mathbf{W}^i\,p(\mathbf{u}(t_k)) + \mathbf{W}^r\,p(\mathbf{x}(t_k)) + \mathbf{b}]$ and the output is given by $\mathbf{y}(t) = \mathbf{W}^o\mathbf{x}(t)$ where the $\mathbf{W}^i$, $\mathbf{W}^r$, $\mathbf{W}^o$ represent the weight matrices (input, reservoir and readout), $\mathbf{b}$ the bias, $g$ the activation function and $a$ the fading (short term memory). The input, reservoir weight matrices and the bias are set to random and not changed and only the readout one is trained with linear regression. In this way predictions on large multidimensional chaotic datasets can be done.[215]

With a similar set up, a spin glass dynamics was optically simulated exploiting the wave-front shaping offered by the spatial light modulator (a micromirror device) to play the role of the spin variables.[216] Using the interference downstream of a scattering material to implement the random couplings between the spins, and measuring the light intensity on a number of target pixels to retrieve the energy of the system, it was possible to solve a Hopfield system.[89]

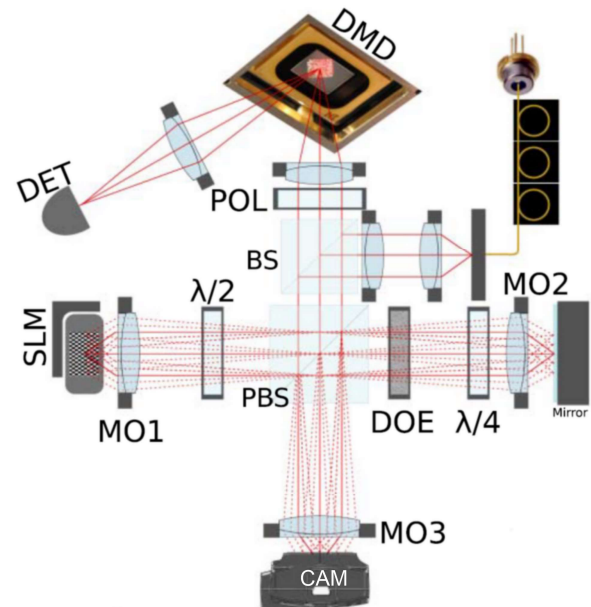

**Figure 30.** Experimental setup of a RNN based on diffractively coupled nodes. The NN state is encoded on the SLM; and it is imaged onto the camera (CAM) through a diffractive optical element (DOE). A digital micromirror device (DMD) creates a spatially modulated image of the SLM's state. The integrated state, obtained via superimposing the modulated intensities, is detected, creating the photonic RNN's output. Adapted from Bueno Maktooby, et al. Optica 2018, 5, 756

#### 4.4.1.3 Computing with (nonlinear) waves

Conventional digital signal processing suffers from two main problems, one originating from the data bottleneck and resulting in low processing speed and one originating from the analogue-to-digital conversion (and vice versa) resulting in high power consumption and hardware complexity: discretization of signal is time consuming and cannot be performed in a massively parallel way. This has incentivized analogue wave-based computation shown to be possible with optical waves in special platforms such as metamaterials.[217] Photonic over acoustic are the metamaterials of choice because not only can they operate at the speed of light but also perform many operations at the same time relying on Fourier optics, Green's function or metasurface approaches.[218]

Linear manipulation of waves can perform all sorts of linear transformations. But more complex computations can also be tackled such as solving linear integral equations as demonstrated with metamaterial structures.[219] Since an integral equation, is an inverse problem, of the form $f(x) = I_{in}(x) + \lambda \int_a^b K(x,t)\phi(t)dt$, a feedback mechanism (such as a metamaterial whose scattering matrix corresponds to the kernel $K(x,t)$ excited with the input signal $I_{in}(x)$) can be used to generate this solution. Like in the case of 3D printed diffractive elements this system is not reconfigurable. Therefore, wave-based computation systems, relying on materials with carefully tailored scattering properties operating on an incident wave front, require material fabrication and (re)configuration for each computation. This hindrance could be bypassed by using a random medium instead, but reconfiguring the wave-front.[220]

The *nonlinearity* in the information processing is a necessary condition for the network to become a *universal* approximator, which means it can represent any function when given appropriate weights. Systems governed by the non-linear Schrödinger equation under highly nonlinear conditions produce solutions of the soliton, rogue, or shock wave type. These nonlinear waves operating as reservoir computing systems have been shown to be universal approximators.[209] The "amount" of nonlinearity was found to present a fundamental threshold for universal interpolation which is dependent on training data. The operation of this approach is based on a computing reservoir where a wave, $\psi_{in}(\xi,\mathbf{x})$, moulded according to the input vector **x** (coding) and biased, evolves obeying a nonlinear partial differential equation $i\,\partial_t\psi + \partial_\xi^2\psi + \kappa|\psi|^2\psi = 0$. The readout layer decodes the output in the usual manner by sampling the non-linearly-evolved wave, $\psi_{out}(\xi,\mathbf{x})$, in a reduced number of channels ($\xi$'s). This formulation confirms that a linear operation ($\kappa = 0$) produces a non-universal approximator, however, not all nonlinear evolutions lead to universal approximators.

#### 4.4.2. Optical fibres

Optical fibres are used in basically two ways: using time-coded information by making pulses interact as they propagate in loops or by using a spatially coded information by using the interaction of multiple transversal modes.

#### 4.4.2.1 Single mode fibre

The power of RC largely depends on a high number of ANs whereby input is nonlinearly converted to a high-dimensional space. In addition, a short-term memory is required to make the current state depend only on its recent past and process temporal sequences.

This can be readily achieved in a system where a laser is self-coupled through a single-mode fibre loop (see **Fig. 31**). Pulse-coded data is introduced by means of a second (tunable) laser during delay time (the time it takes for a pulse to circle the ring).[221] In this way, the operation of the laser is delay-modulated by the input pulses: the state of the laser depends on every input pulse introduced in the loop as far back as the delay time. Effectively, this single-neuron time multiplexing is equivalent to many ANs reservoir in synchronous operation.[222] An alternative implementation uses the wavelength of the laser as the dynamical variable.[223] In a configuration where the two orthogonal polarizations from a vertical cavity surface emitting laser are injected in a fibre loop, a time-delay optical feedback reservoir computing system was used to solve two problems simultaneously.[224]

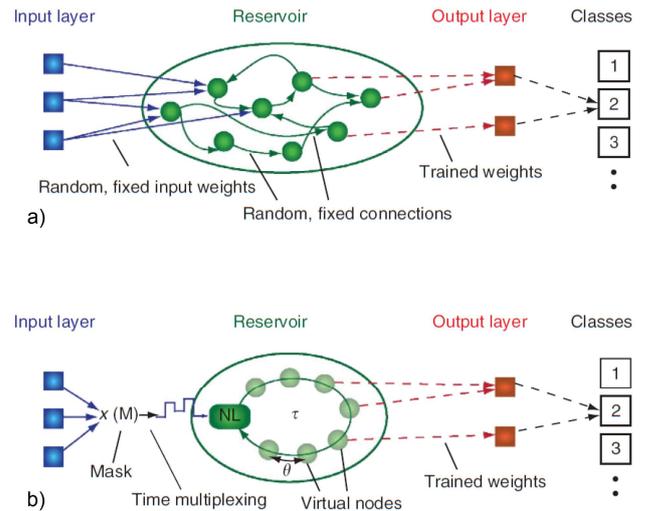

**Figure 31.** a) In a classical RC scheme the input is coupled into the reservoir via a randomly connected input layer to spatially distinct nodes in the reservoir. The connections between reservoir nodes are randomly chosen and kept fixed (the reservoir is untrained). b) A RC utilizing a nonlinear node with delayed feedback is obtained by dividing the delay loop into equal intervals and using time multiplexing. The input state bits (sampled and held for the duration feedback loop) are multiplied by a mask, added to the delayed state of the reservoir at the previous round and, then, fed into the nonlinear node. Adapted from L. Appeltant, M. C. Soriano, G. Van Der Sande, J. Danckaert, S. Massar, J. Dambre, B. Schrauwen, C. R. Mirasso, I. Fischer, *Nat. Commun.* **2011**, *2*, 1.

#### 4.4.2.2 Multimode and multicore fibres

The light intensity pattern at the output of a multimode fibre presents an apparently random distribution owing to the complex propagation of coherent waves that are scrambled giving rise to a speckle. Random as it may seem, careful measurement of the relationship between fields at the input and at the output can yield the complex transmission matrix that can be used to compute the output produced by arbitrary inputs, in particular those handled in endoscopy imaging techniques. These techniques are usually cumbersome since both amplitude and phase need to be considered. Alternatively a DNN can be trained to learn such a relationship and later to predict the output resulting from given images or reconstructing the input at sight of the output even with intensity-only knowledge. [225]

A fully optical method with minimal intervention of ordinary computers was implemented by coupling an external laser to a large area vertical cavity surface emitting laser by means of the transfer matrix of a multimode fibre. The conformation of the large area, vertical cavity, surface emitting laser (VCSEL) as an array of laser leads to the natural interaction of adjacent units and the creation of a recurrent network of ANs operating as a reservoir computer. The binary input is coded with a digital micromirror device (DMD) and the response from the large area laser is read out by a detector through a trainable second DMD.[226]

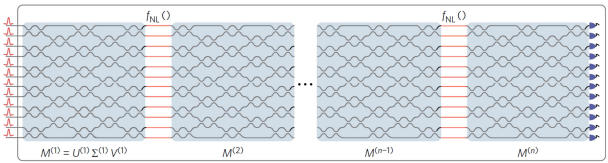

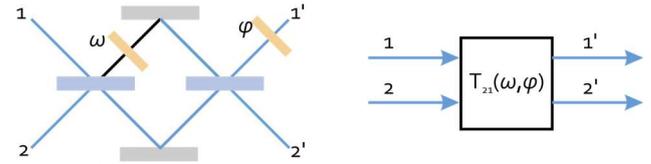

**Figure 32.** Integrated circuit configuration of a DNN showing some layers with the interference and non-linear units separated. Adapted from Y. Shen, N. C. Harris, S. Skirlo, M. Prabhu, T. Baehr-Jones, M. Hochberg, X. Sun, S. Zhao, H. Larochelle, D. Englund, M. Soljačić, *Nat. Photonics* **2017**, *11*, 441.

One of the challenges optical computing needs to overcome is the inefficient use of nonlinear functionalities due to the high intensity requirements. One way to mitigate this may arise from the advantageous combination of the linear and non-linear ANN parts in a non-linear multimode optical fibre that simultaneously provides tight lateral confinement and long interaction length.[227]

The lateral interaction between neighbouring cores in a multi-core optical fibre can emulate the synaptic operation in ANNs. In order to transfer signal between parallel cores amplification with active doping ions is provided by a pumping beam though selected cores that act as weights.[228] This conceptual implementation is restricted (input channels, hidden layers and outputs) by the limited number of cores available. Using multimode fibres instead demonstrated the ability to improve the accuracy in processing colour images.[229] Here images are projected on the entrance of the fibre which decomposes them in the multimode space on exiting the fibre. An ANN processed the actual source images and those after being scrambled by the multimode fibre to confirm that the latter process reduced the number of nodes required to achieve the same error rate.

### 4.4.3. Integrated photonics

Integrated photonic circuits present the advantage of mechanical stability but, more importantly profit from the ability to use coherent light, hence, electromagnetic fields (complex) rather than intensities (real). Operating on complex numbers essentially doubles the internal degrees of freedom as compared with real-valued networks, leading to roughly twice the size.

#### 4.4.3.1 Mach-Zehnder circuits

A notable proposal of a DNN in a nanophotonic Si integrated circuit relying on *coherent light* is proposed in [230]. In this realization each of the two layers of the DNN is composed by two units dedicated respectively to *optical interference* —that takes care of the matrix multiplication associated with the multiplication of the weights by the inputs to the layer— and the *optical non-linearity* output associated with the activation function (see **Fig. 32**). The interference unit operation relies on two facts. One, the *singular value decomposition* of matrices whereby an $m{\times}n$ real matrix $\mathbf{M}$ can be expressed as $\mathbf{M} = \mathbf{U}\,\boldsymbol{\Sigma}\,\mathbf{V}^\dagger$ where $\boldsymbol{\Sigma}$ is an $m{\times}n$ diagonal matrix and $\mathbf{U}$ and $\mathbf{V}$ are $m{\times}m$ and $n{\times}n$ *unitary* matrices respectively. And two, that a unitary transformation as those represented by $\mathbf{U}$ and $\mathbf{V}$ (rotations in $m$- and $n$-dimensional spaces) can be mapped by the cascaded operation of beam splitters with a phase shifter in one arm, or Mach-Zehnder interferometers (see **Fig. 33**) in this particular case.[199] Since $\boldsymbol{\Sigma}$ is diagonal it won't mix channels and can be implemented by line attenuators or, more practically, amplifiers. The optical non-linearity unit can be simply implemented with saturable absorption or bistability, as often used in photonic circuits, and acting on single channels (represented by a diagonal matrix) although in this work it has been only *emulated* by performing its job in an ordinary computer. This DNN was tested for speech recognition with vowels and on written digit recognition. In this kind of implementations research is still required in order to optimize robustness which may come at the expense of expressivity.[231]

**Figure 33 | Mach-Zhender** The unitary transformation represented by $\begin{pmatrix} k'_1 \\ k'_2 \end{pmatrix} = \begin{pmatrix} e^{i\phi}\sin\omega & e^{i\phi}\cos\omega \\ \cos\omega & -\sin\omega \end{pmatrix}\begin{pmatrix} k_1 \\ k_2 \end{pmatrix}$ can map the operation of a beam splitter or a Mach-Zehnder where $\sin\omega = \sqrt{R}$ and $\cos\omega = \sqrt{T}$ and $\phi$ represents an external phase shifter in one arm.

A fully *passive* integrated photonic circuit enjoys the advantage that, not containing optical amplifiers, its speed is not limited by carrier lifetimes. However, it lacks the nonlinear part and must be operated in reservoir computing mode. It will thus benefit from the nonlinear response of the readout detectors to the complex amplitudes circulating in the network.[232] An equally passive operation can be expected from a ingeniously designed optical cavity (quarter stadium in a photonic crystal) if almost chaotic field mixing dynamics are properly sustained.[233]The cavity *quality* determines the modes' lifetimes which, in turn, provides a fading memory necessary for dynamic reservoir computing.

#### 4.4.3.2 Semiconductor optical amplifiers (SOA)

Optical signal processing in a RC configuration was proposed with a reservoir using SOA and an electronic read out, which often is a simple linear discriminant whose training depends on mathematical calculations like matrix pseudo-inversion that are carried out by a traditional computer.[234] This proposal responded to the fact that the power transfer of an SOA is similar to the (positive half) hyperbolic tangent activation function used in ANNs. The topology in this proposal is restricted in order to be accommodated in a 2D optical circuit without crossings and each node receives input from one or two other nodes and feeds it forward to one or two as well.

An implementation was realized with three hidden layers using a four-dimensional feature vector encoded in four wavelengths with InP optical amplifiers[235]. Their multiplexed light is broadcast to ANs comprised of four (different colours) inputs, four amplifiers to weight the four data signals individually and a multimode interferometer-based combiner.

#### 4.4.3.3 Micro-ring, micro-disk resonators

A simple lattice of micro-ring resonators coupled through a network of waveguides was proposed to realize a reservoir computing system. Here the activation functions relied on the nonlinear recoupling between the waveguides and the rings, owing to thermal effects and generated free carriers.[236]

Micro-ring *resonator banks* were proposed[237] to realize convolutional DNNs based on the *broadcast-and-weight* protocol soon after it was experimentally demonstrated to enable reconfigurable analogue ANNs in silicon photonic integrated circuits.[238] Briefly, multiple wavelengths (that represent the input feature vector) are distributed (broadcast) to all input neurons (comprising an input- and an output-waveguide coupled through a row of micro-ring resonators) that pick up and sum contributions from different wavelengths. The signal, collected by a photodiode, is used to modulate a distinct laser wavelength which is bundled with those of all other ANs and sent to the next layer. The multiplication is carried out by the micro-rings by tuning their drop coefficients to characteristic wavelengths; the accumulation is carried out by the photodiodes.

Micro-disks instead of micro-rings were proposed for a CNN where convolutions are performed by photonic integrated circuits while the other process operations are handled by CMOS circuits.[239] The photonic processing units comprise a photonic matrix-vector multiplier that relies on an array of micro-disk resonators, photodetectors and a wavelength multiplexer for parallel computations; a photonic ripple carry adder consisting of microdisk-based electro-optic full adders; and a photonic binary shifter consisting of a square mesh of microdisk resonators. The analogue to digital and digital to analogue conversions required between photonic an electronic processing constitute a hindrance in the performance.

### 4.4.3.4    Phase change materials

Photonic circuit integration requires solutions for weight storage that can be obtained from the monolithical integration of materials such as BaTiO₃. Its ferro-electric switching can enable the realization of non-volatile phase shifting elements when in combination with applied electric fields. Photorefractive effect can store synaptic weights in GaAs refractive index gratings, created by interfering optical beams, that constitute weights matrices. A smart configuration that permits to access the photorefractive weight matrix from orthogonal directions gives direct access to the transposed matrix speeding up the backpropagation algorithm.[240]

Micro-ring resonators incorporating phase change material elements had been proposed[241] and their functionality has been demonstrated.[242]

A fully optical —including all-optical synapses— integrated and scalable platform implementing SNN was realized using phase change material cells to store weights.[243] See **Fig. 34**. PCM cells (GST in this case) are integrated on top of the waveguides so that they can alter the propagating mode in a controlled manner: when in the crystalline phase they absorb most of the light inhibiting transmission. Input summation is performed by wavelength division multiplexing (integrate) and nonlinear activation is achieved through PCM-loaded micro-ring resonators, where the incoming summed signal serves to change the ring resonance and activate the switching when above the corresponding power threshold (fire). In order to add unsupervised learning capability to this architecture part of the neuron's output is channelled back through a waveguide.

### 4.4.3.5    Superconductors

Loop neurons (inductively coupled loops of supercurrent) are optoelectronic spiking systems where synapses are comprised of a Josephson junction in parallel with a superconducting single photon detector. Photons —used to rate-encode signals— are received by single photon detectors and integrated in supercurrents and stored in superconducting loops. The amount of current added by such events is determined by the weight, which is dynamically controlled by another such pair, and the stored current proportional to the number of detected photon arrival events. The activation function is performed by a *receiving loop* that is inductively coupled to all synaptic loops, thus integrating the currents from all of them. When the current exceeds certain (learnable) value, a current pulse —fluxon— initiates an amplification cascade that ultimately drives a semiconductor light emitting diode. Photons produced are used as input to downstream neurons.[245]

### 4.4.4.    Spiking excitable systems

Excitable systems are nonlinear dynamical media with the capacity to propagate one perturbation of some kind but which cannot support the passing of another until a certain time has elapsed (known as the *refractory* time). Biological neurons are an example of such sys-tems. A powerful model shown to faithfully reproduce this behaviour is the Ermentrout-Kopell or *theta* model[246] which makes very weak assumptions satisfied by Hodgkin-Huxley-like models of membranes.[119]

When a laser is working in the excitable regime[247] it presents dynamics analogous to those of biological neurons, only several orders of magnitude faster. If integrated, a network of such lasers promises the capacity for ultrafast computing.

One such implementation makes use of the extraordinary properties of graphene as a saturable absorber.[248] A fibre ring laser configuration including a cavity with chemically synthesized graphene sandwiched between two fibre ends and a stretch of highly erbium-doped fibre as gain medium can be used in temporal pattern detection. Circulation direction is selected with an isolator and the ring is pumped with a diode laser. When energy in the ring reaches the threshold for absorption saturation a pulse is emitted into the ring — performing the activation function whose output can be extracted with a coupler— and the system enters a refractory regime where it cannot respond a second pulse. This device is capable to process rate coded information, pulse coincidence, clustering, correlation etc.

The close relationship between *implementation* and *application* of AI is best reflected when the device used to execute AI function is, at the same time, the subject of study. A living three-dimensional brain tumour was demonstrated to operate as a random optical reservoir computing learning machine for the investigation of cancer morphodynamics.[249]

## 5.    Applications: advanced materials

Given a material in some morphology, its properties can be computed using purely theoretical (mathematical) tools or approximated using numeric methods. This process from the system to the properties can be described as forward modelling.

In chemistry this is performed by solving Schrödinger equation from first principles, which requires numerical methods. The fast development of computers in the second half of the 20th century prepared the ground for this approach to tackle ever increasing complexity compounds and led J. Pople to create a program—Gaussian 70—that could perform *ab initio* calculations that predicted the behaviour of molecules of modest size, purely from the fundamental laws of physics.[250]

This became a tool that allowed masses of researcher access to quantum chemistry so that, with little theoretical background, they could deal with systems comprising immense numbers of ions and electrons. In this way computation can anticipate the properties of a compound before it is synthesised. Density functional theory[251] added to this techniques an ability to describe the properties of large sets of existing and potential structures including molecules, crystals, etc. All that remains is wide availability of data, software and education.[252]

Nowadays applications of AI have become so widespread that one hardly realizes the daily impact it has on our activity. However, despite its current wide acceptance, only last century, the use and potential of AI in chemistry were met with reservations.[253]

Apart from ordinary computing and communications, mostly software-related, there are other areas, especially in science, including quantum many-body problems[254] or even particle physics and cosmology, quantum computing, and chemical and material physics, where ML is having a phenomenal impact.[255] Materials science and AI have a strong relationship in that both benefit from each other, since many materials and their properties and applications can be discovered through AI and ML methods and, as has been discussed,

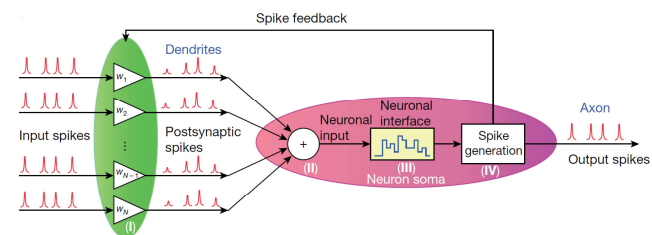

**Figure 34.** Integrated circuit configuration of an SNN several pre-synaptic input neurons and one post-synaptic output neuron connected via PCM synapses. Adapted from J Feldmann Nature 569, 208 (2019).

AI hardware depends on current and future materials for implementations.

This alliance is being perceived as the beginning of a human-machine partnership between expert scientists and materials-savvy digital assistants.[256]

## 5.1.    The materials challenge

Conservative calculations, using only 98 elements from the periodic table, estimate the number of possible materials as a googol ($10^{100}$), which is more than the number of atoms in the known universe.[257]

About 500 naturally occurring amino acids are known (though only 20 appear in the genetic code) and can be classified in many ways. Chains of these form polypeptides that can be used to code the synthesis of proteins. The enormity of the number that can be considered is hardly conceivable, let alone synthesized and screened. The size the problem can be appreciated by realizing that if one such chain could be synthesised every second since the universe started, only those up to seven-long could have been produced.

Combinatorial chemistry, an archetypal brute-force method to create many possible compounds in a single process that may lead to tens, up to millions, registered by A Furka in the early eighties, is an example of massive production of compounds that helps shorten the screening of pharmaceutical compounds.[258] The number of small molecules possibly synthesizable in laboratories may be placed somewhere between $10^{24}$ to $10^{60}$, far beyond any imaginable database to feed ML algorithms.

At sight of this landscape, algorithmical von Neumann computing, based on sequential logic progression is revealed impractical. But the alternative, ML techniques, require well curated data to feed on. Data regarding reactions available to train ML is even more scarce and difficult to use than that generated by *ab initio* techniques. This gives a context to the potential and task ahead of ML in the area of chemistry with fruit harvested in vastly diverse areas such as quantum mechanically derived energy and force evaluation,[259] molecular dynamics,[260] inverse design of molecular crystals,[261] large proteins,[262] chemical reactivity strategies[263][264] or intelligent experimental data evaluation and model assessment based on experiments *carried out* on public and proprietary data sets.[265] The nature and amount of data available to the ML techniques is decisive in choosing the method. In this sense, methods that are intrinsically suitable for small size data benefiting from transfer learning[266] and those capable to curate data by completing missing data are receiving much attention.

### 5.1.1.    Materials informatics
Data-driven *materials informatics*[267] has emerged as a discipline in itself encompassing different strategies sharing the same purpose, namely, to produce rapid predictions based purely on past data.[268] Materials informatics is becoming useful to access properties hard or impossible to measure or to compute (for the time or resources required) but for which sufficient pre-existing or easily to generate data is available. As for any approximation procedure, the capacity of these techniques is not only interpolative (like finding a new material among a class) but extrapolative (predicting properties beyond the class from which data was provided). The statistical rigour of the ML techniques must always provide estimates of the uncertainty of the prediction.

Despite their incredible success, these techniques are not without issues and, like the knowledge they are meant to uncover, they still may profit from further development.[269]

A full review of its applications to materials science is a difficult task and a taxonomy depends much on approach.[270] A dissection by the nature of the materials could become too intricate and contribute only to restrict the view and the message that ML can benefit essentially all materials research areas. Those interested in how these techniques apply to their materials of choice just need to search material-ML in bibliographic search engines to obtain a more material circumscribed literature. Here is one based more on how and what AI can do for materials and less on materials types and their functionalities.

## 5.2.    Discovery and design

Simple perceptrons capable to perform basic regression and classification but with just a few layers, shallow (as opposed to deep) ANN soon were capable to predict physicochemical properties such as solubility or thermodynamic properties.[271] The accuracy of most models seems similar the experimental measurements, when assessed with a test set selected from the working database.

In chemistry, in order to map input data, $\{\mathbf{x}\}$, into labels, $\{\mathbf{y}\}$, ANNs build mathematical relations $h(\mathbf{x}) = \mathbf{y}$ where $\mathbf{x}$ is a "representation" of a set of atoms forming the molecule under study and $\mathbf{y}$ is the chemical property under evaluation. The difficulty to find a satisfactory mapping depends on the complexity of the relationship and normally relies heavily on an adequate coding of the data into chemical feature representations.

The feature representation (atomic numbers, chemical bonding, atomic coordinates, etc.) from which ML feeds is thus crucial and best practice in constructing descriptor requires *uniqueness* (for example descriptor should be invariant under the symmetries of the system while sensitive to asymmetries, *e.g.* chirality), *universality* (descriptors adaptable to any system), and *efficiency* (key to make ML competitive against classical computing methods).[272] Generating these descriptors has become an expertise in itself.[273] This kind of data is *denser* than experimental one, which is naturally sparser, therefore, completing one with another helps improve the databases.[274] A wise balance of chemical descriptors, size of training data, and choice of ML algorithm helped improve predictions of nuclear magnetic resonance chemical shift for complex molecular structures.[275]

The great power of surrogate models can be used to create new materials or compounds based on learnt data or, in a more useful and work-saving way, to optimise for more than one property at the same time.

### 5.2.1.    Active learning
Surrogate models, relying on available data, can provide candidates for the fulfilment of a sought property that will, later on, be validated experimentally in the laboratory or the computer. Thus, the model is used in a passive way with no control over the data-dependent error in the predictions. *Active* learning adopts an iterative procedure whereby initial available data is used to start a data learning-data generation cycle so that the data generated after a cycle, when fed back to the ML model maximizes the improvement (either as success in identifying best performers or in perfecting the predictions).[276] This procedure helps make better data selection and can be combined with laboratory hardware to implement robot-assisted materials optimization.[277]

### 5.2.2.    Generative design
At variance with conventional screening, including active learning, where the data space is essentially discrete, *generative* ML models build a continuous feature vector space by mapping the training data. From that *latent* space new data can be readily extracted on demand. These models typically use variational autoencoders (which reconstruct the input after being compressed into a few latent variables) or GANs.[270] Since the underlying structure-property relationships are frequently nonlinear in complex functional materials, the materials produced by these models can be very different from those on which they are generated and offer greater potential for discovery in fields as diverse as biomedicine[278] and photonics.[279] Success is more limited in the area of solids for the added difficulties brought about by periodicity and the need for periodic boundary conditions.

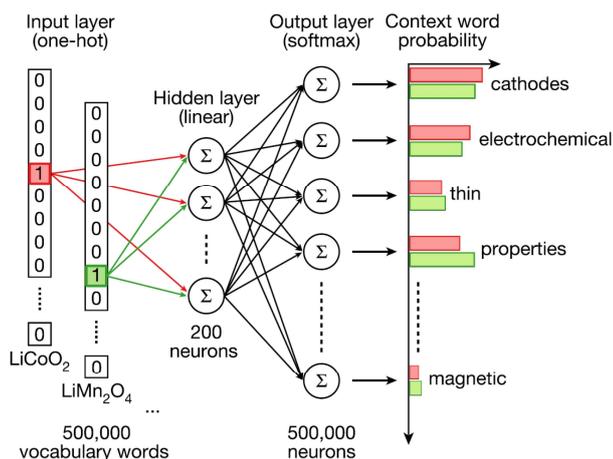

**Figure 35.** Target words are represented as vectors with ones (1) at their corresponding vocabulary indices and zeros (0) everywhere else (one-hot encoding). These vectors are used as inputs for a NN with a single hidden layer, which is trained to predict all words mentioned within a certain distance (context words) from the given target word. Adapted from V. Tshitoyan, J. Dagdelen, L. Weston, A. Dunn, Z. Rong, O. Kononova, K. A. Persson, G. Ceder, A. Jain, *Nature* **2019**, *571*, 95.

### 5.2.3. Data mining

*Data mining* can be considered a step in high throughput workflow. The mining process (screening) refers to the successive application of increasingly strict constraints to the database in order to select the best candidates according to the pursued attributes. In this way, materials complying with the attributes are kept and those that fail are discarded. The attributes over which the database is filtered, called *descriptors*, need not be physical quantities and may involve a good deal of insight to define.[280]

Unfortunately, in comparison with other fields, the corpus of materials data is much smaller and sometimes more diverse (heterogenous in terms of usability) which redounds in detriment of the ML application. In general, the predictive *accuracy* of a ML model increases with increasing training data abundance, seemingly following a power law[281] and, at the same time, it eases pattern detection. It appears that the influence of data size on predictive accuracy is mediated by the model complexity (degrees of freedom).

### 5.2.4. Text mining

The vast majority of available scientific knowledge is in the form of text describing phenomena, materials and methods. This information format makes traditional statistical analysis difficult even for modern ML methods. For these methods, the main source of structured, readily processable data is contained in property databases[282] which only hold a small fraction of the accumulated knowledge. Much of this knowledge, containing interpreted relationships between materials and properties, synthesis descriptions etc, is in the form of natural language which, to be useful, should be labelled by human intervention in order to create training data.

Recently it was shown that such knowledge can be encoded — information-dense embedding—without supervision.[283] This technique was shown to be capable to recommend materials for certain functionalities even years prior to their discovery, suggesting that much more knowledge is contained in literature than is apparent. Briefly, ML algorithms like GloVe or Word2vec, upon training on suitable text samples, produce vectors —several hundred-dimensional— representing words that lie at *smaller* cosine distance when the words refer to *closer* concepts so that their syntactic and semantic relationships are translated to the high-dimensional vector space. See **Fig. 35**. This method does not include or attach expert knowledge in the process and yet predictions are consistent with scientific intuition. Furthermore, vector representation permits to do algebra (addition, dot product, etc) with words to perform analogies. Once again, this work also showed that data selection is critical by comparing accuracy when choosing paper abstracts only or full Wikipedia as text corpus for the training; the latter including a lot more unnecessary or domain unspecific data.

Processing information contained in hundreds of millions of scientific papers not only demonstrated that the knowledge was already available to suggest materials for new functionalities but, surprisingly, these results showed, too, that new applications for known materials can be found automatically by extrapolating the domain knowledge contained in literature.[283] Similar strategies could predict synthesis and preparation routes optimization for functional materials.

## 5.3. Molecular simulation

### 5.3.1. Density functionals

Density functional theory[251] is currently among the most effective in computational cost for accuracy quantum mechanics-based methods of material property prediction. Briefly this theory states that, in a system with $N$ electrons, the external potential $V(\mathbf{r})$, felt by the electrons is a unique functional of the electronic density $n(\mathbf{r})$ and that the ground state energy $E(n)$ is minimal for the exact density. That is, the knowledge of the electron density is enough to obtain the energy of the ground state. This theory relies on some critical approximations involving electron-electron exchange which, in some relevant cases fails, prompting an endless search for better approximations.[284]

The accumulation of tens and hundreds of thousands of density functional theory calculations in large databases[285] from which it is possible to train ML algorithms is contributing to accelerating materials design and discovery.

### 5.3.2. High throughput design

High throughput computational materials design[280] is based on the mixing of computational quantum-mechanical–thermodynamic approaches and a multitude of techniques rooted in database construction, apt for intelligent data mining in emulation of the experimental counterpart.[286]

The actual procedure comprises three steps dealing with i) virtual materials growth, where thermodynamic and electronic properties are computed; ii) rational materials storage in structured databases and iii) materials characterisation and selection to extract novel materials or physical insights.

The logical sequence is to profit from density functional theory and high-throughput approaches to generate massive amounts of data that ultimately constitute the input on which data mining, screening and other ML methods feed.[287] With these tools success has accompanied the search for binary (se for instance **Fig. 36**) and ternary compounds, solar materials, water photosplitting and catalysis, scintillating materials, carbon capture and gas storage, topological insulators, piezo and thermoelectric materials, etc.

### 5.3.3. Interatomic potentials

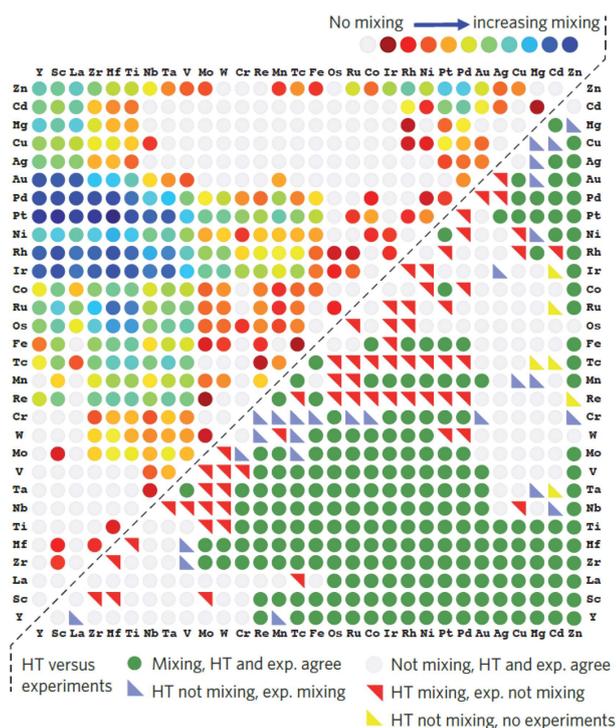

**Figure 36.** High throughput analysis of intermetallic compounds. The top left triangle shows the mixing tendency of the 435 d-electron binary intermetallic compounds (a measure of the strength of a mixture to oppose disorder). High throughput techniques are capable to predict these mixtures stability. The bottom right triangle report the success of the prediction with circles (grey or green) mark the agreement cases. Adapted from S. Curtarolo, G. L. W. Hart, M. B. Nardelli, N. Mingo, S. Sanvito, O. Levy, *Nat. Mater.* **2013**, *12*, 191.

Most physical properties of materials derive directly from the total potential energy; in fact, the equilibrium structure (molecule or crystal) is that of minimum energy. Departure from the equilibrium configuration leads to higher energies and, in this way, force constants, elastic, vibrational (phonons), and many other properties can be derived from the curvature of the potential energy surface as activation energies. In addition phase transitions or reactions can be associated to *passes* between local minima of reactants and products. First principles methods based on quantum mechanics, founded on the Born-Oppenheimer approximation, can provide such surfaces but can be computationally intensive with systems of more than one thousand atoms. A circumvention to that problem comes from the use of empirical *interatomic potentials*, also called force fields, instead of solving the Schrödinger equation. Naturally this reduces computing requirements at the same time as accuracy and applicability.[288]

ML methods offer a shortcut to potential-energy surfaces by fitting large data sets from electronic structure calculations (*e.g.* from density functional theory) to analytic functional relationships and dropping any physical meaning. Progress in this area has depended on two essential aspects.

On the one hand, the structural description —and the associated selection of feature vectors carrying the structural inputs— revealed critical in dealing with high-dimensional systems unlike small molecules and the need to support symmetry invariances.

The other crucial factor is the application of ML techniques such as NNs, support vector machines, etc.[289]

One of the most challenging difficulties ML faces in this area is the exponentially growing complexity chemical composition introduces.

### 5.3.4. Crystal structure prediction

Predicting the crystal structure that a given chemical formula will adopt entails finding the unit cell and atomic positions that minimize the relevant thermodynamic quantity. The number of formula units per primitive cell, a crucial parameter, is unavailable for unknown formulas and makes it necessary to find the most stable configuration for unit cells containing all possible formula units and selecting the smallest. A colossal challenge. In practical terms, this is tackled by computing the potential energy surface for different atomic arrangements and finding the minimum energy in a seemingly straightforward route. A class of efficient implementations are gathered under the denominations *evolutionary* and *genetic* algorithms that, unlike purely random searches, explore the whole energy surface while simultaneously zooming in on the most promising parameter regions.[290] Since this search is history aware they are considered examples of AI.

These techniques, like interatomic potentials, can be used to generate a database of structures at great computational cost. Alternatively mining a database of known crystal structures (data mining structure predictor) can provide probabilities that a given formula presents certain crystal structures.

Using this model, experimentally observed structures appeared in the top five of the list of most likely structure types 90% of the times.[291] Unfortunately, since the model feeds on existing data on known structures, it cannot predict the existence of a structure never seen before.

### 5.3.5. Lattice models

Apart from obvious solid-state crystals, a number of physics problems can be stated in terms of periodic arrays like those dealing with alloys, spins interactions, etc. One remarkable example is the Ising model of spin glasses. In this kind of models, certain magnitude, often the energy, is computed as a sum of pair interactions that can be expanded considering groups of spins in the so-called *cluster expansion* model.

Various approaches have been developed to apply cluster expansions that could be interpreted in a Bayesian framework that considers a prior probability distribution of effective cluster interaction parameters.[292]

### 5.4. Property prediction

*Inverse design* aims to discover tailored materials from the starting point of a particular desired functionality.[261] Inverse design reverses the usual research-development cycle where an idea ends in a product, by starting with the desired functionality and searching for the best molecular structure that fulfils the required function. This technique was initiated with the high throughput virtual screening methods that produced so many advances in molecules and materials for photovoltaics, batteries, organic diodes, catalysis etc. Optimization methods are preferable since they start with a smaller number of configurations. ML-derived *generative* models pursue to determine the *joint* probability $P(y,x)$ that property $y$ appears with material/molecule configuration $x$; at variance with *discriminative* models concerned with computing the *conditional* probability $P(y|x)$ that property $y$ appears given configuration $x$.

Using composition, structure, morphology, processing etc. to generate the target property or performance criteria requires careful selection of training data, cross-validation of parameter selection, and testing on unseen data to ensure reliability. All this careful and conscientious training pays off later in computing time saved.

This strategy has demonstrated its value in a plethora of materials comprising inorganic (metals and alloys, ceramics and composites), organic (polymers), hybrid etc. where physical and chemical properties as well as functionalities were successfully predicted.[270]

### 5.4.1. Structure-property relationship

One of the most intensive use of ML techniques in materials science is its deployment as a substitute —surrogate models— for the full experimental laboratory work linking the processing, structure, property and performance prediction cycle based on widely accessible numerical representations.[293] ML algorithms permit to chart the composition, structure, morphology landscape in order to find best-suited candidates to fulfil a technological requirement without the need for actual experimental work. Obviously, only a judicious selection of training data and a rigorous testing can grant a reliable prediction.

Applications successfully cover property prediction of every kind including physical (mechanical, electronic, thermal, optical, magnetic, etc.) and chemical (reactivity, catalytic, etc.)[270] and have been combined with robotic systems to proceed in an autonomous fashion.[294]

Modelling disordered materials in general and amorphous ones in particular poses a tremendous challenge to numerical techniques because their lack of periodicity precludes the use of most of well-established models. Here it becomes necessary to resort to artificially imposed periodicity so that the material is approximated as having a pseudo-amorphous structure with large amorphous unit cells repeated periodically. Numerous disordered regions, comprising randomly generated disordered clusters, can be used to extract optimized feature vectors that, once learnt by a ML algorithm, can later make predictions.[295]

### 5.4.2. From quantum mechanical computations

Computation of materials properties based on quantum mechanics is nowadays accurate, versatile, fairly general, and not tied to concrete systems. However, solving the Schrödinger equation, can turn computationally costly and time consuming. This is a typical scenario where ML can be advantageous by training an algorithm on a reduced set of computed data so that it predicts the rest.

Firstly, materials are distilled into numerical *fingerprints* (feature vectors) that are subsequently mapped to properties. Then these mappings are used for inference.

For example the fingerprint for a molecule class can be the vector of ordered eigenstates of zero-padded Coulomb matrices.[296] This pioneering fingerprint model has been extended in various ways to suit different applications and solve issues like those arising from symmetry in solid state, for instance, and leading to a wide variety of descriptors.[297] These fingerprints are used to train a ML algorithm that learns to associate fingerprints to certain magnitudes of interest (*e.g.* atomization energy). The trained model can then, at negligible computational cost, predict the properties of unknown molecules.

## 5.5. Materials characterization

Materials characterization often involves routine tasks related to data interpretation, image or spectrum analyses, etc. some of which can be tackled favourable y by ML protocols.

### 5.5.1. Image techniques

Apart from the most well-known applications concerned with, for example, face recognition there are a few areas where image analysis, tending to generate diagnosis and interpretation of images in general, have an important niche in the application of ML techniques. Some involve microscopies of several kinds and affect the processing stage to accelerate diagnosis and reduce human intervention.

For materials science, the power of image analysis resides, not in the possibility of algorithmically to recognizing the object represented in an image, but in the opportunities if offers to automate image-dependent techniques such as the various microscopies, echographies or even magnetic resonance imaging.[298]

For microstructure identification, quantification of spatial arrangement and shape distribution are the principal components. In a typical task of micrograph classification the general approach involves feature extraction (detection of local regions of pixels that include relevant parts of a microstructure, such as edges, corners, indents, protrusions) and feature selection (through dimensionality reduction) to acquire the feature vectors that are then used for training, validating, and finally testing various classification models.[299]

Scanning electron microscopy characterization involving multiple length scales require very large areas (centimetre-scale) with nanometre-scale detail and are, thus, all but impossible to realize in practice using inherently slow conventional SEMs. Single-beam high throughput scanning electron microscopy can handle this situation producing enormous amounts of graphical data that only a ML algorithm can process.[300]

Adding an extra dimension to ordinary imaging techniques renders human interpretation, even of few micrographs, all the more tedious which makes the use of ML methods even more advantageous. This is the case with X-Ray tomography, for instance, where AI is bolstering the study of battery materials.[301]

### 5.5.2. Diffraction

Multiple techniques such as X-ray,[302] neutron[303] and electron[304] *diffraction* are routinely employed to obtain three-dimensional spatially resolved crystallographic characterization of different morphologies in materials. Massive data analysis can only be carried out with the aid of high throughput methods relying on ML.

### 5.5.3. Phase diagram determination

Materials crystalline *phases* are best revealed from their X-ray diffraction properties considering the Fourier transform relationship between the reciprocal-space diffraction pattern and the real-space electron density. However, identification of constitutive phases in a multiphase mixture can become a complicated task when using conventional rule-based data analysis tools such as commercially available computational packages.[305]

Interpretation of data when many hypotheses need to be tested, is a ground where AI finds a direct application. Using XRD spectra as vectors, it is possible to employ clustering analysis to group them and associate similar ones to different phases in a collection arising from a mixed phases sample. This technique is best applied in ternary, quaternary, or more components alloys grown on a large substrate that offer a wide range of compositions.[306]

Recent CNN developments can fully dispense with human intervention in inference and even avoid feature engineering for training. Although the training uses simulated data, tests on experimental XRD data return a very high accuracy of nearly 100% for phase identification.[307]

### 5.5.4. Synthesis route design

Text mining has been used with success in identifying materials *synthesis routes*. Deployment of natural language processing methods with ML is capable to extract from the experimental sections of journal articles the synthesis parameters (temperatures, synthesizing actions like "heating", "stirring", "calcination", etc., and material names) required to prepare target materials.[308] Feature vectors may contain a *similarity* dimension providing information both from similarity to other routes (synthesis parameter-based similarity) and/or other materials (material-based similarity).

Computational screening of materials synthesis parameter sets is far behind other screening applications like drug discovery. The probable reason might be data sparsity and data scarcity because, too often, synthetic routes strongly depend on certain parameter in a sparse, high-dimensional space that is difficult to explore directly. In addition, at least for some materials of interest, only scarce volumes of reported syntheses are available.

Furthermore, these techniques can also be used to transfer learning to predict the result of synthesis procedures for materials outside the training data set.

### 5.5.5. Materials processing

ML techniques have been extensively used to predict design and characterize materials properties closely related to their atomic or electronic structure. But they have also enjoyed success in tasks that can be regarded as far from the elemental matter composition in manufacturing and process control like estimating flow of melting steel or modelling abrasion in machining process.

Magnitudes like flow stress, friction, hardness strength, fatigue or fracture are relatively far from fundamental properties and become very difficult to compute from fist principles.[309]

Additive manufacturing can be considered a technique even further away from basic (materials chemistry) synthesis. It is a disruptive technology that uses digital manufacturing to create 3D objects usually in a layer by layer fashion and guided by a computer aided design. Because it follows a 3D printing strategy it can build complicated topologies impossible to machine even if the materials allowed.[310] Since it allows materials in all research areas to be built (biomedicine, mechanical engineering, metallurgy, photonics, etc.) its interaction with machine learning techniques is granted.[311] Apart from the important contribution to materials design, ML still has further application in additive manufacturing because it can contribute to the design, the process and the production. The majority of applications are related to processing, however, and more in particular to parameter optimization, in-process monitoring and quality inspection.

Like in many areas of application, the complex relationships between design parameters —such as materials properties, product design, process parameters, process signatures, post fabrication processes, etc.—on the one hand and the goals pursued, embodied in the final product quality on the other, is the principal obstacle for widespread industrial adoption because it limits process robustness, stability and repeatability. Novel approaches adopt a *digital twin*-enabled collaborative data management framework where different product lifecycle stages are considered.[312]

# 6.    Applications: advanced structures

Applications of AI in materials science mostly concentrate on discovering or enhancing chemical compounds or composite materials for their properties or in optimizing their production methods and characterization techniques.

The popularization of nanotechnologies brought the widespread awareness that matter properties of nanosized objects may amply differ from those of bulk materials; this marked the flourishing of *nanomaterials*.[313] This, in turn, awakened another current of research based on the knowledge that structured material systems, where chemical composition is only half of the description, can present properties quite far from those of their components.

In many cases, materials are intended as the ground where physical interactions play and the vast majority of these interactions concern some kind of *wave* propagation. When the length scale of the structuring is much shorter than the characteristic lengths of the involved properties the systems are referred to as *metamaterials* which have acquired a status of *new research area*.[314] This highlights the scientific and technological importance wave materials (essentially electromagnetic and acoustic) have acquired.

Nevertheless, even in the case where structuring feature size is much larger than both the atomic scale or the metamaterial building block, properties and performance prediction/enhancement can be a tough challenge in computational terms. For this reason, all these regimes are areas where AI techniques can be a real booster. In addition, future composites and structures need to be tuneable, adaptive, and multipurpose. Optimization of more than one performance mark complying with multiple constrains is impractical under traditional experimentally guided trial-and-error processes where the search for untapped regions of the solution space is impossible.

## 6.1.    Traditional Inverse engineering

Inverse-design problem-solving works by optimizing a number of parameters describing the behaviour of a system to attain a desired response. Forward calculations in electromagnetic and elastic systems while moderately complex are well understood; on the contrary, inverse design usually involves more lengthy calculations and presents challenging issues.

There are essentially two ways to go about to solve inverse problems: reversing the equations describing the system (*adjoint* method) or searching the parameter space step by step optimising a function (*genetic* algorithm). These methods are often used in conjunction with ML ones —especially with generative ones— so it is difficult to make a clear separation between classic and DL algorithms.

### 6.1.1.    Adjoint inverse engineering

The adjoint method computes the gradient, $\nabla_p f$, of cost function, $f(\mathbf{x}, \mathbf{p})$, (sought-after property) in order to optimise a set of parameters, $\mathbf{p}$, that minimize the function with the constraint of the governing equation $g(\mathbf{x}, \mathbf{p}) = 0$ (underlying physics) whose solution gives $\mathbf{x}$ for input $\mathbf{p}$. In order to do that the transpose of the Jacobian matrix $g_x^\top$ is used, hence the name.

As an example, for electromagnetic materials, $f$ represents a performance and $g$ the Maxwell's equations; while $\mathbf{x}$ can be the electromagnetic fields and $\mathbf{p}$ the distribution of materials (shapes, dielectric functions, etc). Solution is iterative as for each $\mathbf{p}$, $\mathbf{x}$ must be computed in order to calculate $f$ and, subsequently, its gradient. In comparison with genetic algorithms, this method is more efficient by far but it often requires profound knowledge of the physics involved and is greatly non-trivial.

Some implementations, like those involving shape optimization applied to electromagnetic design, enjoy further advantages in that the iterative scheme can profit from commercial forward solvers.[315]

In the case of non-linear photonics even though the problem is a non-linear one minimisation of the cost function requires the determination of its gradient which is still a linear problem.[316]

### 6.1.2.    Genetic optimization methods

*Genetic* algorithms[317] explore the parameter space of a complex design by randomly combining single elements (genes) from the two sets of parameters (chromosomes) describing a *parent* couple's individuals and adding some mutations to give rise to the next generation individual parameters set in an iterative way. When the performance of the next generation is improved the chromosome is saved and used to produce further generations, otherwise it is ditched. In this way successive generations present better performance regarding the sought-after properties in the same way evolution selects the apt individuals. This approach is hard on computation power and time as the problem grows exponentially with the number of parameters.

This approach has been widely used in electromagnetic wave materials[318] as well as in acoustics.[319]

## 6.2. Electromagnetic systems

Electromagnetic systems deserve a section of their own; such is their variety and technological impact. The underlying governing laws are just four well know equations. Yet, the fact that materials respond so differently at different wavelength ranges and the influence the object size has, even for given material and wavelength, opens a vast field of opportunity in creating functionalities or applications.

Advances in nanotechnology, making use of careful engineering of photonic structures to levels smaller than optical wavelengths, permits nowadays to manipulate light —transmittance, polarization, helicity, frequency, etc.— in unprecedented manners. The potential for application of systems designed in this way is so tremendous that the problem lies rather on the engineers' imagination.

Inverse design methods have a long and successful tradition in photonics where they contributed to the design of many devices such as power splitters, light trapping structures, dielectric nanoantennas and many others. Arguably, these devices could hardly have been discovered or optimized with simple intuition.[320] However, acknowledged pitfalls of traditional optimization, like local minima, solution non-uniqueness, and expensive computation, spurred the use of data-based techniques such as deep learning to boost the power and efficiency of these methods. [321]

DL typically relies on large amounts of training data that, in turn, requires significant amounts of computational resources to be generated (through simulation). This cost is, however, incurred just once. Classical optimization requires the same amount of simulations for *each* design. The advantage for data-driven methods is thus clear.

It is accepted good practice to deploy *discriminative* methods (supervised learning as in classification and regression) when dealing with moderately complex systems (high number of degrees of freedom). On other hand, *generative* models (related to unsupervised classification) are reserved for systems of added complexity (extremely high degrees of freedom). This helps prevent overfitting. The training data should always be more abundant that the degrees of freedom something only possible to fulfil for moderately complex systems. These methods are usually deployed in combination with others such as *black box* or *evolutionary* methods (not dependent on gradients).[322]

### 6.2.1. ML forward design

Notwithstanding what has been mentioned, the high complexity of systems fulfilling certain functionalities has made allegedly simple forward modelling an actual challenge and optimizing them a formidable one.

Forward modelling addresses the computation of electromagnetic properties of given structures (described in terms of the materials concerned and their geometric distributions) *simply* by solving Maxwell's equations. Analytical and numeric approaches have long tradition and a plethora of methods have been developed. The *transfer matrix method* can describe light propagation in stratified media by obtaining analytical expressions for the complex fields at the interfaces and hence reflection and transmission. For periodic structures, similar results can be obtained, working in the Fourier space, with the *rigorous coupled wave analysis* method. However, when the structures acquire more complexity than stratification or when periodicity is forsaken, and more so especially in three dimensions, numeric methods become indispensable. They come in two principal flavours: *finite difference time domain* and *finite element* methods. Both are based on the discretization of the system (space and time) and have demonstrated formidable power in solving nanoscopic-featured structures. However, the source of their power is at the same time their weak point: they can become excessively demanding in computing power. This is especially critical, for instance, in plasmonic systems where the metal feature size needs deeply sub-wavelength detail whereas the total size of the system can be many orders of magnitude larger. For these reasons even in the case of forward modelling, there is a place for data-driven techniques like ML.

At any rate, most ML optimization tasks whose chief aim is the inverse process often include a forward design part, sometimes the same as the inverse working in the opposite direction.

#### 6.2.1.1 Particles and multilayers

ML and especially DL can speed up the prediction of optical response of complex systems by approximating Maxwell's equations instead of solving them. In a pioneering example, in order to compute the scattering cross section of $SiO_2/TiO_2$ *multilayer particles*, a DNN was trained with the exact scattering cross section spectra of several thousand examples of randomly chosen layer thicknesses. See **Fig. 37**. The input was a vector containing the shells' thicknesses and the output one containing the exact cross section at several wavelengths. The training was validated with similar (but unknown to the DNN) exact spectra and then the DNN was used to predict the spectrum of an arbitrary shell structure in fast time.[323] The results show that the DNN is not interpolating or using some fitting procedure but discovering the underlying pattern linking structure to response so that it can solve problems never encountered.

In the previous example, the data on which the learner trains is easy to obtain and, therefore, abundant; so the more expensive part of ML is secured. However, in many cases where training data is hard to come by, the economy made in the predictive or inferring part of ML is hardly compensated if the data acquisition cost is taken into account. This is a perfect scenario for transfer learning where the ML methods are trained on tasks with abundant data and then used for the truly targeted one by transferring the knowledge. In practical terms knowledge transfer is implemented by i) training a DNN on the abundant or inexpensive data, ii) preserving certain number of hidden layers, iii) randomly initializing the rest and finally iv) training on the scarce or expensive data. In [324] a DNN is trained on data concerning a multilayer system of a given number of layers and the knowledge is transferred to learn a physically *similar* systems with a different number of layers. Alternatively, this transfer can be performed between physically *different* systems: training on scattering cross sections of multi-layered core-shell nanoparticles and transferring the knowledge to learn transmission spectra of multi-layered thin films. It is even possible to realize multitask learning by preserving a number of NN layers and training the rest for specific scarce data tasks. In this case, each target task benefits from the knowledge acquired by the others.

#### 6.2.1.2 Mesostructures

While the above structures can be fully described with very few parameters, (*e.g.* a few shell thicknesses) the predictive power of DL forward modelling is even more clearly exhibited when dealing with arbitrary shapes (requiring pixel definition).

Computing the diffraction of metagratings[325] (flat surfaces containing an arbitrary shape, periodically repeating pattern antenna) requires many parameters to be described: in fact an image. The large dimensionality of the input data in the problem is similar to computing the absorption by irregular shape plasmonic structures where forward modelling by DL aided by the image processing capabilities of CNNs clearly excels.[326]

### 6.2.2. MLInverse design

The advantage of ML inverse design stems from the power of computation to exploiting hidden information contained in big data. Well-designed and trained ML algorithms overpower human expertise allowing human creativity to open new fields. The tools offered by ML are varied and, typically, the employed methods combine several machines into a mixed architecture. Usually bidirectional DL

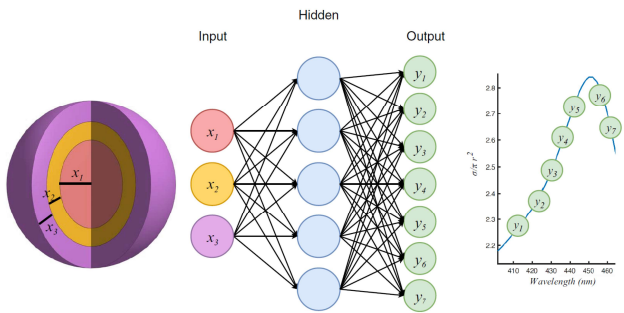

**Figure 37.** A trained NN can predict the response of a shell particle with acceptable precision in much faster time than a numerical calculation. Adapted from J. Peurifoy, Y. Shen, L. Jing, Y. Yang, F. Cano-Renteria, B. G. DeLacy, J. D. Joannopoulos, M. Tegmark, M. Soljačić, *Sci. Adv.* **2018**, *4*, eaar4206.

models are combined with simulation components and conventional optimization methods are also included.[327]

ML aided inverse design has its own ways but a natural one is the reversal of a forward operating ML algorithm. The procedure in [323] was reversed to demonstrate that just like it computed the optical properties of a test structure working in one direction it could choose the specifications required to offer certain properties when working in the other.

### 6.2.2.1 Scattering

The colour provided by *scattering* from nanoparticles is attractive for high-resolution colour printing. Plasmonic nanostructures would be good candidates but for the losses and low saturation they produce. Dielectric nanostructures can be ideal since resonances can be controlled by shape and size. While optimizing using other ML models struggles to find solutions outside the training data parameter space reinforcement learning does not suffer from this weakness.

A reinforcement learning machine was trained to find the radii, spacings and thicknesses and the antireflective coating thickness for Si cylinders to produce reflectance spectra of pure primary colours. From these geometrical configurations, reflectance spectra can be obtained by FDTD simulations which can be translated into CIE coordinates. A reward based on the closeness to pure red, green and blue in the L*a*b* colour space was established leading to result improved as compared to conventional optimization.[328] If instead of optimizing scattering intensity for a target wavelength the system is optimized for phase instead of saturated colour images, high efficiency holograms can be printed.[329]

### 6.2.2.2 Multilayer structures: 1D

As stated earlier, one severe issue in inverse design is the fact that very different configurations can produce very similar responses leading to poor convergence and stimulating the search for diversions to circumvent it.

For example, a forward operating DNN trained to return *multilayer* transmission spectra when presented with the material geometry descriptions can be coupled to another reverse-operating DNN working in front. When the front DNN is provided with a target spectrum, it returns design parameters that the second one takes as input to yield a simulated spectrum. Using the difference between input and output spectra as a loss function the tandem machine can generate design parameters to satisfy any desired functionality.[330]

### 6.2.2.3 Metasurfaces: 2D

*Metasurfaces* are, essentially, complex diffraction gratings where the simple 1D saw-tooth shaped grooves are extended to a 2D plane where wisely arranged, arbitrary shapes, built with metals and dielectrics combined can create any imaginable diffraction-based manipulation of electromagnetic wave fronts. However, this is more easily said than done: the enormous degree of freedom in such structures along with the complex response of the materials involved may turn even conventional inverse design methods ineffective.

A metasurface composed of repeating metal features easily described by a handful of parameters can be given to a discriminative learning machine with two networks working together. A *geometry predicting* ANN (that after training outputs good parameters for an input spectrum) is coupled to a transmittance *spectrum computing* ANN (that after training outputs spectra for given parameters without the hassle of actual exact computation). Together they operate both as a modelling and an optimizing procedure.[331]

Unsupervised learning was shown to be capable to deal with these systems. In particular, GANs are especially appropriate to manage high dimensional data like pixelated images (metasurface unit cell) that are *generated* by one part of the learning machine, which is trained to match the desired spectrum. The *discriminating* part is in charge of discerning if the patterns are feasible in fabrication terms, if they are in the real geometric space and the performance. The competition with the discriminator forces the generator to produce legit patterns with the desired function. In this way, the generator learns what is important in the system to produce the target performance.[279]

If a part of the data is labelled, the ML need not be fully unsupervised. In fact, such a semi-supervised model[332] (based on a variational autoencoder) not only was shown to offer reliable results which but, thanks to its ability to handle labelled data, can help to exploit a researcher's experience while easing the burden of numerical simulations. A 2D latent space showing several geometry groups (corresponding to the shapes used in training) emerged from the training adding interpretability and confirming the ability of the autoencoder to learn shapes.

Very often the application aims at obtaining good passive performance, such as diffraction. Other properties like optimization of *thermal* radiation involve acting on the spectral distribution of electromagnetic radiation.[333] Here the goal is the design of a metasurface to be coupled to a hot element whose radiation must be reshaped to match the absorption of a photovoltaic cell for generating electrical power by radiative heat transfer. The virtue of the method, that complements *adjoint* topology pattern optimization with an adversarial autoencoder, is that the latent space is densely filled with a broader variety of encoded designs. A further refining step is applied finally to reach reshaping efficiencies close to unity.

### 6.2.2.4 Metamaterials: 3D

Fundamental principles like symmetry requirements and intuition can guide the design of functional metamaterials but all too often, that falls short of the design detail needed. The chiroptical response, for instance, bears a complex relationship with geometrical chirality for a performing metamaterial to be designed from meta-atoms. A pioneering work using ML with two bidirectional DNNs operating in parallel was capable to generate design parameters for a 3D metamaterial composed of stacked pairs of twisted split ring resonators showing the targeted circular dichroism.[334] As in many cases the forward modelling is provided in addition to the chief objective, inverse design.

### 6.2.2.5 Disorder design with AI procedures

Disorder comes in so many flavours and is so intrinsically attached to artificial materials that it attracts attention from many perspectives, bot theoretical and experimental. For example, complex problems like optimizing the performance of a demultiplexer based on a disordered Si photonics circuit was demonstrated using classical optimization techniques. Solutions were obtained by minimizing the violation of Maxwell's equations by desired output fields and also by minimizing the difference between the output field distribution and the desired one resecting Maxwell's equations.[335]

In wave materials, the use of deep CNNs has permitted to identify physical relationships between disordered structures and wave localization.[336] This relationship can be used to design structures with tailored properties for wave control such as in lasing, energy storage or bandgap materials. In certain cases, scale-free properties appear that help combat the risk of disorder inherent to fabrication.

In the engineering terrain, the frequent DL strategy of two coupled bidirectional by networks working together, was efficient to determine the best arrangements of holes to be opened in a squared matrix in order to produce a 2D in-plane power-splitting filter in an SOI photonic circuit.[337] Another example of the search for imperfect or non-symmetric structures is provided by the optimization of microcavities in photonic crystals where after many years of improvements guided by experience, human intuition yields to ML.[338]

### 6.2.2.6    Topological systems
*Topological* materials derive their properties from the special environment they offer for propagating wave perturbations. Initially electronic (quantum Hall effect), they soon were applied to photonics and phononics, elastic waves.[339] The origin of their properties lies in the global rather than on the local configuration, manifests itself through lower dimensional states (surface in 3D and edge states in 2D) and above all leads to states that are immune to scattering by disorder (*protected*). The latter is the most appealing property since it can lead to improved performance devices. In photonics, topological states derive from the complex optical response of materials geometrical structure whose design is still a challenge.

The Zak phase—the Berry phase gained by going across the Brillouin zone— is an invariant characterizing the topological properties of a band. Even for a 1D structure, its relationship to structural parameters cannot be given by simple mathematical expressions. Hence, inverse designing a material structures responding to some topological requirements is far from trivial. A ML network can be easily trained to learn the topological properties from design parameters (easily numerically computed in an example involving a photonic crystal of Si and $SiO_2$ [340]). By coupling an inverse design learner in tandem to the forward one it is possible to train it to overcome the problem of 'multiple configurations satisfying one request' that gives rise to divergence.[330] When geometrical parameters are stationary upon the action successive operation of the forward and inverse networks the inverse design network is trained and ready to produce a source of direct relationship between phases and structures.

Adding an operator parameter space to the problem enables this model to be extended  beyond the training data space and yield accurate predictions in phase transitios.[341]

A generalized selfconsistent extension of this approaches can handle more complex geometries (cylindrical waveguides) and higher dimensionalities.[342]

### 6.2.2.7    Integrated quantum photonics
The next step in computing will be enabled by quantum technologies and integrated quantum photonics will be a key step toward scalable quantum applications.

Obviously, the *design* and optimization of photonic systems as platforms for quantum technologies depends and can profit as much as any other applications on ML inverse design. The data-driven nature of ML techniques is well adapted to the usual statistical processing in quantum *measurements* in which data sparsity is characteristic. For instance, the goodness of a quantum emitter can be obtained from a classification process based on sparse autocorrelation measurements; something typically assigned to classical ML regression or classification algorithms. Finally, *automation* of quantum experiments will be necessary if full exploitation of quantum technologies is to be extended, beyond *simple* processing, measurement and metrology, to the design of more complex experiments. See **Fig. 38**. Here active and reinforcement learning have a central role in supplementing human insight.[343]

### 6.2.2.8    Whole devices
The usual virtual screening carried out for materials discovery can be extended to a whole *fabrication process* where all the steps involved in selecting a molecule in consideration of its photonic properties, synthesis constraints, performance, and fabrication compatibility are taken into account.

Starting with a 1.6 million candidates, theoretical insight, quantum and synthetic chemistry, characterization, industrial expertise and even fabrication and device performance were all integrated in the search for an organic light emitting diode molecule and the subsequent chain of processes leading to a performing device.[344] A trained ML model was relied upon to predict the thermal activated delay fluorescence decay rate and assist to build a library wherefrom further select candidate molecules for subsequent narrowing down steps on other parameters. After computation, the design space was further reduced with human intervention and then experimental validation carried out (synthesis and device fabrication). See **Fig. 39**. Close to one thousand untested molecules are expected to match or exceed the performance of those demonstrated to have state of the art figures.

### 6.2.2.9    Ultrafast Photonics
The realm of fast optical phenomena has both fundamental and applied relevance for the study of physical and chemical processes leading to the generation of optical pulses and also for the opportunities such technologies offer in the study of the dynamics of a multitude of problems. Therefore it is only natural that AI and its methods contribute to the understanding and development of ultrafast light sources and their applications.[345]

These techniques have an impact on several aspects beginning with the design and operation of laser devices. Here, realigment, self-tuning or autonomous operation, for instance, are achieved with algorithmic (genetic) feedback loop optimization; or the rich landscape of nonlinear dynamics and instabilities that require the use of ML. Likewise, control coherent dynamics with extra-cavity shaping, pulse compression etc. are usually achieved with adaptive algorithms.

DNNs have been applied in the inverse design of temporal pulse reconstruction and models of hidden physics are arising where *physics-informed* NNs are used to identify governing laws.

### 6.2.2.10    Communications and systems
The ultimate step in the hierarchy that begins with the substance, continues with the material, and rises through the component and the device, is the *system*, where multiple apparatuses operate coordinated. The system level is nowadays reaching the level difficulty Weaver assigned to *organized complexity* for which he anticipated "Science must, over the next 50 years, learn to deal with".[40] Although, as shown, AI and its techniques have a widespread use in lower levels it will become indispensable here. A natural example is optical networks.

In order to optimize operation and improve performance of optical communication systems and networks several scenarios can be broadly defined. In situations, *e.g.* routing, where the network is deterministic, observable, static, and fully understood *search* algorithms and optimization theory is suitable.

For increased complexity, *e.g.* prohibitive size network, *local search* methods and metaheuristics is needed (evolutionary and teaching-learning optimization). Optical network systems can be endowed with intelligent *agents* whose actions may have an impact on each other's decisions. This is the typical scenario where *game* theory comes into play. Advanced agents incorporating knowledge, reasoning and planning make use of shared knowledge bases keeping information of the system call for the use of cognitive optical networks.

The handling of stochastic events, a common occurrence that the agents encounter, requires that agent deal with uncertainty in a robust way, for which Bayesian networks are a useful resource. Agents can handle uncertainty using stored rules with probability information but advanced ones should learn the probability from experience, something *bayesian* learning methods can provide. Supervised, unsupervised, and reinforcement learning have been widely used in multiple combinations. A broad review of methods and applications can be found in ref. [346].

## 6.3.    Mechanical systems

As with any functional material, properties of the material itself may undergo variation when a superstructure is imprinted on the material. This change is in the origin of the metamaterials concept. Heat transfer, linear elastic constants, nonlinear viscosity, nonlinear regime induced by their high deformability, even density, and other properties may be altered by structuring. This potential for modulation, hence optimisation, triggers the search of novel or improved performances.

This search was even further invigorated by the advent and popularization of additive manufacturing that opened the way to so-called *microarchitectured* (sometimes *cellular*) materials, where a superstructure is created from the material by adding it at designated locations with (sub)micrometric resolution.

In this way, hierarchical composites offer themselves as one of the most promising strategies to improve the mechanical properties and functionalities of synthetic materials. They can be crudely separated in several classes attending to the production process (additive manufacturing of self-assembly) or the principal function they serve, most often the wave they support (electromagnetic, acoustic).

The inverse design and optimization methods follow routes parallel to those widely adopted in other fields, especially electromagnetic. All are largely based on *topology optimization* techniques,[347] initially design for mechanical structures, where the emphasis was on optimizing mechanical strength with the minimum amount material. The strength of a structure can hardly be better than that of its containing envelope yet its weight can be large reduced without much loss of the former.[a]

### 6.3.1.    Single structures

Computational data-driven approaches can substitute for experimentally guided methods in designing mechanical structures with optimized performances in different target properties.

A typical process, based on Bayesian ML starts with a phase where *experiments* to sample the input variables space (geometry of the structure, elastic constants of the materials used) are designed. Next, a nonlinear finite element analysis is carried out on selected data points from which the output database is generated. This constitutes the training data for a ML algorithm that establishes input–output relationships where optimization permits to extract optimum designs depending on the desired performance.[348]

### 6.3.2.    Metamaterials

Finite elements and conventional data-driven models usually address the optimization of structural designs rather than metamaterial designs.

DL models have acquired the power to optimize metamaterials for *maximizing* mechanical properties such as the bulk modulus, the shear modulus, or *minimizing* the Poisson's ratio to the extreme of creating negative values (auxetic materials). Being metamaterials hence homogenized composites, training data can be obtained from large sets of homogenization inverse-problem solutions with random geometrical features (constituent materials, volume fraction, architecture).[349]

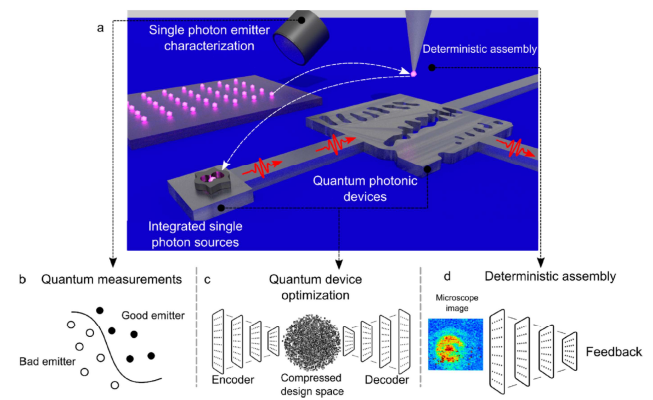

**Figure 38**. ML for quantum technology design. Adapted from Z. A. Kudyshev, V. M. Shalaev, A. Boltasseva, *ACS Photonics* **2021**, *8*, 34.

Of the features to be optimized, topology is, perhaps, the worthiest of study for the impact on performance. However, only recently full liberty in design has been made possible by the advances in additive manufacturing.[310] Gradient-free (genetic) approaches are easier and enjoy the advantage of dispensing with the computation of gradients but require relatively large amounts of data —more so considering the large design space to probe— and have slower convergence rates. In this way, by training a DL model with data generated using the energy-based homogenization method and periodic boundary conditions, 2D metamaterials topology can be non-iteratively optimized for various mechanical properties.[349]

### 6.3.3.    Spinodal composites

Customarily, metamaterials (sometimes also called cellular materials) are based on periodic cells, formed by rods and plates of various descriptions, repeated in several directions. While this simplifies design, mechanically the joints between these components give rise to stress concentration detrimental for their performance. To confront this shortcoming, metamaterials based on smooth topologies have gained pull since they avoid points of stress concentration while still showing excellent performance by leveraging the advantages of doubly curved surfaces.[350]

Spinodal topologies are natural formations resulting from self-assembly processes like in various kinds of phase separation[351] that present smooth surfaces and fulfil such requirements. Despite being non-periodic, their theoretical parametrization is surprisingly simple so that costly simulations can be neglected in favour of simple homogenization to map a rich and continuously tuneable property space. Curiously, the lack of periodicity of conventional metamaterials widens the property space and opens the ability to produce seamless functional grading composites where directionality is absent so that anisotropy can be tailored. Ingenious DNN techniques can mine this rich data quarry to produce, for instance, functionally graded materials (see **Fig. 40**) or biomimetic artificial bone architectures.[352]

### 6.3.4.    Acoustic metamaterials

Acoustic metamaterials are the accepted designation for mechanical metamaterials when focus is placed on elastic or acoustic wave response.[353] Their study usually concerns the optimization of amplitudes and frequencies of stop and pass bands in the spectra.

They have been the focus of attention for some years now and, the many unconventional properties they present —negative refraction, lensing, cloaking— show promise in technological applications like directional waveguides, mechanical filters, sub-wavelength edge detectors, invisibility cloaks, ultrasound focusers, mechanical energy propagators or subwavelength imaging devices.[354]

---

[a] Robert le Ricolais: "The art of structure is where to put the holes."

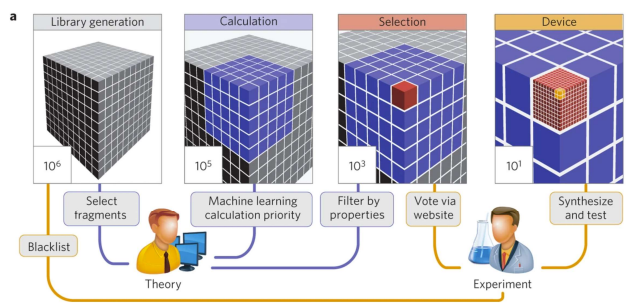

**Figure 39**. Diagram of the collaborative approach in the discovery pipeline for an OLED device. The search space decreases by over five orders of magnitude as the screening progresses. Adapted from R. Gómez-Bombarelli, J. Aguilera-Iparraguirre, T. D. Hirzel, D. Duvenaud, D. Maclaurin, M. A. Blood-Forsythe, H. S. Chae, M. Einzinger, D.-G. Ha, T. Wu, G. Markopoulos, S. Jeon, H. Kang, H. Miyazaki, M. Numata, S. Kim, W. Huang, S. I. Hong, M. Baldo, R. P. Adams, A. Aspuru-Guzik, *Nat. Mater.* **2016**, *15*, 1120.

Conventional modelling based on theoretical/numericali analysis becomes difficult (computing cost, partial derivatives of objectives functions) and effective medium methods can be helpful. Nevertheless, the latter only provides direct problem solutions, that is, predicts the properties given the metamaterial parameters. The inverse problem remains a difficult one only ML can tackles with guarantees.

In these circumstances, a Gaussian-Bayesian[355] model working in a cycle of training and prediction can make the best of a small initial training data set. This model directly creates mapping relationships between parameters and functionalities permitting to inversely adjust the parameters for a particular desired functionality without the need for physical models. It showed success in optimizing an acoustic metamaterial for low-frequency ventilated sound absorption.[356]

In acoustic band gap optimization problems the computational effort can be unsustainable and ML supported by a surrogate optimization approach can offer a respite. Suitable *surrogate* objective functions constructed applying ML techniques, produce well-performing suboptimal solutions with a smaller computational effort.[357] The training data is previously generated by evaluating the true objective function on a finite subset within the design parameter space. ML trained on these data can produce inter/extrapolated on the whole design space with little computational cost.

## 6.4. New science from AI

Data pertaining to some material or materials class can be seen as static pictures —possibly from different angles— from which human expertise or machine learned algorithms can identify the material or its class and, subsequently, infer additional properties. This kind of learning is the most familiar one where the algorithm is a mere black box that won't provide insight.

### 6.4.1. Law discovery

However, from a more popular—and possibly more demanding— point of view, artificial intelligence is expected to acquire expertise in guessing the future, the outcome of dynamical phenomena given sufficient prior information.

In more scientific terms this can be expressed by hoping that an intelligent system can guess the equations of motion and compute the evolution. Similarly, it could be expected that a properly trained intelligent agent should be capable to discover the laws governing the relationship between the set of variables describing the system at hand, that is, uncovering the underlying physical laws that, very often adopt the form of partial differential equations. It should be noted that many complex systems have so far eluded being properly described by a suitable set of variables let alone quantitative analytical modelling. Notable examples can be picked from biology (neuroscience, epidemiology, ecology), physical science and engineering (power grids) or social science (economy, finance).

A possible formulation of the problem could be finding the function $N(\cdot)$ of observed physical property, $u$, the variables, $x$, and various partial derivatives including time derivative, $u_t$, $u_x$, $u_{xx}$, etc, that provides the time evolution of the property $u$: $u_t = N(u, u_x, u_{xx}, \ldots, x, \mu)$. The objective is to find $N$ based on time series snapshots, $u(x,t)$, of the dynamical evolution at given spatial positions, $x$, with a key assumption: the function should have a small number of terms (only a few partial derivatives). For instance, for a harmonic oscillator $u_t = N = -i\mu x^2 - ihu_{xx}/2$. A method using parse regression, a library of potential functional forms, and parsimonious model selection was capable to learn both Hamiltonian and dissipative nonlinear dynamic systems ranging from periodic to chaotic even in the presence of Gaussian noise and survived severe subsampling.[358]

In the quantum physics area, formulating and testing candidate Hamiltonians for quantum systems based on experimental data is hard if the interactions at play are unknown. Thus, showing that it is possible, using unsupervised ML, to infer the Hamiltonian governing a nitrogen V centre[359] opens the development of ML agents for testing hypotheses under limited prior assumptions. This constitutes a great step towards the interpretation of large quantum systems.

### 6.4.2. Truly quantum ML

Different kinds of problems require different adapted computing approaches.

Benioff showed that all a quantum computer needed to realize the computational power of a Turing machine was reversible unitary evolution.[28] This proves that, computationally, quantum mechanics is at least as powerful as classical computing. This was suggested by Feynman too.[29] But the first to explicitly ask the question whether computing is more efficiently done using quantum mechanics was Deutsch.[31] Some problems can quickly be solved exactly by quantum computers, while classical computers can only solve them quickly with high probability and the aid of a random number generator.

Then Shor proposed the factoring algorithm but until *quantum supremacy* was declared more than twenty years elapsed.[33]

#### 6.4.2.1 Q-Classical relationship

Most AI in general and ML developments have been applied to and benefitted from classical physics but they are not restricted to it. Their extension to the quantum domain is, again, a two-way road.[360] As quantum computing develops, more and deeper applications to ML can be anticipated. Although mostly theoretical for the time being, the intersections between quantum information and ML manifests in topics such as employing quantum computing in the development of learning algorithms and the use of ML for the design and account of quantum experiments. The inherently massively *parallel* character of ML seems ideally suited to quantum information processing. The *noisy* character of the type of data for which ML and AI were developed is also just a natural ingredient of quantum information technologies.

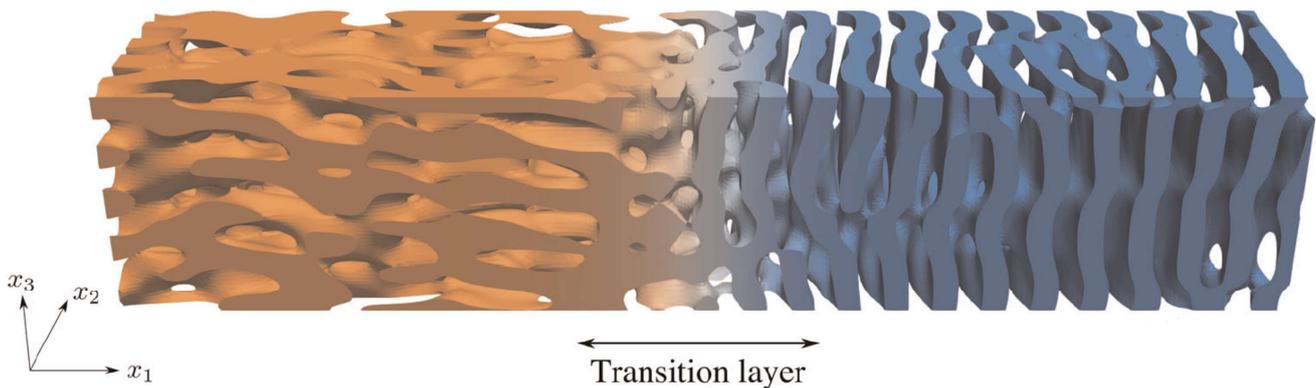

**Figure 40.** Functional grading. A spatially varying architecture transitioning between columnar and lamellar topologies optimized by ML. Adapted from S. Kumar, S. Tan, L. Zheng, D. M. Kochmann, *npj Comput. Mater.* **2020**, *6*, 1.

#### 6.4.2.2 Who's better

The characteristic that quantum systems generate atypical patterns generally absent in classical ones, invites to postulate that quantum computers may outperform classical ones in ML tasks. The field of quantum ML aims at producing quantum software that makes quantum ML the faster. Although the hardware and software challenges are still substantial, the building blocks of ML quantum algorithms are being produced.[361]

The questions that remain to answer, though, are what true advantages quantum will provide to ML even at theoretical level and how they can be implemented. The most powerful and versatile known manifestation of AI is the brain and apparently, it does not rely on quantum for its operation beyond the quantum nature of matter. Much of the answer depends on how fast quantum information technologies will actually develop beyond mere proofs of concept of quantum supremacy, nowadays subject of debate and controversy.

Current progress suggests the possibility that quantum technology can speed-up *some* ML solutions. However, their extent and, consequently, their impact on practical problems, remains an open question. Some still opine that current state of the art does not yet allow us to conclude that quantum techniques can obtain an exponential advantage.[362]

#### 6.4.2.3 Strategies

In one approach to make practical implementations, effort has been put into the building of quantum versions of ANNs. The main strategies in development to tackle with quantum ML are to find quantum computing algorithms that can replace classical ML ones for a given problem and alternatively make use of the probabilistic description underlying quantum theory in order to describe stochastic processes. Others propose to run classical ML algorithms on a quantum computer hardware.[363]

A notable example profits from the realization that quantum computing presents a surprising similarity to the kernel or expansion methods in ML. To wit, the efficient manipulation of infinite-dimensional Hilbert space in quantum computing is akin to the transformation of feature vectors into large dimensional spaces like support vector machines where data become separated in classes by a clear boundary once translated to the higher dimensional space.[364] This can open a new opportunity for the design of quantum ML algorithms where inputs are encoded in a quantum state and left for a quantum computer to analyse.

At a fundamental level, however, very few studies help answer the big question of how *learning* can be simulated in quantum systems. Thus, a quantum theory of learning is, at best, hopefully impending.

#### 6.4.2.4 Applications

The task of classification of up to eight-dimensional vectors into clusters using a small-scale photonic quantum computer has been experimental demonstrated. This demonstrates the working principle of a core mathematical routine in ML.[365]

The power of AI can also help design experiments. An expert system using ML can reveal useful in quantum experiment design. Judging from recent results, intelligent machines could have a more creative role in future research. An AI system was trained to create a variety of high dimensional multiphoton entangled states with improved efficiency. Without being demanded, the system autonomously finds experimental techniques which are just becoming standard in current quantum optical experiments.[366]

A topic busying many researchers is the fundamental issue of quantum generalizations of learning and other AI concepts. This concerns the very meaning of *learning* and *intelligence* in a world that is only properly described by quantum mechanics.[360]

### 6.5. Life sciences

A system based on AI trained using thousands of retina images can achieve physician-level sensitivity and specificity in diagnosing diabetic retinopathy. This is single example but the number of areas where these techniques are nowadays helping progress are fast growing. Initial *rule-based* approaches have led to automated interpretation of electrocardiograms, clinical practice, translational and basic biomedical research among other but are both costly to create and fragile in their application requiring frequent curation of knowledge by experts. Thus, only more modern approaches that, leveraging ML methods, better integrate complex interactions, do deserve the AI label.[367]

The aspects of biomedicine more related to advance materials where AI can have an impact concern, for instance, genome interpretation or biomarkers discovery, in the biochemical area. Other aspect, more device oriented, concerns robotic surgery, or patient monitoring and wearable devices where materials for interfacing need development. On a more fundamental key, AI can be of great help in modelling living systems and so, strengthen their understanding.

#### 6.5.1. Immune system inspiration

While AI and ML have been inspired by natural systems its power and that of networks has been deployed to model complex mechanisms such as the immune system where a different natural strategy to learn is deployed. Like the central nervous system, the immune system is capable of complex information processing. It recognizes foreign molecules, learns their structure, retains a memory of its amino acid sequence and retrieves it in future occasions.

Antibodies are molecules attached to the lymphocytes' membrane capable to detect foreign material identified as toxic and mark it for elimination, a task for lymphocytes and phagocytes.[a] Composition determines the shape and chemical reactivity and therefore makes antibodies capable to selectively *fit* complementary antigens. For an organism to survive it must have or be in a position to synthesise an antibody for any antigen it becomes exposed to. When a lymphocyte detects an antigen, it is stimulated to produce more like it and secrete antibodies in a chain reaction. In addition to antigen detecting parts (paratopes) antibodies chemical structure contains certain distinctive parts (epitopes) that can be recognised by other antibodies. The recognition not being strictly univocal opens the possibility for a series of antibodies to create a loop of recognition: paratope $p_i$ recognizes epitope $e_{i-1}$ in response to one antibody $p_1$ reacting to a foreign antigen $e_0$. Under equilibrium conditions this loop can be sustained in time once antigen with $e_0$ is eliminated thus creating a *memory* that can withstand the introduction of new types of antigens but can also be discarded over time in what can be termed *forgetting*.

Assuming a network model where antibodies are represented by chains of epitopes and paratopes with recognition rules and rewards, the balance between antigens and antibodies can be formulated mathematically in sets of equations showing the similitude of the immune system to a network with learning capability.[368] The capability to learn, store and recognise patters can be matched to machine learning models that without supervision can perform parallel computing.[369]

A good effort in creating artificial immune systems has been devoted to simulation and modelling in order to understand immune system operation. However, a fruitful reverse path has provided advances in strengthening operating systems' security for instance inspired by the ability of the immune system to distinguish *self* from alien. ML techniques like genetic algorithms, too, benefited from knowledge of the immune system but, apparently, sticking too close to natural systems without the motivation they obey can hinder the development of artificial replicas. Capitalizing on self's stationary character, originating the ability to distinguish alien, certain developments offer potential in learning data containing temporal characteristics.[370]

### 6.5.2. Connection to Neuroscience

Both fundamental, scientific insights in the nervous system operation and medical developments have come as fruits of advances in neural interface technologies. In turn, they have often come hand in hand with the development of novel materials or material platforms like implantable neural interface systems.

Materials aspects such as mechanical compliance, biocompatibility (bioinert materials) or lasting biofluid barriers are basic requirements to be met by potential materials. On the other hand, micromanufacturing is essential owing to the requirements of the tissues in which these devices will be implanted and the maturity and availability of the technologies. This is aimed at the grand challenge: the development of high-density interfaces with populations of neurons across entire tissue systems in animals, obviously including humans.

A special place among the artificial nervous cells is reserved for sensing. Although simple physical phenomena such as sound, light and, even touch, are well provided as far as detection is concerned, smell and flavour present a different kind of sensing problem (just because they respond to chemical species).

A broad overview of neural *interfaces* (materials and structures) and *systems* (high density neural interfaces, bioresorbable or long-lasting encapsulation) can be found in ref.[371]

#### 6.5.2.1 Optoelectronic synapse

Materials for neuromedicine applications should include ordinary neurons but specialized ones, such as vision ones, are urgently desired in order to enhance and eventually replace human vision in scientific and then industrial environments.

If in addition to the sensing, processing and storing capabilities could be added to the devices a great economy of means would be achieved thus accomplishing the long soughtafter integration of volatile photodetectors with nonvolatile storage devices.

A titanate-based heterojunction was demonstrated to operate as an artificial optoelectronic synapse that, through combined light and electric field modulation of the Schottky barrier, responds to visible light in a neuromorphic manner. It allows synaptic facilitation, short/long-term memory, and learning behaviour for handling optical information.[372]

#### 6.5.2.2 Artificial neurons

Tactile sensor have been realized integrating sensing, processing and memory. One of the most relevant information provided by tactile signals: texture, is harder to realize. And the hardest is providing sensors with learning capabilities.

A resistive pressure sensor, a soft ionic cable, and a synaptic transistor can be assembled into a neuromorphic tactile processing system capable of integrate and differentiate the spatiotemporal features of *touched* patterns for recognition is shown. The system comprises the sensing, transmitting, and processing components that are parallel to those found in a sensory neuron. This kind of investigation, profiting from advances in skin-like materials[373]can hand neuromorphic sensing to robotics and prosthetics.[374]

## 7. Outlook

The differences between various computing strategies, classical, AI and quantum, need to be understood and spread in order to avoid a futile competition. There seems to be a general consensus that the realm of AI is in high complexity problems, obviously not restricted to materials science, where it will subsume classical and quantum approaches.[244]

The limits of AI as analysed in the management area[375] can provide good insight for other areas such as the considerations regarding *visible* elements of the system under study. Circumstances or interactions that are not detected cannot be considered and render the model incomplete. These considerations advocate the complementarity of first principles or classical methods and AI to progress hand in hand.

In the near future, AI in general and ML in particular (with their many incarnations) need to reach the general materials science community for the many advantages they may bring for its advancement and the yet to discover new materials and systems capable to expedite the execution of ML algorithms.

Thus, prospects for AI in materials science come in the two tacitly implied areas. They involve development of materials and techniques to implement AI-running devices and the use of AI to discover new materials for other applications.

---

[a] Mammals typically dispose of on the order of $10^7$-$10^8$ different antibody types, each with its own unique chemical composition. Since the number of antigens is too large it is not possible for the organism to store recipes in its DNA to generate antibodies for any new antigen. All it can do is keep building blocks that, combined, can direct the synthesis of many useful antibodies that, like master keys, can fit entire classes of antigens. Fitting of antigens on antibodies is not strictly specific; on the contrary, a given antigen can be recognised by up to a thousand antibodies. This versatility may become adverse if it turns the immune system against itself and it's thought that a natural process eliminates those antibodies once produced.

Probably one of the most important obstacles for the future development of AI methods is availability of high-quality data. Information regarding most functional properties generated via direct experimental measurements is rather limited and the cost of generating it by different computation methods can be an expensive trade off in AI economy. Thus, online repository sharing of data, data acquisition techniques, and advanced mining methods constitutes a crucial change in research collaborative ways. At the same time, making ML algorithms smarter so that they can directly make use of available knowledge (physical laws and principles, constraints, symmetries, etc.) will redound in more efficient training with smaller datasets.

Beyond but not apart from materials science, AI has several fronts where it impacts society that concern the development of a *good AI society*. This involves both governments and private sectors, the research community and academia, and citizens at large. In the opinion of some, the general disposition of the world main political players comes short of establishing the vision and long-term strategy for such a goal.[376]

## 8. Acknowledgements

This work was possible thanks to the Spanish MICIN under projects RTI2018-093921-B-C41, SMOOTH and PID2021-124814NB-C21 (SPhAM). I wish to thank F. Mompeán, M. C. Soriano, and L. Martín-Moreno for useful suggestions and careful proofreading.

## 9. Conflict of interest

The author declares no conflict of interest.

## 10. Keywords

artificial intelligence, machine learning, neural networks, materials discovery, materials design, neuromorphic computing, artificial neuron